\documentclass[12pt]{article}

\usepackage{latexsym}
\usepackage{amssymb,amsfonts,amsmath}
\usepackage{graphicx} 
\usepackage{indentfirst}
\usepackage{bbm}
\usepackage{amssymb}
\usepackage[normalem]{ulem}
\usepackage{verbatim}
\usepackage{amsmath, amsthm, amssymb}
\usepackage{mathrsfs}
\usepackage{hyperref}
\usepackage{amsfonts}
\usepackage{dsfont}
\usepackage{cite}
\usepackage{xcolor}
\usepackage{enumitem}  

\parskip=\medskipamount

\newcommand {\cD}{{\cal D}}
\newcommand {\cE}{{\cal E}}
\newcommand {\cF}{{\cal F}}

\newcommand {\cN}{{\cal N}}

\newcommand {\cR}{{\cal R}}

\newcommand {\cU}{{\cal U}}
\newcommand {\cV}{{\cal V}}

\def\a{\alpha}

\def\b{\beta}
\def\c{\chi}
\def\d{\delta}

\def\g{\gamma}
\def\G{\Gamma}

\def\l{\lambda}
\def\m{\mu}

\def\o{\omega}

\def\q{\theta}

\def\s{\sigma}

\def\z{\zeta}
\def\D{\Delta}
\def\F{\Phi}
\def\J{\Psi}
\def\L{\Lambda}
\def\O{\Omega}

\def\S{\Sigma}
\def\U{\Upsilon}

\def\rd{{\rm d}}
\def\ri{{\rm i}}
\def\re{{\rm e}}

\newcommand{\ad}{{\dot{\alpha}}}                           
\newcommand{\bd}{{\dot{\beta}}}                            
\newcommand{\ve}{\varepsilon}                            

\renewcommand{\aa}{{\a\ad}}
\newcommand{\bb}{{\b\bd}}
\newcommand{\pa}{\partial}                           
\newcommand{\hf}{\frac12}

\newcommand{\vf}{\varphi}

\newcommand{\be}{\begin{equation}}
\newcommand{\ee}{\end{equation}}
\newcommand{\bea}{\begin{eqnarray}}
\newcommand{\eea}{\end{eqnarray}}
\newcommand{\non}{\nonumber}
\newcommand{\bm}[1]{\mbox{\boldmath$#1$}}

\def\double #1{#1{\hbox{\kern-2pt $#1$}}}

\newcommand{\gd}{{\dot\g}}
\newcommand{\dd}{{\dot\d}}

\newif\ifdtup

\newcommand{\bsubeq}{\begin{subequations}}
\newcommand{\esubeq}{\end{subequations}}

\numberwithin{equation}{section}

\newcommand{\sSU}{\mathsf{SU}}
\newcommand{\sSL}{\mathsf{SL}}

\newcommand{\sSO}{\mathsf{SO}}

\newcommand{\sU}{\mathsf{U}}

\newcommand{\mc}{\mathcal}
\newcommand{\mf}{\mathfrak}
\newcommand{\ms}{\mathscr}
\newcommand{\mb}{\mathbb}

\makeatletter
\DeclareFontFamily{OMX}{MnSymbolE}{}
\DeclareSymbolFont{MnLargeSymbols}{OMX}{MnSymbolE}{m}{n}
\SetSymbolFont{MnLargeSymbols}{bold}{OMX}{MnSymbolE}{b}{n}
\DeclareFontShape{OMX}{MnSymbolE}{m}{n}{
    <-6>  MnSymbolE5
   <6-7>  MnSymbolE6
   <7-8>  MnSymbolE7
   <8-9>  MnSymbolE8
   <9-10> MnSymbolE9
  <10-12> MnSymbolE10
  <12->   MnSymbolE12
}{}
\DeclareFontShape{OMX}{MnSymbolE}{b}{n}{
    <-6>  MnSymbolE-Bold5
   <6-7>  MnSymbolE-Bold6
   <7-8>  MnSymbolE-Bold7
   <8-9>  MnSymbolE-Bold8
   <9-10> MnSymbolE-Bold9
  <10-12> MnSymbolE-Bold10
  <12->   MnSymbolE-Bold12
}{}

\let\llangle\@undefined
\let\rrangle\@undefined
\DeclareMathDelimiter{\llangle}{\mathopen}%
                     {MnLargeSymbols}{'164}{MnLargeSymbols}{'164}
\DeclareMathDelimiter{\rrangle}{\mathclose}%
                     {MnLargeSymbols}{'171}{MnLargeSymbols}{'171}
\makeatother

\begin{document}

\begin{titlepage}
\begin{flushright}
August, 2022 \\
\end{flushright}
\vspace{5mm}

\begin{center}
{\Large \bf 
Conformal interactions between matter and \\ higher-spin (super)fields}
\end{center}

\begin{center}

{\bf Sergei M. Kuzenko, Michael Ponds and Emmanouil S. N. Raptakis} \\
\vspace{5mm}

\footnotesize{
{\it Department of Physics M013, The University of Western Australia\\
35 Stirling Highway, Perth W.A. 6009, Australia}}  
~\\
\vspace{2mm}
~\\
Email: \texttt{ 
sergei.kuzenko@uwa.edu.au, michael.ponds@uwa.edu.au, emmanouil.raptakis@research.uwa.edu.au}\\
\vspace{2mm}

\end{center}

\begin{abstract}
\baselineskip=14pt

In even spacetime dimensions, the interacting bosonic conformal higher-spin (CHS) theory can be realised as an induced action. The main ingredient in this definition is the model $\mathcal{S}[\varphi,h]$ describing a complex scalar field $\varphi$ coupled to an infinite set of background CHS fields $h$, with $\mathcal{S}[\varphi,h]$ possessing a non-abelian gauge symmetry. Two characteristic features of the perturbative constructions of $\mathcal{S}[\varphi , h]$ given in the literature are: (i) the background spacetime is flat; and (ii) conformal invariance is not manifest. In the present paper we provide a new derivation of this action in four dimensions such that (i) $\mathcal{S}[\varphi , h]$ is defined on an arbitrary conformally-flat background; and (ii)  the background conformal symmetry is manifestly realised. Next, our results are extended to the $\mathcal{N}=1$ supersymmetric case. Specifically, we construct, for the first time, a model $\mathcal{S}[\Phi, H]$ for a conformal scalar/chiral multiplet $\Phi$ coupled to an infinite set of background higher-spin superfields $H$. Our action possesses a non-abelian gauge symmetry which naturally generalises the linearised gauge transformations of conformal half-integer superspin multiplets. The other fundamental features of this model are: (i) $\mathcal{S}[\Phi, H]$ is defined on an arbitrary conformally-flat superspace background; and (ii) the background $\mathcal{N}=1$ superconformal symmetry is manifest. Making use of $\mathcal{S}[\Phi, H]$, an interacting superconformal higher-spin theory can be defined as an induced action.

\end{abstract}
\vspace{5mm}

\vfill

\vfill
\end{titlepage}

\newpage
\renewcommand{\thefootnote}{\arabic{footnote}}
\setcounter{footnote}{0}

\tableofcontents{}
\vspace{1cm}
\bigskip\hrule

\allowdisplaybreaks

\section{Introduction}

The origin of conformal higher-spin (CHS) theory goes back 
to the 1985 review article by Fradkin and Tseytlin on the classical and quantum aspects of conformal supergravity \cite{FT}.
These authors proposed free gauge-invariant actions  for CHS gauge fields (spin $s>2$) in Minkowski space ${\mathbb M}^4$ as a generalisation of the well-known models for massless spin-1 gauge field, conformal gravitino (spin 3/2) and conformal graviton (spin 2). Cubic couplings for CHS fields of all integer spins $s\geq 2$ were constructed by Fradkin and Linetsky in 1989 \cite{FL}, and
 a year later  their results were extended 
 to the superconformal case \cite{FL-4D}. The Fradkin-Linetsky approach was based on gauging the infinite-dimensional CHS algebra 
 $\mathfrak{hsc}^\infty (4) $ introduced in \cite{FL-algebras}  and its superconformal extension  $\mathfrak{shsc}^\infty (4|1) $ also constructed in \cite{FL-algebras}.\footnote{The CHS superalgebras discovered in \cite{FL-algebras} provide a natural extension of 
 the  anti-de Sitter (AdS) higher-spin superalgebras pioneered by Fradkin and Vasiliev \cite{FV1,FV2,Vasiliev88}.} 
 In three dimensions, the gauging of the $\cN$-extended CHS superalgebra $\mathfrak{shsc} (\cN | 3) $ 
(being isomorphic to the four-dimensional AdS higher-spin superalgebra 
$\mathfrak{shs}(\cN|4)$ \cite{Vasiliev88})
naturally leads to the superconformal higher-spin (SCHS) theory in closed form \cite {FL-3D} since the corresponding action is of the Chern-Simons type.\footnote{The infinite-dimensional 
 CHS superalgebra $\mathfrak{shsc} (\cN | 3) $ contains the three-dimensional $\cN$-extended superconformal algebra $\mathfrak{osp}(\cN|4; {\mathbb R})$ as a maximal finite-dimensional subalgebra. The gauge theory of $\mathfrak{osp}(\cN|4; {\mathbb R})$ is the theory of $\cN$-extended conformal supergravity developed in \cite{vanN85} for $\cN=1$, in \cite{RvanN86} for $\cN=2$ and in 
 \cite{FL-3D,LR89} for $\cN>2$. The gauging of $\mathfrak{osp}(\cN|4; {\mathbb R})$ in superspace was carried out in \cite{BKNT-M1}, which has resulted in the three-dimensional superconformal tensor calculus and off-shell actions for conformal supergravity \cite{BKNT-M2}. }
 The gauging of the bosonic higher-spin algebra $\mathfrak{hsc} ( 3)  \equiv \mathfrak{shsc} (0 | 3) $ 
was independently carried out by Pope and Townsend \cite{PopeTownsend} who also proposed the Chern-Simons action for this algebra as a generalisation of three-dimensional conformal 
gravity.\footnote{Conformal gravity in three-dimensions as a gauge theory was also studied in 
\cite{vanN85, Horne:1988jf,BKNT-M1}.} 

Despite being fully successful in three dimensions  \cite {FL-3D}, so far the Fradkin-Linetsky geometric formalism in four dimensions \cite{FL-4D} has not been extended beyond the cubic approximation.  
An alternative approach was put forward by Segal in 2002 \cite{Segal} 
(soon after the prowerful
proposal by Tseytlin \cite{Tseytlin}) who constructed the unique bosonic gauge theory of interacting symmetric traceless tensor fields of all ranks
in a spacetime of any even dimension $d\geq 4$. Nowadays, Segal's theory is often referred to as CHS gravity, see e.g. \cite{BBJ,Bekaert:2022poo}.
Perhaps the simplest definition of CHS gravity in even dimensions is as an {\it induced action}; that is, as  the logarithmically divergent part of the effective action associated with the model $\mc{S}[\varphi, h]$ of a conformal complex scalar field $\varphi$ coupled to background CHS fields $h$ \cite{Segal, Tseytlin, BJM}. 
Explicit computations of lowest-order perturbative corrections to the CHS gravity action have been given in \cite{Segal,BJM,Bonezzi}. Such calculations are performed about a flat background spacetime.

Diffeomorphisms and Weyl transformations belong to the gauge group of CHS gravity. 
Therefore, the choice of a flat background spacetime implies the presence of a corresponding background conformal symmetry. 
However, in the framework advocated in \cite{Segal,BJM}, the conformal symmetry of the model $\mc{S}[\varphi,h]$, which is the central ingredient in the induced action approach, is hidden. 
One of the goals of our paper is to provide a new formalism in which background conformal symmetry is manifest at all stages, and employ it to construct $\mc{S}[\varphi,h]$. Such a formalism would be useful, for example, when computing the induced action, since it can then be recast as a problem of calculating correlation functions involving primary currents, allowing one to use standard CFT techniques. 

In their gauging of the CHS superalgebra $\mathfrak{shsc}^\infty (4|1) $, Fradkin and Linetsky \cite{FL} did not identify off-shell $\cN=1$ superconformal analogs of the free CHS models \cite{FT}. Free $\cN=1$ SCHS gauge theories 
were constructed relatively recently in Minkowski superspace \cite{KMT},
and also in arbitrary conformally-flat backgrounds \cite{KP, Kuzenko:2020jie}.\footnote{Recently these results have been generalised to $\cN>1$ in \cite{Kuzenko:2021pqm}, and to $\cN=1$ higher-depth fields in \cite{Kuzenko:2019eni}.}
The second  goal of this paper is to develop a superfield formalism that is required to define  a unique gauge theory of interacting $\cN=1$ superconformal multiplets $H^{\a(s)\ad(s)}$ of all half-integer superspins $(s+\hf)$, where $s=0,1,2, \dots$, (or $\cN=1$ CHS supergravity) as an induced action.\footnote{At the component level, the SCHS gauge prepotential $H^{\a(s)\ad(s)}$ contains CHS fields with spin $\{s+1, s+1/2, s\}$. This supermultiplet is said to carry superspin-$(s+1/2)$. } 
In particular, we construct the model $\mc{S}[\F,H]$ describing a chiral multiplet $\F$ coupled to an infinite tower of background gauge prepotentials  $H^{\a(s)\ad(s)}$, such that $\mc{S}[\F,H]$ possesses a non-abelian gauge symmetry.

This paper is organised as follows. 
Some elementary facts regarding the (super)field Noether procedure are discussed in section \ref{NPapp}, which serves to establish some notations and set the scene for the main results. 
In subsections \ref{secCSLinVar} and \ref{CSsec1}, the Noether procedure is employed to construct cubic interactions between a conformal scalar field and a tower of bosonic CHS fields, consistent to first order in the latter. In subsections \ref{CSsec2} and \ref{CSsec3} we utilise these results, in tandem with the methods pioneered in \cite{Segal}, to promote the CHS fields to background fields in a manifestly conformally-covariant fashion. Finally, in subsection \ref{secCSrigid} we provide a detailed discussion on the rigid symmetries of the free matter action.
The formalism developed in section \ref{secCS} is generalised to $\mc{N}=1$ superspace in section \ref{masslesschiral} and applied to the case of massless chiral matter interacting with half-integer superspin SCHS gauge multiplets. 
In particular, we start by constructing cubic interactions consistent to first order in the gauge superfields, and then promote the latter to background superfields for the first time. 
In section \ref{section4} we investigate the problem of constructing superconformal cubic interactions between $\mc{N}=2$ matter and integer superspin $\mc{N}=1$ SCHS gauge multiplets.  A concluding discussion is given in section \ref{secDiscuss}. 

The main body is accompanied by six appendices. Appendix \ref{AppCSS} provides a brief review on the geometry of conformal (super)gravity. 
In appendix \ref{AppRigid} we highlight the similarities between the Noether and Segal procedures for coupling matter to gauge fields.
The purpose of appendix \ref{Appendix A} is to show the equivalence between two methods of constructing conformally-covariant differential operators, one of which is heavily used in sections \ref{secCS} and \ref{masslesschiral}. In appendix \ref{AppendixAdjoint} we define the Hermitian adjoint of a differential operator, which requires special care in the supersymmetric case.  In appendix \ref{Appendix B} we show that gauge transformations of the (S)CHS fields in the proposed formalism are a non-abelian extension of their corresponding free gauge symmetry. Lastly, in appendix \ref{AppE} we present new higher-spin conformal supercurrents which may be useful for future work.

 We note that throughout this entire paper we make use of the convention that indices denoted by the same symbol are to be symmetrized over, e.g. 
\begin{align}
U_{\a(m)} V_{\a(n)} = U_{(\a_1 . . .\a_m} V_{\a_{m+1} . . . \a_{m+n})} =\frac{1}{(m+n)!}\big(U_{\a_1 . . .\a_m} V_{\a_{m+1} . . . \a_{m+n}}+\cdots\big)~, \label{convention}
\end{align}
with a similar convention for dotted spinorial indices. 


\section{Noether procedure for matter coupled to (S)CHS fields}\label{NPapp}

In this section we recall some basic elements of the Noether procedure as applied to the construction of consistent interactions between (super)conformal matter and an infinite tower of (S)CHS fields. The former is denoted $\Sigma^I$ and the latter is collectively denoted $\mf{h}^A = \{\mf{h}^{\a(s)\ad(s)}\,|\,s\geq 0\}$, where $\mf{h}^{\a(s)\ad(s)}=\mf{h}^{(\a_1\dots\a_s)(\ad_1\dots\ad_s)}$ could represent either an integer spin-$s$ conformal field $h^{\a(s)\ad(s)}$ or a half-integer superspin-$(s+\frac12)$ conformal superfield $H^{\a(s)\ad(s)}$. 
The following discussion is easily generalised to their Grassmann-odd counterparts. 

There are two main approaches to the Noether procedure encountered in the literature, each with slightly different but complimentary set-ups and philosophies. First we outline the approach advocated in this work, which is similar in spirit to e.g. \cite{BBvD,BJM, BJM2}, then we briefly comment on the second. Since we are dealing with conformal interactions, below we make use of conformal (super)space, restricting ourselves to conformally-flat background (super)spaces. 

The starting point of the standard Noether procedure is the supposition that there exists a complete non-linear action $\mc{S}[\Sigma^I,\mf{h}^A]$ describing the aforementioned interactions, whose power series expansion in the gauge fields begins as follows
\begin{align}
\mc{S}[\Sigma^I,\mf{h}^A] = \mc{S}_0[\Sigma^I]+g\mc{S}_1[\Sigma^I,\mf{h}^A]+g^2\mc{S}_2[\Sigma^I,\mf{h}^A]+\mc{O}(\mf{h}^3)~. \label{NPappNLact}
\end{align}
Each summand is bilinear in the matter field and the corresponding power of $\mf{h}^A$ coincides with the associated power of the book-keeping parameter $g$. In particular, $\mc{S}_0[\Sigma^I]$ is the free matter action and $\mc{S}_1[\Sigma^I,\mf{h}^A]$ is the cubic Noether coupling, which may always be written in the form
\begin{align}
\mc{S}_1[\Sigma^I,\mf{h}^A] = c_A\int \text{d}\mu\, \mf{h}^A\mf{J}_A~. \label{NPappCubicAct}
\end{align}
Here $\text{d}\mu$ is the appropriate (super)space measure, $c_A$ are some undetermined coefficients and the composite tensor $\mf{J}_A$ dual to $\mf{h}^A$ is bilinear in $\Sigma^I$. 

By assumption, action  $\mc{S}[\Sigma^I,\mf{h}^A]$ is invariant under: (i) (super)conformal transformations of the background (super)space; and (ii) some non-linear gauge transformations $\delta_{\xi} \Sigma^I$ and $\delta_{\xi} \mf{h}^A$. 
The former means that $\Sigma^I$ is a primary (super)field with weight $\Delta_{\Sigma^I}$ and $\sU(1)_R$ charge $q_{\Sigma^I}$, and similarly for $\mf{h}^A$.\footnote{This is not necessarily the case; one may have conformal invariance without the fields being primary. This is the case e.g. in Metsaev's ordinary derivative formulation of the free bosonic CHS actions \cite{Metsaev2, Metsaev}. } Within the framework of conformal superspace described in Appendix \ref{CSSappendix},
these statements read
\begin{subequations} \label{NPappSCprop}
\begin{align}
K^B\Sigma^I&=0~,\qquad \mathbb{D} \Sigma^I = \Delta_{\Sigma^I}\Sigma^I~,\qquad \mathbb{Y}\Sigma^I = q_{\Sigma^I}\Sigma^I~, \label{NPappSCpropa}\\
 K^B\mf{h}^A&=0~,\qquad \mathbb{D} \mf{h}^A = \Delta_{\mf{h}^A}\mf{h}^A~,\qquad \mathbb{Y} \mf{h}^A = q_{\mf{h}^A}\mf{h}^A~.\label{NPappSCpropb}
\end{align}
\end{subequations}
Here $K^{B}=(K^b, S^\b ,\bar S_\bd)$ are the generators of special conformal and $S$-supersymmetry transformations, whilst $\mathbb{D}$ and $\mathbb{Y}$ are the dilatation and $\sU(1)_R$ generators.
In the non-supersymmetric case, the generators $K^B$ 
should be replaced with $K^b$, and the generator $\mathbb Y$ is absent.
The weights and charges in \eqref{NPappSCprop} are model dependent and are such that \eqref{NPappNLact} has  conformal weight zero and $\sU(1)_R$ charge zero. 

The gauge transformations $\delta_{\xi} \Sigma^I$ and $\delta_{\xi}\mf{h}^A$ are parameterised by gauge parameters collectively denoted  $\xi^A$, and may also be expanded in a power series as follows
\begin{subequations}\label{NPappGT}
\begin{align}
\delta_{\xi} \Sigma^I &= \delta^0_{\xi}\Sigma^I+g\delta^1_{\xi}\Sigma^I+g^2\delta^2_{\xi}\Sigma^I+\mc{O}(\Sigma^3)~, \label{NPappGTa}\\
\delta_{\xi} \mf{h}^A&=\delta^0_{\xi}\mf{h}^A+g\delta^1_{\xi}\mf{h}^A+g^2\delta^2_{\xi}\mf{h}^A+\mc{O}(\mf{h}^3) ~.\label{NPappGTb}
\end{align}
\end{subequations}
Each summand in \eqref{NPappGT} is linear in $\xi^A$ and the superscript $i$ in $\delta_{\xi}^i$ indicates the corresponding power of the field being varied. In this work, the (super)conformal matter under consideration does not have a zeroth-order gauge variation, and we hereby set $\delta^0_{\xi}\Sigma^I=0$. 
All terms in \eqref{NPappGT} must preserve the properties \eqref{NPappSCprop},\footnote{In the supersymmetric case there may be other off-shell constraints (e.g. chirality) which must be preserved. } which places strong constraints on their possible structure. For instance, we will see that the linear variation $\delta^1_{\xi}\Sigma^I$, which may always be written as
\begin{align}
\delta^1_{\xi}\Sigma^I = -\hat{\mc{U}}^I_{\xi}{}_J\Sigma^J \label{NPappLinVarMatter}
\end{align}
for some differential operator $\hat{\mc{U}}^I_{\xi}{}_J$ parameterised by $\xi^A$, is completely fixed by \eqref{NPappSCpropa}. 

That \eqref{NPappNLact} is invariant under \eqref{NPappGT} leads to consistency conditions at each order in $g$. For example,  the first non-trivial constraint, which occurs at order $g^1$, is
\begin{align}
0=\frac{\delta\mc{S}_0}{\delta \Sigma^I}\delta_{\xi}^{1}\Sigma^I+\frac{\delta\mc{S}_1}{\delta \mf{h}^A}\delta_{\xi}^{0}\mf{h}^A~.\label{NPappConsistency}
\end{align}
The Noether procedure refers to the process of rebuilding the full action \eqref{NPappNLact} order-by-order using invariance under the gauge symmetry \eqref{NPappGT}, and the consistency conditions which result, as the guiding principle. 

Reconstruction of the cubic sector \eqref{NPappCubicAct} is simple, and begins with the observation that if $\Sigma^I$ satisfy the free equations of motion, $\frac{\delta\mc{S}_0}{\delta \Sigma^I} \approx 0$, then constraint \eqref{NPappConsistency} becomes
\begin{align}
0\approx\frac{\delta\mc{S}_1}{\delta \mf{h}^A}\delta_{\xi}^{0}\mf{h}^A~. \label{NPappConsistencyOS}
\end{align}
Here `$\approx$' represents equality modulo the free equations of motion for $\Sigma^I$ (which we also refer to as being `on the mass shell'). This is equivalent to a conservation equation on $\mf{J}_A$. 

Hence, the first step is to construct conserved (super)currents $\mf{J}_A$ and couple them to the dual gauge fields $\mf{h}^A$ via \eqref{NPappCubicAct}. Clearly, the action $\mc{S}_1$ is invariant under $\delta_{\xi}^{0}\mf{h}^A$ on the mass shell. To lift the gauge symmetry off the mass shell, one deduces the linear transformation of the matter field \eqref{NPappLinVarMatter}, and uses eq. \eqref{NPappConsistency} to determine the coefficients $c_A$ in \eqref{NPappCubicAct}. It follows that $\mc{S}_0 + \mc{S}_1$ is invariant under \eqref{NPappGT} to first order in the gauge fields,
\begin{align}
\delta_{\xi}\big(\mc{S}_0 + \mc{S}_1\big)= \mc{O}(\mf{h})~. \label{NPappCubicInv}
\end{align}

In this approach the rigid symmetries of $\mc{S}_0$ do not play an important role, but may be easily deduced from this analysis. Specifically, from \eqref{NPappCubicInv} it is clear that if the gauge fields are set to zero then $\delta_{\xi}^1\mc{S}_{0}=-\delta_{\xi}^0\mc{S}_{1}$ and hence $\delta_{\xi}^1\mc{S}_{0}=0$ if $\delta_{\xi}^0\mf{h}^{A}=0$. Parameters $ \xi^A_0$ satisfying $\delta_{\xi_0}^0\mf{h}^{A}=0$ are known as reducibility parameters (see e.g. \cite{BekaertRigid}). As discussed in the proceeding sections, the latter are closely related to (super)conformal Killing tensors of the background (super)space.

Upon determining the cubic-sector, one then advances the process to the next order in gauge fields (i.e. quartic-order in all fields), where non-trivial issues arise. 
For instance, a priori, only the free (zeroth-order) gauge transformations in \eqref{NPappGT} are known. Determining higher-order deformations to the free transformations is a tricky part of the procedure, since they must be  algebraically consistent  and form a closed gauge algebra modulo trivial symmetries,\footnote{Given a functional $\mc{S}[\U^I]$ of some fields $\U^I$, a transformation $\delta \U^I$ is said to be a trivial symmetry if it may be expressed in the form $\delta \U^I=A^{IJ}\frac{\delta \mc{S}}{\delta \U^J}$ (here we are using DeWitt's condensed notation). This is possible if and only if $\delta\mc{S}=0$ and $\delta\U^I\approx 0$ (see e.g. \cite{HenneauxT}). Here $A^{IJ}$ is some differential operator satisfying $A^{IJ}=-A^{JI}$.
 } see e.g. \cite{BBvD}. This will be a focal point of our analysis in sections \ref{secCS} and \ref{masslesschiral}, and will motivate our transition to the formulation suggested by Segal \cite{Segal}.

Finally, though it is not actively used in this work, it is instructive to briefly recall the salient features of the second approach to the Noether procedure.  
 Here, one starts with the free matter action, deduces all of its rigid symmetries and attempts to gauge them. 
 The method is best illustrated by an example, for which we use the complex scalar field $\Sigma^I=(\bar{\varphi},\varphi)$ with action \eqref{CSact}. In this case the rigid symmetries $\delta_{\xi_0}^1\varphi=-\hat{\mc{U}}_{\xi_0}\varphi$ are parametrised by conformal Killing tensors $\xi^{\a(s-1)\ad(s-1)}_0$ which satisfy $\nabla^{\a\ad}\xi_0^{\a(s-1)\ad(s-1)}=0$. 
To gauge these symmetries, the conformal Killing condition is relaxed and $\xi_0$ is replaced with unconstrained $\xi$. One then finds that $\delta_{\xi}\mc{S}_0 = -\sum_s\int \text{d}^4x \, e \, \nabla^{\a\ad}\xi^{\a(s-1)\ad(s-1)}j_{\a(s)\ad(s)}$ for some tensor $j_{\a(s)\ad(s)}$ bilinear in $\Sigma^I$. This contribution may be compensated for by introducing dual gauge fields $h^{\a(s)\ad(s)}$ which transform as $\delta_{\xi}^0h^{\a(s)\ad(s)}=\nabla^{\a\ad}\xi^{\a(s-1)\ad(s-1)}$ and which couple to $\Sigma^I$ according to \eqref{NPappCubicAct} with $c_A=1$ (the currents have the correct normalisation by construction). This new coupling will result in higher-order non-zero contributions which should be compensated for by deforming the gauge transformations of $h^{\a(s)\ad(s)}$ and including a quartic sector in the action, and so on. At this point the above approach converges with the other. 
For a more general discussion on this approach and its relation to Segal's method \cite{Segal}, see appendix \ref{AppRigid}. 
 We also refer the reader to the article  \cite{BekaertRigid} for a detailed account of rigid symmetries, and a discussion on gauging them.

The second Noether approach is commonly found throughout the supersymmetric literature, see e.g. \cite{VanN81}, 
 and is useful when, a priori, it is not known which gauge fields the matter can couple to, or the (super)currents $\mf{J}_A$ are unknown. 
 For example, such ideas have been employed to determine the structure of 
 the Weyl multiplets for conformal supergravity in diverse dimensions
 by using the method of supercurrent multiplets \cite{HST,OS,BdRdW}.
 It is well known that the multiplets of conformal currents $j_A$ furnish off-shell representations of conformal supersymmetry. To determine the universal structure of $j_A$, one considers a simple free superconformal theory and works out the multiplet of the energy-momentum tensor.
 Once $j_A$ is known,
 the associated gauge multiplet of fields $\U^A$ (the Weyl multiplet) is determined via the Noether coupling $S_{\text{NC}} = \int \text{d}\mu\, \U^A j_A$.
This procedure is concisely described by Bergshoeff et {\it al.} \cite{BdRdW}.
In superspace, the Noether procedure was used, e.g., in \cite{Osborn,MSW,K10} 
for ordinary supersymmetric theories, and more recently in \cite{Koutrolikos1,Koutrolikos2,Buchbinder:2018wzq,Koutrolikos3,Gates:2019cnl} and \cite{BIZ2} for $\cN=1$ and $\cN=2$ supersymmetric higher-spin theories in Minkowski superspace respectively.
Refs. \cite{Koutrolikos1,Koutrolikos2,Buchbinder:2018wzq,Koutrolikos3,Gates:2019cnl}  made used of the off-shell massless higher-spin gauge multiplets introduced in \cite{KSP,KS}, see \cite{BK} for a review.


\section{Complex scalar field coupled to bosonic CHS fields} \label{secCS}

The theory of an infinite tower of interacting CHS fields is most easily defined as an induced action \cite{Segal, Tseytlin, BJM}. More specifically, given the action $\mc{S}[\varphi,h]$ for a conformal complex scalar field $\varphi$ coupled to background 
higher-spin fields $h^A$, invariant under gauge transformations $\delta\varphi$ and $\delta h^A$, an effective action $ \Gamma[h]$ may be defined according to
\begin{align}
\re^{\ri \Gamma[h]}=\int\mc{D}\varphi\mc{D}\bar{\varphi}\, \re^{\ri \mc{S}[\varphi,h] }~. \label{EffAct}
\end{align}
As $\mc{S}[\varphi,h]$ is bilinear in $\varphi$, the latter may be integrated out in \eqref{EffAct}, and the CHS action may be identified with the logarithmically divergent part of $\Gamma[h]$, which is local and invariant under the gauge transformations $\delta h^A$.\footnote{The invariance of the logarithmically divergent part can be seen
explicitly if one uses, for instance, dimensional regularisation, see e.g. \cite{Buchbinder} for details.}

Clearly, the core ingredient in this definition\footnote{In the original work \cite{Segal}, Segal put forward another definition for CHS theory, which is often referred to as the `formal operator' approach, see e.g. \cite{BBJ} for a summary. }  of CHS theory is the action $\mc{S}[\varphi,h]$. A solution to the problem of consistently coupling $\varphi$ to background bosonic CHS fields was first proposed in \cite{Segal}, and elucidated in \cite{BJM} (for a discussion regarding Majorana fermion matter, see \cite{GrSk}). However, in these approaches 
the conformal symmetry, which should be present, is either not discussed or, at the very least, not manifest.\footnote{To our knowledge, conformal invariance was discussed only in \cite{Segal}, and proven only at the linearised level.} This is partly due to a formulation in terms of the so called `dressed' CHS fields, which are non-primary.

In this section we provide a new formalism to construct $\mc{S}[\varphi,h]$ such that conformal symmetry is manifest at all stages. 
Our strategy is to first make use of the Noether procedure, where $h^A$ are treated as infinitesimal sources, and then  promote them to background fields by transitioning to the approach suggested by Segal. To maintain control of the conformal symmetry we (i) deal directly with only primary, or `undressed', CHS fields (see appendix \ref{Appendix B} for the relation between (un)dressed CHS fields in our context); and (ii) employ the framework of conformal (super)space (see appendix \ref{AppCSS} for a summary). Simultaneously, the latter point  allows us to complete the program on all conformally-flat backgrounds in a covariant fashion.\footnote{The approaches \cite{Segal, BJM} begin with the free conformal scalar in Minkowski space. Once it is successfully put on a background of higher-spin fields, one can interpret the resulting model as living in a non-trivial gravitational field perturbed about a flat background (since the conformal graviton resides within the tower). In our case $\mc{S}[\varphi,h]$ may be interpreted as living in a gravitational field perturbed about a conformally-flat background.  } 

We recall the action for a free massless complex scalar field $\varphi$ propagating on a general gravitational background
\begin{align}	
	\mc{S}_{0}[\varphi]= \int\text{d}^{4}x \, e \,  \bar{\varphi}\, \Box \, \varphi~, \qquad \Box := \nabla^a \nabla_a ~.\label{CSact}
\end{align}
Here $e:=\text{det}(e_m{}^{a})$ and $\nabla_a$ is the background conformally covariant derivative. The latter encodes curvature dependent terms which only become apparent after `degauging', see appendix \ref{AppCSS} for more details on this process. In particular, upon degauging, the above action becomes
\begin{align}
\mc{S}_{0}[\varphi]= \int\text{d}^{4}x \, e \,  \bar{\varphi}\, \big(\mc{D}^a\mc{D}_a+\frac{1}{6}\mc{R}\big)\, \varphi ~,
\end{align}
where $\mc{D}_a$ is the usual (torsion-free) background Lorentz covariant derivative with $\mc{R}$ its associated Ricci scalar.
 For the remainder of this work we make exclusive use of $\nabla_a$, and do not comment on the corresponding degauged expressions.

Varying this action with respect to $\bar{\varphi}$ yields the equation of motion
\begin{align}
	\Box \varphi \approx 0~.
	\label{33}
\end{align}
Action \eqref{CSact} is known to be conformal if $\varphi$ is a primary field with weight $1$
\begin{align}
	K_a \varphi = 0~, \qquad \mb{D}\varphi=\varphi, \label{phiprop}
\end{align}
where $K_a$ and $\mathbb{D}$ are the special conformal and dilatation generators, respectively.

\subsection{Differential operators on the space of primary scalar fields} \label{secCSLinVar}

As a first step in building interactions involving $\varphi$, we analyse the possible transformation rules consistent with its kinematic properties. 
In accordance with the discussion in section \ref{NPapp}, we start by proposing the following linear transformation rule for $\vf$
\begin{align}
	\delta \varphi=-\hat{\mc{U}}\varphi~,\qquad \hat{\mc{U}}=\sum_{s=0}^{\infty}\hat{\mc{U}}^{(s)}, \label{CSgt}
\end{align} 
where $\hat{\mc{U}}^{(s)}$ is a differential operator of maximal order $s$ 
\begin{align}
	\hat{\mc{U}}^{(s)}=\sum_{k=0}^{s}\zeta_{(s)}^{\a(k)\ad(k)}\nabla^k_{\a\ad} ~. \label{GenOpphi}
\end{align}
At this stage, the coefficients $\zeta_{(s)}^{\a(k)\ad(k)}$ are complex. Operator $\hat{\mc{U}}$ must preserve the off-shell properties \eqref{phiprop} of $\vf$, and hence so too must each individual $\hat{\mc{U}}^{(s)}$. As we now show, this implies that all coefficients in \eqref{GenOpphi} may be expressed in terms of a single parameter, which may ultimately be identified with the gauge parameter associated with the spin-$(s+1)$ CHS field. 

It is clear that $\hat{\mc{U}}^{(s)} \varphi$ must have weight 1, which implies
\begin{subequations}\label{CSSymm}
\begin{align}
	\mb{D}\z_{(s)}^{\a(k)\ad(k)}&=- k \z_{(s)}^{\a(k)\ad(k)}~. \label{CSSymm1}
\end{align} 
Further, we require that the transformed field is primary, $K_{\bb}\phantom{.}\hat{\mc{U}}^{(s)}\varphi=0$, which implies that the coefficient of lowest weight (highest rank) is a primary weight $-s$ field, 
\begin{align}
K_\bb \z_{(s)}^{\a(s) \ad(s)} = 0~,\qquad \mathbb{D} \z_{(s)}^{\a(s) \ad(s)}=-s\z_{(s)}^{\a(s) \ad(s)}~,\label{TopCofPrim}
\end{align}
and the remaining coefficients satisfy
\begin{align}
 K_{\b \dot{\b}} \z_{(s)}^{\a(k) \ad(k)}  = 4(k+1)^2 \z_{(s)}^{\a(k)}{}_{\b\bd}{}^{\ad(k)}~,  \qquad 0 \leq k \leq s-1 ~. \label{CSSymm2}
\end{align}
\end{subequations}
The unique ansatz compatible with constraints \eqref{CSSymm} is
\begin{subequations} \label{l=0subsector}
\begin{align}
	\label{ansatzCSSymm}
	\z_{(s)}^{\a(k) \ad(k)} = a_{(s,k)} \nabla_{\bb}^{s-k} \z_{(s)}^{\a(k) \b(s-k) \ad(k) \bd(s-k)} ~, \qquad 0 \leq k \leq s-1 ~,
\end{align}
for undetermined coefficients $a_{(s,k)}$. We see that $\z_{(s)}^{\a(s) \ad(s)}\equiv \z^{\a(s)\ad(s)}$ is the only independent parameter left. The coefficients $a_{(s,k)}$ may be readily computed by inserting \eqref{ansatzCSSymm} into \eqref{CSSymm2},
\begin{align}
a_{(s,k)}=\ri^s \binom{s}{k}^2\binom{2s+2}{s-k}^{-1} ~, \label{CScoeff}
\end{align}
\end{subequations}
which leads to the operator
\begin{align}
	\hat{\mc{U}}^{(s)} &= \ri^s \sum_{k=0}^{s}\binom{s}{k}^2\binom{2s+2}{s-k}^{-1} \nabla_{\bb}^{s-k} \z^{\a(k) \b(s-k) \ad(k) \bd(s-k)} \nabla_\aa^k~. \label{CSOp}
\end{align}

The operator \eqref{CSOp} is the unique operator (up to overall normalisation) of the functional form \eqref{GenOpphi}  which preserves all off-shell properties of $\varphi$. However, it is not the most general operator with this property, which would include terms involving powers of the d'Alembertian $\Box$ in the ansatz \eqref{GenOpphi}. 
In the next section we will see that it is possible to achieve gauge invariance to order $\mc{O}(h)$ without such terms (see eq. \eqref{CSCubic}). However, they turn out to be necessary to close the gauge algebra, which is the subject of section \ref{CSsec2}.

\subsection{Noether procedure to order $\mc{O}(h)$} \label{CSsec1}

From $\varphi$ and its conjugate $\bar{\varphi}$ one may construct the rank $s\geq 0$ composite tensors
\begin{align}
	j_{\a(s)\ad(s)}=\ri^s\sum_{k=0}^{s}(-1)^k\binom{s}{k}^2\nabla_{\a\ad}^k\varphi\nabla_{\a\ad}^{s-k}\bar{\varphi}~  \label{CScurrent}
\end{align}
It may be shown that $j_{\a(s)\ad(s)}$ possesses the following crucial features:
\begin{enumerate}[label=(\roman*)]
	\item Reality
	\begin{subequations}
		\begin{align}
			j_{\a(s)\ad(s)}=\bar{j}_{\a(s)\ad(s)}~;
		\end{align}
		\item Conformal covariance on arbitrary gravitational backgrounds
		\begin{align}
			K_\bb j_{\a(s)\ad(s)} = 0~, \qquad \mb{D}j_{\a(s)\ad(s)}=(s+2)j_{\a(s)\ad(s)}~; \label{CScurrentprop}
		\end{align}
		\item Transverse on-shell when restricted to conformally-flat backgrounds (for $s>0$)
		\begin{align}
			C_{abcd}=0\qquad \implies \qquad \nabla^{\b \bd}j_{\b\a(s-1) \bd \ad(s-1)}\approx 0~. \label{CSTOS}
		\end{align}
	\end{subequations}
\end{enumerate}
To the best of our knowledge, the conformal currents \eqref{CScurrent} in Minkowski space were introduced for the first time by Craigie, Dobrev and Todorov \cite{CDT}, although the method to derive these currents had been described earlier by Makeenko 
\cite{Makeenko}.\footnote{The fermionic CHS currents were first constructed by Migdal 
\cite{Migdal}.}
The conformal currents \eqref{CScurrent} in ${\mathbb M}^4$ differ from those used in 
\cite{BJM}, as will be discussed in the conclusion.

The currents \eqref{CScurrent} with $s=1$ and $s=2$ are actually conserved on-shell on arbitrary gravitational backgrounds. However, it was shown in \cite{BeccariaT}  (building on the analysis 
in \cite{GrigorievT})
that the Weyl tensor $C_{abcd}$ is the obstruction for this property to hold in the case $s=3$. This is also true for spin $s> 3$, and hence for the remainder of this section we restrict ourselves to conformally-flat backgrounds. 

The conformal currents $j_{\a(s)\ad(s)}$ naturally couple to the integer spin-$s$ conformal gauge fields $h_{\a(s)\ad(s)}$ via the cubic Noether coupling\footnote{We should mention that at the bottom of the tower of CHS fields there sits a scalar field $h$. This is a primary weight-two field  (i.e. has the properties of an auxiliary field) without a zeroth-order gauge transformation, and is not typically considered a CHS field. Nevertheless, it plays a crucial role in the full non-linear model, where it acquires a first-order non-abelian gauge transformation. In some cases, when making contact with external models (such as scalar QED or scalar coupled to conformal gravity), it may be identified as a composite field depending non-linearly on the CHS fields with $s\geq 1$. See e.g. \cite{Segal, BJM} for more details. }
\begin{align}	
	\mc{S}_{1}[\varphi,h]=\sum_{s=0}^{\infty}c_s\,\mc{S}^{(s)}_{1}[\varphi,h_{(s)}]~,\qquad \mc{S}^{(s)}_{1}[\varphi,h_{(s)}]=\int \text{d}^{4}x \, e \, h^{\a(s)\ad(s)}j_{\a(s)\ad(s)}~,\label{CSNoether}
\end{align}
for undetermined real coefficients $c_s$.
The real field $h_{\a(s)\ad(s)}$ has the conformal properties
\begin{align}
	K_\bb h_{\a(s)\ad(s)}=0~,\qquad \mb{D}h_{\a(s)\ad(s)}=(2-s)h_{\a(s)\ad(s)}~.\label{CHSprop}
\end{align}
By virtue of the above properties, the cubic action
\begin{align}
	\mc{S}_{\text{cubic}}[\varphi,h]=\mc{S}_{0}[\varphi]+\mc{S}_{1}[\varphi,h] \label{CSCubicact}
\end{align}
is clearly conformal.\footnote{Owing to property \eqref{CScurrentprop}, action \eqref{CSCubicact} is actually conformal on arbitrary backgrounds.}
Furthermore, on account of \eqref{CSTOS}, the action \eqref{CSNoether} is invariant under the zeroth-order gauge transformations
\begin{align}
	\delta_{\ell} h_{\a(s)\ad(s)}&= \nabla_{\aa}\ell_{\a(s-1)\ad(s-1)} ~,\qquad s\geq 1~,\label{CHSgt}
\end{align}
provided $\varphi$ is on-shell $\delta_{\ell}\mc{S}_1\approx 0$. Here $\ell_{\a(s-1)\ad(s-1)}$ is a real and unconstrained primary field,
\begin{align}
K_{\b \bd} \ell_{\a(s-1)\ad(s-1)}=0~,\qquad \mathbb{D} \ell_{\a(s-1)\ad(s-1)} = (1-s)\ell_{\a(s-1)\ad(s-1)}~.\label{CHSgtParProp}
\end{align}

In order to elevate the gauge symmetry \eqref{CHSgt} off the mass shell, it is necessary to endow $\varphi$ with its own transformation rule. For this we use the rule \eqref{CSgt} with \eqref{CSOp} derived in the previous subsection, and impose the reality condition $\bar{\z}_{\a(s) \ad(s)} = - \z_{\a(s) \ad(s)}$. Then, the properties \eqref{TopCofPrim} and \eqref{CHSgtParProp} allow us to make the field identification  $\z^{\a(s)\ad(s)}\equiv \ri \ell^{\a(s)\ad(s)}$, where $\ell_{\a(s)\ad(s)}$ is the gauge parameter associated with $h_{\a(s+1) \ad(s+1)}$.
Now we may fix the coefficients $c_s$ in \eqref{CSNoether} by requiring the cubic action \eqref{CSCubicact} to be invariant up to terms linear in gauge fields,
\begin{align}
c_s=\frac{(-1)^{s}}{2} \binom{2s}{s}^{-1} \qquad \implies \qquad	\delta_{\ell}\mc{S}_{\text{cubic}}[\varphi,h] = \mc{O}\big(h\big)~. \label{CSCubic}
\end{align} 
It should be emphasised that this statement holds off-shell.


\subsection{Closing the gauge algebra}\label{CSsec2}

As it stands, the cubic coupling of $\vf$ to the CHS fields $h_{(s)}$ described in the previous section is actually not consistent at order $\mc{O}(h)$. This can be seen from the fact that the gauge transformations \eqref{CSgt} with \eqref{CSOp} do not form a closed algebra (modulo trivial symmetries), as their commutator involves terms proportional to the operator $\Box$.
Therefore, to ensure that the algebra of gauge transformations is closed, one must include terms in $\hat{\mc{U}}$ involving all powers of $\Box$. The introduction of such terms must be done in a manner consistent with the conformal properties of $\vf$, which is a non-trivial task. To maintain control of the conformal symmetry, there are two (related) ways to proceed: (i) employ the same procedure that allowed us to deduce \eqref{CSOp}; or (ii) make use of a conformal compensator to build primary operators. 
Below we make use of approach (ii), whilst in appendix \ref{Appendix A} we use the first method and elaborate on the relation between the two. 

Let us try to build primary operators from those already at our disposal. The operator $\hat{\mc{U}}$, given in \eqref{CSgt} with \eqref{CSOp}, takes a weight one primary scalar field to a weight one primary scalar field. The operator $\Box$ takes a weight one primary scalar field to a weight three primary scalar field. In order to form their composition, we introduce a nowhere-vanishing conformal compensator field $\eta(x)$ with the properties
\begin{align}
 \bar{\eta}=\eta~,\qquad \mb{D}\eta=-2\eta~, \qquad K_{\a\ad}\eta=0~. \label{CScompprop}
\end{align}
It follows that the order $s$ operator $\hat{\mc{U}}^{(s-2)} \eta \phantom{.} \Box$ takes a weight one primary scalar field to a weight one primary scalar field. Continuing this process, we can build primary operators involving powers of $\Box$ according to\footnote{When there is a chance of ambiguity, we place an arrow over a derivative to indicate that it acts on everything to the right. For example, we have $(\eta\overrightarrow{\square})^2=\eta^2\Box^2+\eta(\Box\eta)\Box +2\eta(\nabla^a\eta)\nabla_a\Box$.}
\begin{subequations}\label{genOpCS}
\begin{align}
\hat{\mc{V}}^{(s)}:=\sum_{l=0}^{\lfloor s/2 \rfloor}\hat{\mc{U}}^{(s-2l)}_{[s,l]}\big(\eta\overrightarrow{\square}\big)^l~, \label{genopform1}
\end{align}
where we have used the notation
\begin{align}
\hat{\mc{U}}^{(s-2l)}_{[s,l]}:= \sum_{k=0}^{s-2l}a_{(s-2l,k)} \nabla_{\aa}^{s-2l-k} \z_{[s,l]}^{\a(s-2l) \ad(s-2l)} \nabla_\aa^k~, \label{newGPop}
\end{align}
\end{subequations}
with $a_{(s,k)}$ given by \eqref{CScoeff}.
The superscript $(s-2l)$ in $\hat{\mc{U}}^{(s-2l)}_{[s,l]}$ indicates the operator order and is determined by the two subscript numbers $[s,l]$.
The operator $\hat{\mc{U}}^{(s-2l)}_{[s,l]}$ coincides with $\hat{\mc{U}}^{(s-2l)}$ in \eqref{CSOp} except we have labelled the new gauge parameters $\z_{[s,l]}^{\a(s-2l) \ad(s-2l)}$, which parametrise the terms in \eqref{genopform1} proportional to $\eta^l$, by an integer $l$ with $0\leq l \leq \lfloor s/2 \rfloor$. These gauge parameters have the conformal properties
\begin{align}
K_{\b\bd}\z_{[s,l]}^{\a(s-2l) \ad(s-2l)}=0~,\qquad \mb{D}\z_{[s,l]}^{\a(s-2l) \ad(s-2l)}=-(s-2l)\z_{[s,l]}^{\a(s-2l) \ad(s-2l)}~. \label{ConfpropCSgp}
\end{align}
The original gauge parameters in \eqref{CSOp} correspond to those with $l=0$, $\z_{[s,0]}^{\a(s)\ad(s)} \equiv  \z^{\a(s)\ad(s)}$.

Using $\hat{\mc{V}}^{(s)}$ we propose the following gauge transformation for the matter field $\vf$
\begin{align}
\delta \vf = -\hat{\mc{V}}\vf~, \qquad \hat{\mc{V}}:= \sum_{s=0}^{\infty}\hat{\mc{V}}^{(s)}\vf~.\label{genvarCS}
\end{align}
By construction the operator $\hat{\mc{V}}^{(s)}$ is consistent with the conformal properties of $\vf$
\begin{align}
\mb{D} \hat{\mc{V}}^{(s)}\vf= \hat{\mc{V}}^{(s)}\vf ~,\qquad K_{\a\ad}\hat{\mc{V}}^{(s)}\vf=0~,
\end{align}
and hence so too is the variation \eqref{genvarCS}.

It is well known \cite{Deser,Zumino} that every gravity-matter system can be realised as a coupling of conformal gravity to the original matter fields and a conformal compensator $\c(x) $, which is a nowhere vanishing primary scalar field of non-zero weight $\D_\c \neq 0$. If the resulting action explicitly depends on the compensator, then the original gravity-matter theory is not conformal. In other words, the conformal symmetry is spontaneously broken. This is not the case in our approach. The compensator $\eta$ is just a book-keeping device, which is introduced in order to simplify the explicit form of the higher-derivative operators. The compensator  $\eta$ may be absorbed into the coefficients of the operator \eqref{genOpCS}, see appendix \ref{Appendix A}.

\subsection{Inclusion of auxiliary gauge fields} \label{CSsec3}

In the previous subsection we constructed the most general linear transformation rule for $\vf$ which included the operators $ \Box^l$ that were necessary to close  the gauge algebra on $\vf$.
With these new structures came new gauge parameters $\z_{[s,l]}^{\a(s-2l)\ad(s-2l)}$ (the old ones correspond to $l=0$ whilst the new ones correspond to $1\leq l \leq \lfloor s/2 \rfloor$) which parametrised them. 
This in turn means that we must introduce new gauge fields\footnote{These auxiliary gauge fields correspond to the trace parts of the undressed CHS fields in \cite{Segal, BJM}.} $h_{[s,l]}^{\a(s-2l)\ad(s-2l)}$ into the model that are associated with the new gauge parameters and find how they transform under them. 
These gauge fields enter the action at the cubic level and the currents they couple to will involve $\Box$ operators, in contrast to the currents \eqref{CScurrent} which do not.

To formulate such structures in a conformally covariant way, we make use of the conformal compensator method introduced in the previous section. We start with the standard observation that after integrating by parts, the action \eqref{CSCubicact} may be rewritten in the form\footnote{Here we have absorbed the constants $c_s$ into the definition of $h_{\a(s)\ad(s)}$, which means the gauge transformations \eqref{CHSgt} will be scaled accordingly, see \eqref{CHSdeformedGT}. } 
\begin{align}
\mc{S}_{\text{cubic}}[\vf,h]=\int \text{d}^{4} x \, e \, \bar{\vf}\phantom{.}\hat{H}\phantom{.}\vf~,\qquad \hat{H}=\Box+\sum_{s=0}^{\infty}\hat{H}^{(s)}~, \label{IBPCSact}
\end{align}
where $\hat{H}^{(s)}$ is the following operator of maximal order $s$
\begin{align}
\hat{H}^{(s)}=\sum_{k=0}^{s}b_{(s,k)}\nabla^{s-k}_{\a\ad}h^{\a(s)\ad(s)}\nabla_{\a\ad}^k~,\qquad b_{(s,k)}=(-\ri)^s\binom{s}{k}\binom{s+k}{k}~. \label{CScubicOp}
\end{align}
The latter takes a weight one primary scalar field to a weight three primary scalar field,
\begin{align}
\mb{D} \hat{H}^{(s)} \vf = 3\hat{H}^{(s)} \vf~,\qquad K_{\b\bd}\hat{H}^{(s)}\vf =0~.  \label{HopProp}
\end{align}
In addition $\hat{H}^{(s)}$ is Hermitian, $(\hat{H}^{(s)})^{\dagger}=\hat{H}^{(s)}$,  which follows from the identity
\begin{align}
\sum_{p=k}^{s}(-1)^p\binom{p}{k}b_{(s,p)}=(b_{(s,k)})^* ~.
\end{align}
Consequently $\hat{H}$ is also Hermitian, $\hat{H}^{\dagger}=\hat{H}$, which reflects the reality of the action \eqref{IBPCSact}.
In appendix \ref{AppendixAdjoint} we discuss what is meant by the Hermitian adjoint of an operator (which can be a subtle issue, particularly in the supersymmetric case). The adjoint $\hat{H}^{\dagger}$ of $\hat{H}$ should be computed with respect to definition \eqref{CSadjointA}.
It is important to emphasise that $\hat{H}^{(s)}$ is actually the unique operator of the functional form \eqref{CScubicOp} satisfying the properties \eqref{HopProp} and $\hat{H}^{\dagger}=\hat{H}$. Hence it may be derived independently and without knowledge of the Noether coupling \eqref{CSNoether}. 

Using $\hat{H}^{(s)}$ and the conformal compensator $\eta$, we can build the order $s$ Hermitian operator
\begin{align}
\hat{H}^{(s-2)}\eta\square + \overrightarrow{\square}\eta\hat{H}^{(s-2)} \non
\end{align}
which takes a weight one primary scalar field to a weight three primary scalar field. Continuing this procedure for higher powers in $\Box$, one arrives at the following Hermitian operator\footnote{In principle, one could instead consider the operator $\hat{G}^{(s)}:=\sum_{l=0}^{\lfloor s/2 \rfloor} \sum_{q=0}^{l}g_{(s,l,q)}\big(\overrightarrow{\square}\eta\big)^q\hat{H}^{(s-2l)}_{[s,l]}\big(\eta\overrightarrow{\square}\big)^{l-q}$. For Hermiticity, the constants should satisfy $\big(g_{(s,l,l-q)}\big)^*=g_{(s,l,q)}$, but are otherwise arbitrary. However, such an operator is equivalent to \eqref{Hboxop} in a sense similar to the equivalence of operators discussed in appendix \ref{Appendix A}. }
\begin{subequations}\label{RandomEquationLabel}
\begin{align}
\hat{G}^{(s)}:=\frac{1}{2}\sum_{l=0}^{\lfloor s/2 \rfloor}\Big[\hat{H}^{(s-2l)}_{[s,l]}\big(\eta\overrightarrow{\square}\big)^{l}+ \big(\overrightarrow{\square}\eta\big)^l\hat{H}^{(s-2l)}_{[s,l]}\Big]~,\qquad \big(\hat{G}^{(s)}\big)^{\dagger}=\hat{G}^{(s)}\qquad  \label{Hboxop}
\end{align}
where we have used the notation\footnote{The superscript $(s-2l)$ in $\hat{H}^{(s-2l)}_{[s,l]}$ indicates the operator order and is determined by the two subscript numbers $[s,l]$ which label the gauge fields.}
\begin{align}
\hat{H}_{[s,l]}^{(s-2l)}:=\sum_{k=0}^{s-2l}b_{(s-2l,k)}\nabla^{s-2l-k}_{\a\ad}h_{[s,l]}^{\a(s-2l)\ad(s-2l)}\nabla_{\a\ad}^k~,\qquad \big(\hat{H}^{(s-2l)}_{[s,l]}\big)^{\dagger}=\hat{H}^{(s-2l)}_{[s,l]}~,
\end{align}
\end{subequations}
with $b_{(s,k)}$ given in \eqref{CScubicOp}. From this we can define the manifestly real and conformal action
\begin{align}
\mathcal{S}[\vf, h]=\int \text{d}^{4} x \, e \, \bar{\vf}\phantom{.}\hat{G}\phantom{.}\vf~,\qquad \hat{G}=\Box+\sum_{s=0}^{\infty}\hat{G}^{(s)}~, \label{CSgenact}
\end{align}
which is equal to the original cubic action \eqref{CSCubicact} plus terms involving the new gauge fields. In this form, the action \eqref{CSgenact} is manifestly invariant under the gauge transformations
\begin{align}
\delta \vf = -\hat{\mc{V}}\vf~,\qquad \delta \hat{G} = \hat{\mc{V}}^{\dagger}\hat{G}+\hat{G}\hat{\mc{V}}~, \label{CSNLGT}
\end{align}
where $\hat{\mc{V}}$ is given by \eqref{genvarCS} and  \eqref{genOpCS}. The  adjoint operators in \eqref{RandomEquationLabel} are to be computed with respect to definition \eqref{CSadjointA}, whilst $\hat{\mc{V}}^{\dagger}$ in \eqref{CSNLGT} should be computed with respect to \eqref{CSadjointB}. 

A few comments are in order. 
The operator $\hat{H}^{(s-2l)}_{[s,l]}$ coincides with $\hat{H}^{(s-2l)}$ in \eqref{CScubicOp} except we have labelled the new gauge fields $h_{[s,l]}^{\a(s-2l)\ad(s-2l)}$, which parametrise the terms in \eqref{Hboxop} proportional to $\eta^l$, by an integer $l$ with $0\leq l \leq \lfloor s/2 \rfloor$. The gauge fields $h_{[s,l]}^{\a(s-2l)\ad(s-2l)}$ are real and have the conformal properties
\begin{align}
K_{\b\bd}h_{[s,l]}^{\a(s-2l)\ad(s-2l)}=0~,\qquad \mb{D}h_{[s,l]}^{\a(s-2l)\ad(s-2l)}=(2-s+2l)h_{[s,l]}^{\a(s-2l)\ad(s-2l)}~.
\end{align}
The original CHS gauge fields correspond to those with $l=0$, $h_{[s,0]}^{\a(s)\ad(s)} \equiv  h^{\a(s)\ad(s)}$.

The gauge variation of $h_{\a(s)\ad(s)}$ may be extracted from \eqref{CSNLGT}. In appendix \ref{Appendix B} we show that at the lowest order in gauge fields the free gauge transformations 
\begin{subequations}\label{CHSdeformedGT}
\begin{align}
\delta h_{\a(s)\ad(s)}&= \nabla_{\a\ad}\ell_{\a(s-1)\ad(s-1)}+\mc{O}(h)~,\label{CHSdeformedGTa}\\
\ell_{\a(s-1)\ad(s-1)}&\equiv c_s\ri \big(\z_{\a(s-1)\ad(s-1)}-\bar{\z}_{\a(s-1)\ad(s-1)}\big)~,\label{CHSdeformedGTb}
\end{align}
\end{subequations}
are recovered. Here $\z^{\a(s-1)\ad(s-1)}\equiv \z_{[s-1,0]}^{\a(s-1)\ad(s-1)}$ are the original gauge parameters which parametrise $\hat{\mc{U}}^{(s-1)}_{[s-1,0]}$ and the constants $c_s$ are given in \eqref{CSCubic}. Importantly, we note that the variation \eqref{CSNLGT} has translated into non-abelian gauge transformations of $h_{\a(s)\ad(s)}$ represented by the $\mc{O}(h)$ in \eqref{CHSdeformedGTa} (whose explicit form we do not determine here). The variation \eqref{CSNLGT} also encodes non-linear gauge transformations of the extra gauge fields $h_{[s,l]}^{\a(s-2l)\ad(s-2l)}$, which  may in fact be shown to begin algebraically:
\begin{align}
\delta h_{[s,l]}^{\a(s-2l)\ad(s-2l)} \propto \eta^{-1}\big(\z_{[s-2,l-1]}^{\a(s-2l)\ad(s-2l)}+\bar{\z}_{[s-2,l-1]}^{\a(s-2l)\ad(s-2l)}\big)+\cdots~,\qquad 1\leq l \leq \lfloor s/2 \rfloor~. \label{CSalgebraic}
\end{align}
It follows that the real part of the gauge parameters may be used to impose the gauge condition
\begin{align}
h_{[s,l]}^{\a(s-2l)\ad(s-2l)}=0~, \qquad s\geq 0 ~\text{ and }~ 1\leq l \leq \lfloor s/2 \rfloor~.\label{CSGfix}
\end{align}
Then, as suggested by \eqref{CHSdeformedGTb}, the only remaining gauge parameters are the imaginary parts of $\z^{\a(s-1)\ad(s-1)}$ (the real part is used to gauge away $h_{[s+1,1]}^{\a(s-1)\ad(s-1)}$). 

Hence the gauge fields $h_{[s,l]}^{\a(s-2l)\ad(s-2l)}$  with $l>0$ are auxiliary, and in the gauge \eqref{CSGfix} only the original ones $h^{\a(s)\ad(s)}$ with $l=0$ remain. 
In order to preserve the gauge \eqref{CSGfix}, it is necessary to supplement the variation \eqref{CSNLGT}, and hence the variation of both $\vf$ and $h^{\a(s)\ad(s)}$, with an $h^{\a(s)\ad(s)}$ and $\ell^{\a(s-1)\ad(s-1)}$ dependent gauge transformation. However, the corresponding expressions are complicated and we do not attempt to provide them in this work.  

Finally, we would like to point out that after integrating by parts, action \eqref{CSgenact} becomes
\begin{subequations}
\begin{align}
\mathcal{S}[\vf,h]&=\int \text{d}^{4} x \, e \, \Big\{ \bar{\vf}\Box\vf + \sum_{s=0}^{\infty}\sum_{l=0}^{\lfloor s/2 \rfloor} h_{[s,l]}^{\a(s-2l)\ad(s-2l)}j_{\a(s-2l)\ad(s-2l)}^{[s,l]} \Big\}~,\label{GenCura}\\
j_{\a(s-2l)\ad(s-2l)}^{[s,l]}&= \frac{1}{2}\sum_{k=0}^{s-2l}d_{(s-2l,k)}\Big(\nabla_{\a\ad}^{s-2l-k}\bar{\vf}\nabla_{\a\ad}^k\big(\eta\overrightarrow{\square}\big)^{l}\vf + \nabla_{\a\ad}^{k}\vf \nabla_{\a\ad}^{s-2l-k}\big(\eta\overrightarrow{\square}\big)^{l}\bar{\vf}\Big)~, \label{GenCur}
\end{align}
\end{subequations}
with $d_{(s,k)}:=\ri^{s}(-1)^k\binom{s}{k}^2$. The current $j_{\a(s-2l)\ad(s-2l)}^{[s,l]}$ with $l=0$ coincides with \eqref{CScurrent}. Thus we see that through this procedure we have introduced a coupling between the auxiliary gauge fields and trivial currents (trivial in the sense that they vanish on the free equations of motion, and are hence trivially conserved on-shell).

\subsection{Rigid symmetries of the free matter action} \label{secCSrigid}

In accordance with the general discussion in section \ref{NPapp}, the reducibility parameters of the gauge field $h^{\a(s)\ad(s)}$ define rigid symmetries of the free action. 
Specifically, from \eqref{CSCubic} one may deduce that
$\delta_{\ell}\mc{S}_{0}=-\delta_{\ell}\mc{S}_{1}\big|_{h=0}$,
from which it follows that 
\begin{align}
	\nabla_\aa \ell_{\a(s-1) \ad(s-1)} = 0 \quad \iff \quad \d_\ell h_{\a(s) \ad(s)} = 0 \quad \implies \quad \delta_{\ell}\mc{S}_{0}=0~. \label{CSRigidSym}
\end{align}
For $s> 1$, this identifies $\ell_{\a(s-1) \ad(s-1)}$ as a conformal Killing tensor (CKT). For $s=1$ it implies $\ell$ is constant. Therefore, the transformations \eqref{CSgt} with \eqref{CSOp} parameterised by CKTs constitute rigid symmetries of the free action, and consequently of its equation of motion \cite{ShSh, Eastwood}.

Rather than deduce the rigid symmetries of the free action from knowledge of the cubic coupling \eqref{CSCubic}, one may instead arrive at the same conclusion via direct computation. Specifically, let us suppose that the matter field transforms according to the rule \eqref{CSgt}, 
\begin{align}
\delta \vf = -\hat{\mc{U}}^{(s)}\vf~,\qquad \hat{\mc{U}}^{(s)} = \sum_{k=0}^{s}a_{(s,k)} \nabla_{\aa}^{s-k} \z^{\a(s) \ad(s)} \nabla_\aa^k~,\label{CSrigidgt}
\end{align}
with $a_{(s,k)}$ given in \eqref{CScoeff}.  The corresponding variation of the free action is 
\begin{align}
\delta \mc{S}_{0}[\vf,\bar{\vf}] = -\int \text{d}^4x \, e \, \bar{\vf} \Big\{ \big(\hat{\mc{U}}^{(s)}\big)^{\dagger}\Box + \Box\hat{\mc{U}}^{(s)} \Big\}\vf~, \label{CSrigidVar}
\end{align}
where the adjoint (with respect to \eqref{CSadjointB}) of  $\hat{\mc{U}}^{(s)}$ is 
\begin{align}
\big(\hat{\mc{U}}^{(s)}\big)^{\dagger} = \sum_{k=0}^{s}a_{(s,k)}\frac{(s+1)(s+2)}{(k+1)(k+2)} \nabla_{\aa}^{s-k} \bar{\z}^{\a(s) \ad(s)} \nabla_\aa^k~.
\end{align}
If the parameter $\z^{\a(s) \ad(s)}$ is split into two real fields according to $\z^{\a(s) \ad(s)}= R^{\a(s)\ad(s)} +\ri I^{\a(s)\ad(s)} $, then one may show that \eqref{CSrigidVar} reduces to
\begin{align}
\delta \mc{S}_{0}[\vf,\bar{\vf}] &= -\int \text{d}^4x \, e \, \bar{\vf} \sum_{k=0}^{s}a_{(s,k)}\Big\{ 2\frac{(s+1)(s+2)}{(k+1)(k+2)}\nabla_{\a\ad}^{s-k}R^{\a(s)\ad(s)}\nabla_{\a\ad}^k\Box\notag\\
&\phantom{=} +\Big[\Box\nabla_{\b\bd}^{s-k}R^{\a(k)\b(s-k)\ad(k)\bd(s-k)}\nabla_{\a\ad}^k-\nabla_{\b\bd}^{s-k}\nabla^{\a\ad}R^{\a(k)\b(s-k)\ad(k)\bd(s-k)}\nabla_{\a\ad}^{k+1} \notag\\
&\phantom{=}+\ri\Box\nabla_{\b\bd}^{s-k}I^{\a(k)\b(s-k)\ad(k)\bd(s-k)}\nabla_{\a\ad}^k-\ri\nabla_{\b\bd}^{s-k}\nabla^{\a\ad}I^{\a(k)\b(s-k)\ad(k)\bd(s-k)}\nabla_{\a\ad}^{k+1}\Big]\Big\} \vf~.\label{RigidVarCS}
\end{align}
The first term in this expression vanishes only if $R^{\a(s)\ad(s)}=0$, after which we may write
\begin{align}
\delta \mc{S}_{0}[\vf,\bar{\vf}] = \int \text{d}^4x \, e \, \bar{\vf} &\Big\{ a_{(s,s)}\nabla^{\a\ad}I^{\a(s)\ad(s)}\nabla_{\a\ad}^{s+1} -\ri a_{(s,0)}\Box\nabla_{\b\bd}^sI^{\b(s)\bd(s)}\non\\
 &-\ri\sum_{k=0}^{s-1}a_{(s,s-k-1)}\Big[\frac{(k+1)(2s-k+2)}{(s-k)^2}\Box \nabla_{\b\bd}^{k}I^{\a(s-k)\b(k)\ad(s-k)\bd(k)}\non\\
&-\nabla_{\b\bd}^{k+1}\nabla^{\a\ad}I^{\a(s-k-1)\b(k+1)\ad(s-k-1)\bd(k+1)}\Big] \nabla_{\a\ad}^{s-k}\Big\}\vf~.
\end{align}
The first term in this expression vanishes only if $\nabla^{\a\ad}I^{\a(s)\ad(s)}=0$, or in other words  if $I^{\a(s)\ad(s)}$ is a CKT. For any CKT $I^{\a(s)\ad(s)}$, the following set of identities hold
\begin{align}
\Box \nabla_{\b\bd}^{k}I^{\a(s-k)\b(k)\ad(s-k)\bd(k)} = \frac{(s-k)^2}{(k+1)(2s-k+2)}\nabla_{\b\bd}^{k+1}\nabla^{\a\ad}I^{\a(s-k-1)\b(k+1)\ad(s-k-1)\bd(k+1)}~
\end{align}
for $0\leq k \leq s$, which may be proved via induction on $k$. It follows that $\delta \mc{S}_{0}= 0$, and hence the transformation \eqref{CSrigidgt}, parametrised by  $\z^{\a(s) \ad(s)}= R^{\a(s)\ad(s)} +\ri I^{\a(s)\ad(s)} $, is a rigid symmetry of the free action only if $R^{\a(s)\ad(s)}=0$ and $I^{\a(s)\ad(s)}$ is a CKT. 
If we denote the corresponding operator by $\hat{\mc{U}}_{\text{ckt}}^{(s)}$, then from \eqref{CSrigidVar} we have proved that
\begin{align}
\big(\hat{\mc{U}}^{(s)}_{\text{ckt}}\big)^{\dagger}\Box + \Box\hat{\mc{U}}_{\text{ckt}}^{(s)}  =0~.\label{CSrigidID}
\end{align}

We now briefly comment on the trivial symmetries of the free matter action. Consider the transformation
$ \delta_{\l} \vf = \l \Box \vf$ where $\l$ is a primary complex field with conformal weight $-2$. Splitting $\l$ into two real fields,  $\l=\eta+\ri \o$, it is easy to see that if $\eta=0$ then this transformation is a trivial symmetry of $\mc{S}_0$. Similarly, any transformation of the form $ \delta_{\o} \vf = \ri \big(\o \overrightarrow{\square})^l \vf$, for arbitrary integer $l>0$, is also a trivial symmetry of $\mc{S}_0$. It is well known that the commutator of any trivial symmetry with a rigid symmetry is again a trivial symmetry (see e.g. \cite{BekaertRigid}). In the current case this means that order-$s$ transformations of the type
\begin{align}\label{CStrivckt}
\delta \vf = \big[~\hat{\mc{U}}_{\text{ckt}}^{(s-2l)}, ~\ri\big(\o \overrightarrow{\square})^l ~ \big] \vf~,\qquad 1\leq l \leq \lfloor s/2 \rfloor~,
\end{align}
 are trivial symmetries. This fact may be verified directly by employing the relation \eqref{CSrigidID}.

Next we comment on the algebraic structure of the rigid symmetries described above. As discussed in appendix \ref{AppRigid}, they form a Lie algebra with respect to the commutator. Specifically, given two (non-trivial) rigid symmetry operators $\hat{\mathcal{U}}^{(m)}_{\text{ckt}}$ and $\hat{\mathcal{U}}^{(n)}_{\text{ckt}}$ parametrised by the CKTs $\z^{\a(m)\ad(m)}$ and $\xi^{\a(n)\ad(n)}$, their commutator $\big[\hat{\mathcal{U}}^{(m)}_{\text{ckt}},\,\hat{\mathcal{U}}^{(n)}_{\text{ckt}}\big]$ also defines a rigid symmetry of the matter action. It is expected that this new symmetry is parametrised by several CKTs, with the highest rank parameter defined via the even Schouten-Nijenhuis bracket, see e.g. \cite{HL1,KR19}
\begin{align}
[ \z , \xi ]_{\text{SN}}^{\a(m+n-1) \ad(m+n-1)} &:=  m \z^{\a(m-1)\b \ad(m-1)\bd} \nabla_{\b \bd } \xi^{\a(n) \ad(n)}
 - n \xi^{\a(n-1)\b\ad(n-1)\bd} \nabla_{\b \bd } \z^{\a(m) \ad(m)} ~.
\label{SN1}
\end{align}
The algebra of (non-trivial) rigid symmetries of $\mc{S}_0$, identified with the quotient algebra \eqref{Quotient}, corresponds \cite{Segal} to the conformal higher-spin algebra $\mathfrak{hsc}^\infty (4) $ constructed in \cite{FL-algebras}. See also, e.g., \cite{BJM,BBJ,JNT} for discussions on the global symmetry algebra.

To conclude, it is pertinent to comment on the symmetries of the equation of motion \eqref{33}, 
which were classified in 1992 by Shapovalov and Shirokov \cite{ShSh} and ten years later by Eastwood \cite{Eastwood}. Consider an operator $\hat \cV$ acting on the space of primary weight-1 scalar fields. It is a symmetry of the conformal d'Alembertian if 
$\Box \hat{\cV} \vf_0 =0$ for every solution $\vf_0$  
of \eqref{33}.
It is said to be a trivial symmetry of the conformal d'Alembertian if $\hat{\cV} = \hat\cF \Box$ for some operator $\hat\cF$.
If $\hat \cV^{(s)}$ is an operator of order-$s$, 
then it has the form \eqref{genOpCS}. The only non-trivial contribution in  \eqref{genOpCS}
corresponds to $l=0$, and thus $\hat \cV^{(s)} = \hat \cU^{(s)}$. 
The operator $ \hat \cU^{(s)}$ is a symmetry of the conformal d'Alembertian if it is of the type $\hat{\mc{U}}_{\text{ckt}}^{(s)}$, see e.g. \cite{KLRT} for a direct proof.


\section{Scalar multiplet coupled to half-integer superspin SCHS multiplets } \label{masslesschiral}

In this section we elaborate on superconformal cubic interactions between massless chiral matter and half-integer superspin-($s+\frac12$) conformal superfields $H_{\a(s)\ad(s)}$. We then describe, for the first time, how to promote the latter to fully fledged background fields with a non-abelian extension of their free gauge transformations. Our models will be formulated in $\mathcal{N}=1$ conformal superspace, see appendix \ref{CSSappendix} for a brief review.

The action for a conformal chiral scalar superfield $\Phi$ propagating on a generic supergravity  background is
\begin{align}	
\mc{S}_{0}[\Phi]
=\int\text{d}^{4|4}z \, E \,  \bar{\Phi}\Phi~,\qquad \bar{\nabla}_{\ad}\Phi=0~.\label{WZact}
\end{align}
The corresponding equation of motion for $\Phi$ is
\begin{align}
\nabla^2\Phi \approx 0~,\qquad \nabla^2:=\nabla^{\a}\nabla_{\a}~. \label{ChiralEoM}
\end{align}
Action \eqref{WZact} is invariant under the superconformal gauge group if $\Phi$ is a primary superfield with superconformal weight 1 and $\sU(1)_R$ charge $-2/3$,
\begin{align}
K_A\Phi = 0~, \qquad \mb{D}\Phi=\Phi, \qquad \mb{Y}\Phi=-\frac{2}{3}\Phi~. \label{Phiprop}
\end{align}

\subsection{Differential operators on the space of primary chiral superfields}

As a first step in building interactions involving $\Phi$, we analyse the possible transformation rules consistent with its kinematic properties. 
We propose the linear transformation
\begin{align}
\delta\Phi=-\hat{\mc{U}}\Phi~,\qquad \hat{\mc{U}}=\sum_{s=0}^{\infty}\hat{\mc{U}}^{(s)}, \label{Mattergt}
\end{align} 
where $\hat{\mc{U}}^{(s)}$ is a differential operator of maximal order $s$ 
\begin{align}
\hat{\mc{U}}^{(s)}=\sum_{k=1}^{s}\Big(\Theta_{(s)}^{\a(k)\ad(k)}\nabla^k_{\a\ad}+\Omega_{(s)}^{\a(k)\ad(k-1)}\nabla^{k-1}_{\a\ad}\nabla_{\a}\Big) +\Theta_{(s)} ~. \label{GenOpPhi}
\end{align}
Each operator $\hat{\mc{U}}^{(s)}$ must preserve the chiral nature of $\Phi$ and the superconformal properties \eqref{Phiprop}. These conditions prove to determine all coefficients in \eqref{GenOpPhi} in terms of a single parameter, which may be identified with the gauge parameter of the SCHS field. Let us sketch how this is so.

Enforcing the chirality condition, $\bar{\nabla}_{\ad}\phantom{.}\hat{\mc{U}}^{(s)}\Phi=0$, leads to the following constraints 
\begin{align}
\bar{\nabla}^{\ad}\Theta_{(s)}^{\a(k)\ad(k)}=0~,\qquad \Omega_{(s)}^{\a(k)\ad(k-1)}=\frac{\ri}{2}\frac{k}{k+1}\bar{\nabla}_{\bd}\Theta_{(s)}^{\a(k)\ad(k-1)\bd}~,\qquad \bar{\nabla}^{\ad}\Theta_{(s)}=0~.
\end{align}
We see that the superfield $\Omega_{(s)}^{\a(k)\ad(k-1)}$ is determined by the complex longitudinal superfield $\Theta_{(s)}^{\a(k)\ad(k)}$ and $\Theta_{(s)}$ is chiral. The latter two may be expressed in terms of an unconstrained prepotential as $\Theta_{(s)}=\bar{\nabla}^2\xi_{(s)}$ and $\Theta_{(s)}^{\a(k)\ad(k)}=\bar{\nabla}^{\ad}\z_{(s)}^{\a(k)\ad(k-1)}$. The operator $\hat{\mc{U}}^{(s)}$ becomes
\begin{align}
\hat{\mc{U}}^{(s)}=\sum_{k=1}^{s}\Big(\bar{\nabla}^{\ad}\z_{(s)}^{\a(k)\ad(k-1)}\nabla_{\a\ad}^k+\frac{\ri}{4}\bar{\nabla}^2\z_{(s)}^{\a(k)\ad(k-1)}\nabla_{\a\ad}^{k-1}\nabla_{\a}+\bar{\nabla}^2\xi_{(s)}^{\a(k-1)\ad(k-1)}\nabla_{\a\ad}^{k-1} \Big)~, \label{chiralop}
\end{align}
where for convenience we have explicitly included a third type of term which may be obtained via the field redefinition $\z_{(s)}^{\a(k-1)\ad(k-2)}\rightarrow \z_{(s)}^{\a(k-1)\ad(k-2)} -2\bar{\nabla}_{\bd}\xi_{(s)}^{\a(k-1)\ad(k-2)\bd}$ for $2\leq k \leq s$.

It is clear that in order to ensure $\hat{\mc{U}}^{(s)}$ preserves the superconformal properties of $\Phi$,  $\hat{\mc{U}}^{(s)}$ must be both chargeless and weightless, and hence
 \begin{subequations}\label{prepotprop}
\begin{align}
\mb{D}\z_{(s)}^{\a(k)\ad(k-1)}&=-\big(k+1/2\big)\z_{(s)}^{\a(k)\ad(k-1)}~,\qquad \mb{Y}\z_{(s)}^{\a(k)\ad(k-1)}=\z_{(s)}^{\a(k)\ad(k-1)}~,\\
\mb{D}\xi_{(s)}^{\a(k-1)\ad(k-1)}&=-k\xi_{(s)}^{\a(k-1)\ad(k-1)}~,\qquad ~~\phantom{....}\mb{Y}\xi_{(s)}^{\a(k-1)\ad(k-1)}=2\xi_{(s)}^{\a(k-1)\ad(k-1)}~. 
\end{align} 
\end{subequations}
Furthermore, the only non-trivial solution to the constraint 
\begin{align}
K_{B}\phantom{.}\hat{\mc{U}}^{(s)}\Phi=0 \label{primop}
\end{align}
occurs when the lowest weight parameter, $\z_{(s)}^{\a(s)\ad(s-1)}\equiv \z^{\a(s)\ad(s-1)}$, is primary
\begin{align}
K_{B}\z^{\a(s)\ad(s-1)}=0~,\qquad \mathbb{D}\z^{\a(s)\ad(s-1)}=-\big(s+1/2\big)\z^{\a(s)\ad(s-1)} \label{gtpropid}
\end{align} 
 and the remaining coefficients are descendants of it.
 The only descendants of $\z^{\a(s)\ad(s-1)}$ that are consistent with properties \eqref{prepotprop} are
\begin{subequations} \label{ansatzs}
\begin{align}
\z_{(s)}^{\a(k)\ad(k-1)}&=a_{(s,k)}\nabla_{\b\bd}^{s-k}\z^{\a(k)\b(s-k)\ad(k-1)\bd(s-k)}~\qquad \qquad \qquad 1 \leq k \leq s-1~,\\
\xi_{(s)}^{\a(k-1)\ad(k-1)}&=b_{(s,k)}\nabla_{\b}\nabla_{\b\bd}^{s-k}\z^{\a(k-1)\b(s-k+1)\ad(k-1)\bd(s-k)}~~~~\qquad 1\leq k \leq s~,
\end{align}
\end{subequations}
for some undetermined coefficients $a_{(s,k)}$ and $b_{(s,k)}$. Upon inserting \eqref{ansatzs} into \eqref{chiralop} and enforcing \eqref{primop}, the coefficients are fixed to
\begin{align}
a_{(s,k)}=(2\ri)^s\binom{s}{k}\binom{s-1}{k-1}\binom{2s+1}{s-k}^{-1}~,\qquad b_{(s,k)}=-\frac{\ri}{4}\frac{k}{s+k+1}a_{(s,k)}~. \label{chiralConstas}
\end{align}
This yields the following operator (determined up to an overall normalisation)
\begin{subequations}\label{ChiralPrimalOp}
\begin{align}
\hat{\mc{U}}^{(s)} &= \sum_{k=1}^{s}a_{(s,k)}\Big\{ \nabla^{s-k}_{\b\bd}\bar{\nabla}^{\ad}\z^{\a(k)\b(s-k)\bd(s-k)\ad(k-1)}\nabla^k_{\a\ad}\notag \\
&\phantom{=}+\frac{\ri}{4}\bar{\nabla}^2\nabla^{s-k}_{\b\bd}\z^{\a(k)\b(s-k)\bd(s-k)\ad(k-1)}\nabla^{k-1}_{\a\ad}\nabla_{\a}\notag\\
&\phantom{=}-\frac{\ri}{4}\frac{k}{s+k+1}\bar{\nabla}^2\nabla_{\b}\nabla^{s-k}_{\b\bd}\z^{\a(k-1)\b(s-k+1)\bd(s-k)\ad(k-1)}\nabla^{k-1}_{\a\ad}\Big\}~, \qquad s\geq 1 \label{ChiralPrimalOpa}
\end{align}
which is valid for all $s\geq 1$, whilst for $s=0$ we define 
\begin{align}
\hat{\mc{U}}^{(0)}:=\bar{\nabla}^2\z~, \label{ChiralPrimalOpb}
\end{align}
\end{subequations}
where $\z$ is a complex primary scalar superfield with weight $-1$ and charge $2$.

We should point out that Ref. \cite{Koutrolikos1} analysed higher-derivative differential  operators in Minkowski superspace which map the space of  chiral scalar superfields to itself. However, the issue of superconformal covariance, eq. \eqref{primop}, was not discussed. 

 
\subsection{Noether procedure to order $\mc{O}(H)$}

From $\Phi$ and $\bar{\Phi}$ one may construct the 
following composite operators
\begin{align}
J_{\a(s)\ad(s)}=(2\ri)^s\sum_{k=0}^{s}(-1)^k\binom{s}{k}^2\Big(\nabla_{\a\ad}^k\Phi\nabla_{\a\ad}^{s-k}\bar{\Phi}-\frac{\text{i}}{2}\frac{s-k}{k+1}\nabla_{\a\ad}^{k}\nabla_{\a}\Phi\nabla_{\a\ad}^{s-k-1}\bar{\nabla}_{\ad}\bar{\Phi}\Big)~,
 \label{supercur}
\end{align}
with $s\geq 0$.
It may be shown that $J_{\a(s)\ad(s)}$ possesses the following crucial features:
\begin{enumerate}[label=(\roman*)] 
\item Reality
\begin{subequations}\label{3.15}
\begin{align}
J_{\a(s)\ad(s)}=\bar{J}_{\a(s)\ad(s)}~;
\end{align}
\item Superconformal covariance in an arbitrary supergravity background
\begin{align}
K_B J_{\a(s)\ad(s)} = 0~, \qquad \mb{D}J_{\a(s)\ad(s)}=(s+2)J_{\a(s)\ad(s)}~, \qquad \mb{Y}J_{\a(s)\ad(s)}=0~; \label{currentprop}
\end{align}
\item When restricted to any conformally-flat  background, $W_{\a\b\g}=0$, 
 $J_{\a(s)\ad(s)}$ describes a conserved current multiplet on-shell, for any non-negative integer $s$.
This means that  $J_{\a(s)\ad(s)}$ is transverse linear
for $s\geq 1$,
\begin{align}
 \nabla^{\b}J_{\b\a(s-1)\ad(s)}\approx 0 \qquad  \Longleftrightarrow \qquad  \bar{\nabla}^{\bd}J_{\a(s)\ad(s-1)\bd}\approx 0~,\label{TOS1}
\end{align}
and linear for $s=0$,
\begin{align}
\nabla^2J\approx 0 \qquad \Longleftrightarrow \qquad \bar{\nabla}^2J\approx 0~.\label{TOS2}
\end{align}
\end{subequations}
\end{enumerate}
The cases $s=0$ and $s=1$ correspond to the flavour current multiplet \cite{FWZ}
and the Ferrara-Zumino supercurrent \cite{FZ}, respectively. The CHS supercurrents \eqref{supercur}
with $s\geq2$ were introduced for the first time in Minkowski superspace in \cite{KMT}, and soon after these results were extended to AdS superspace \cite{BHK18}. 

The current multiplets  \eqref{supercur} with $s=0$ and $s=1$ are actually conserved on-shell on arbitrary backgrounds. However, the super-Weyl tensor $W_{\a\b\g}$ proves to be the obstruction for this property to hold in the case $s\geq 2$, and hence for the remainder of this section we restrict ourselves to conformally-flat superspace backgrounds. 

The supercurrents $J_{\a(s)\ad(s)}$ naturally couple to the half-integer superspin-$(s+\hf)$ conformal superfields $H_{\a(s)\ad(s)}$ via a Noether coupling
\begin{align}	
\mc{S}_{1}[\Phi,H]=\sum_{s=0}^{\infty}c_s\mc{S}^{(s)}_{1}[\Phi,H]~,\qquad \mc{S}^{(s)}_{1}[\Phi,H]=\int \text{d}^{4|4}z \, E \, H^{\a(s)\ad(s)}J_{\a(s)\ad(s)}~,\label{Noether}
\end{align}
with $c_s$ some coefficients which are yet to be determined. Under the superconformal gauge group, the real superfield $H_{\a(s)\ad(s)}$ transforms according to
\begin{align}
K_B H_{\a(s)\ad(s)}=0~,\qquad \mb{D}H_{\a(s)\ad(s)}=-sH_{\a(s)\ad(s)}~,\qquad \mb{Y}H_{\a(s)\ad(s)}=0~,\label{SHCSprop}
\end{align}
for all values of $s$. By virtue of the above properties, the cubic action
\begin{align}
\mc{S}_{\text{cubic}}[\Phi,H]=\mc{S}_{0}[\Phi]+\mc{S}_{1}[\Phi,H] \label{Cubicact}
\end{align}
is invariant under the superconformal gauge group.\footnote{Action \eqref{Cubicact} is actually invariant under the superconformal group on an arbitrary background superspace.} 
Furthermore, on account of \eqref{TOS1} and \eqref{TOS2}, it is also invariant under the zeroth-order gauge transformations
\begin{subequations}\label{SCHSgt}
\begin{align}
s\geq 1:&\qquad \qquad \qquad \delta_{\L} H_{\a(s)\ad(s)}=\bar{\nabla}_{\ad}\L_{\a(s)\ad(s-1)}-\nabla_{\a}\bar{\L}_{\a(s-1)\ad(s)}~, \qquad \qquad \label{SCHSgt.a} \\[5pt]
s= 0:&\qquad \qquad \qquad \qquad ~~ \delta_{\L} H = \bar{\nabla}^2\L+\nabla^2\bar{\L} ~, \qquad \qquad \label{3.19b}
\end{align}
\end{subequations}
provided $\Phi$ is on-shell, $\delta_{\L}\mc{S}\approx 0$. The gauge parameters $\Lambda_{\a(s)\ad(s-1)}$ and $\L$ are unconstrained superfields with the properties
\begin{align}\label{SCHSgtparprop}
K_{B}\Lambda_{\a(s)\ad(s-1)}=0~,\qquad \mathbb{D}\Lambda_{\a(s)\ad(s-1)}&= -\big(s+1/2\big)\Lambda_{\a(s)\ad(s-1)}~,\qquad \mathbb{Y}\Lambda_{\a(s)\ad(s-1)}=\Lambda_{\a(s)\ad(s-1)}~,\non\\[5pt]
K_B\L=0~,\qquad \mathbb{D}\L&=-\L~,\qquad \mathbb{Y}\L=2\L~.
\end{align}

In order to elevate the gauge symmetry \eqref{SCHSgt} off the mass shell, it is necessary to endow $\Phi$ with its own transformation rule. For this we may use the rule \eqref{Mattergt} with \eqref{ChiralPrimalOp} derived in the previous section, since the properties \eqref{gtpropid} and \eqref{SCHSgtparprop} allow us to make the identifications  $\z^{\a(s)\ad(s-1)}\equiv \L^{\a(s)\ad(s-1)}$ and $\z\equiv \L$ between the parameters.
 Now we may fix the coefficients $c_s$ in \eqref{Noether} by requiring the cubic action \eqref{Cubicact} to be invariant up to terms linear in gauge fields,
\begin{align}
c_s = (-1)^{s}\binom{2s+1}{s}^{-1}~ \qquad \implies \qquad \delta_{\L}\mc{S}_{\text{cubic}}[\Phi,H] = \mc{O}\big(H\big)~. \label{CubicGI}
\end{align}  
It should be emphasised that this statement holds off-shell.

In the rigid supersymmetric case, the cubic vertices \eqref{Noether} were introduced in \cite{KMT}. However, the gauge transformations of the chiral matter, eqs. \eqref{Mattergt} with \eqref{ChiralPrimalOp}, were not discussed and, therefore, the coefficients 
\eqref{CubicGI} were not fixed.  

A few comments are in order about the gauge transformation law
\eqref{SCHSgt.a}
 in theories in Minkowski superspace.
In the $s=1$ case, the transformation law \eqref{SCHSgt.a} corresponds to linearised 
conformal supergravity \cite{FZ2}. The same transformation of $H_{\a\ad}$ 
occurs in all off-shell models for linearised $\cN=1$ supergravity, 
see \cite{BK} for a review. Such actions involve not only the gravitational superfield $H_{\a\ad}$
\cite{FZ2,OS,Siegel}, but also certain compensators. 
For $s>1$, the gauge transformation law \eqref{SCHSgt.a}
was introduced in \cite{KSP} in the framework of the (two dually equivalent) off-shell formulations for the massless superspin-$(s+\hf)$ multiplet.
The massless actions of \cite{KSP} involve not only 
the gauge prepotential $H_{\a(s) \ad(s)}$
but also certain compensators  (see \cite{BK} for a pedagogical review).


\subsection{Consistency to all orders} \label{secChiralBack}

As was the case in the non-supersymmetric model, the gauge transformations \eqref{Mattergt} with \eqref{ChiralPrimalOp}  do not form a closed algebra. This is because we have restricted our attention to transformation operators which, when acting on $\Phi$, do not vanish on the free equations of motion. To rectify this we need to include operators in \eqref{ChiralPrimalOp} which involve $\nabla^2$ and $\Box$. To maintain control of the superconformal symmetry, we will make use of a superconformal compensator. 

Let us introduce a nowhere-vanishing compensating superfield $\U(z)$ which is chiral,
\begin{align}
\bar{\nabla}_{\ad} \U =0~,
\end{align}
and has the superconformal properties
\begin{align}
\mathbb{D} \U = - \U~,\qquad \mathbb{Y} \U = \frac{2}{3}\U~,\qquad K_{B}\U =0~.
\end{align}
From $\U$ we construct the following two operators
\begin{align}
\Delta:= \U \bar{\nabla}^2~,\qquad \bar{\Delta} := \bar{\U}\nabla^2~.
\end{align}
A simple calculation shows that their product, $\Xi:=\Delta \bar{\Delta}$, preserves all properties of $\Phi$:
\begin{align}
\mathbb{D}\Xi \F =  \Xi\F~,\qquad &\mathbb{Y}\Xi\F = -\frac{2}{3}\Xi\F~,\qquad K_{B}\Xi\F =0~,\qquad \bar{\nabla}_{\ad}\Xi\Phi = 0~.
\end{align}

The operator $\Xi$ is the analogue of the non-supersymmetric operator $\eta\Box$ as it encodes all terms which vanish on-shell, 
\begin{align}
\Xi\Phi = \Big[\big(\U\bar{\nabla}^2\bar{\U}\big)\nabla^2+\big(8\ri \U\bar{\nabla}^{\ad}\bar{\U}\big)\nabla_{\a\ad}\nabla^{\a}+\big(16\U\bar{\U}\big)\Box\Big]\Phi~, \label{chiralEoMcomp}
\end{align}
in a superconformal fashion. It follows that we may generate higher-order primary operators by combining powers of $\Xi$ with \eqref{ChiralPrimalOp}. In particular, the most general transformation rule for $\Phi$ preserving all its kinematic properties is
\begin{subequations}\label{genvarChiral}
\begin{align}
\delta \F = -\hat{\mc{V}}\F~, \qquad \hat{\mc{V}}:= \sum_{s=0}^{\infty}\hat{\mc{V}}^{(s)}\F~,
\end{align}
where we have introduced the maximal order $s$ operator 
\begin{align}
\hat{\mc{V}}^{(s)} = \sum_{l=0}^{\lfloor s/2 \rfloor} \hat{\mc{U}}^{(s-2l)}_{[s,l]}\Xi^l
\label{genvarChiral.b}
\end{align}
and the notation
\begin{align}
\hat{\mc{U}}^{(s-2l)}_{[s,l]} &= \sum_{k=1}^{s-2l}a_{(s-2l,k)}\Big\{ \bar{\nabla}^{\ad}\nabla^{s-2l-k}_{\b\bd}\z_{[s,l]}^{\a(k)\b(s-2l-k)\bd(s-2l-k)\ad(k-1)}\nabla^k_{\a\ad}\notag \\
&\phantom{=}+\frac{\ri}{4}\bar{\nabla}^2\nabla^{s-2l-k}_{\b\bd}\z_{[s,l]}^{\a(k)\b(s-2l-k)\bd(s-2l-k)\ad(k-1)}\nabla^{k-1}_{\a\ad}\nabla_{\a}\notag\\
&\phantom{=}-\frac{\ri}{4}\frac{k}{s-2l+k+1}\bar{\nabla}^2\nabla_{\b}\nabla^{s-2l-k}_{\b\bd}\z_{[s,l]}^{\a(k-1)\b(s-2l-k+1)\bd(s-2l-k)\ad(k-1)}\nabla^{k-1}_{\a\ad}\Big\}~.
\end{align}
\end{subequations}
The gauge parameters introduced above are primary with weights and charges given by
\begin{align}
\mathbb{D}\z_{[s,l]}^{\a(s-2l)\ad(s-2l-1)}=-\big(s-2l+1/2\big)\z_{[s,l]}^{\a(s-2l)\ad(s-2l-1)}~,\qquad \mathbb{Y}\z_{[s,l]}^{\a(s-2l)\ad(s-2l-1)}=\z_{[s,l]}^{\a(s-2l)\ad(s-2l-1)}~.
\end{align}
It is clear that the algebra of gauge transformations \eqref{genvarChiral} now closes.

After integrating by parts,\footnote{Integration by parts in conformal (super)space is a subtle and technical issue. For a detailed discussion of this, we refer the reader to appendix B of \cite{Kuzenko:2020jie}. In summary, beginning with a primary Lagrangian, naive integration by parts (IBP) is valid only when the Lagrangian one ends with is also primary, which should be checked by hand.  Although this IBP rule holds in all cases known to the authors, it still remains a conjecture, and we assume it is valid throughout this work.  } the cubic action \eqref{Cubicact} can be expressed in the form\footnote{Here we have absorbed the coefficients $c_s$ into the definition of $H_{\a(s)\ad(s)}$, which means the corresponding gauge transformations will also be scaled accordingly, see \eqref{SCHSdeformedGT}.}
\begin{align}
\mc{S}_{\text{cubic}}[\Phi,H]=\int \text{d}^{4|4} z \, E \, \bar{\Phi}~\hat{H}~\Phi~,\qquad \hat{H}=1+\sum_{s=0}^{\infty}\hat{H}^{(s)} \label{chiralcubicact2}
\end{align}
where $\hat{H}^{(s)}$ is the following operator of maximal order $s$
\begin{subequations} \label{ChiralHop}
\begin{align}
\hat{H}^{(s)}&=\sum_{k=0}^{s}d_{(s,k)}\Big\{\Big(\nabla_{\a\ad}-\frac{\ri}{2}\frac{s-k}{s+1}\nabla_{\a}\bar{\nabla}_{\ad}\Big)\nabla_{\a\ad}^{s-k-1}H^{\a(s)\ad(s)}\nabla_{\a\ad}^k\notag\\
&\phantom{=\sum_{k=0}^{s}d_{(s,k)}\Big\{\Big(\nabla_{\a\ad}}+\frac{\ri}{2}\frac{k}{s+1}\bar{\nabla}_{\ad}\nabla_{\a\ad}^{s-k}H^{\a(s)\ad(s)}\nabla_{\a\ad}^{k-1}\nabla_{\a}\Big\}~,\\
d_{(s,k)}&=(-2\ri)^s\binom{s}{k}\binom{s+k+1}{k}~.
\end{align}
\end{subequations}
Operator $\hat{H}^{(s)}$ takes any primary chiral scalar of weight $1$ and charge $-2/3$ to a (non-chiral) primary scalar with the same weight and charge
\begin{align}
K_B \hat{H}^{(s)} \Phi = 0 ~, \qquad \mathbb{D} \hat{H}^{(s)}\Phi = \hat{H}^{(s)}\Phi~,\qquad \mathbb{Y} \hat{H}^{(s)}\Phi = -\frac{2}{3}\hat{H}^{(s)}\Phi~, \label{ChiralHprop}
\end{align}
and is Hermitian, $\big(\hat{H}^{(s)}\big)^{\dagger}=\hat{H}^{(s)}$, with respect to the definition \eqref{chiralAdjointA}. In fact, one may deduce $\hat{H}^{(s)}$ as the unique operator of the functional form \eqref{ChiralHop} satisfying the aforementioned properties,  without taking the Noether coupling \eqref{Noether} as the starting point. 

Action \eqref{chiralcubicact2} is not invariant under the gauge transformations \eqref{genvarChiral} and $\delta \hat{H} = (\hat{H} \hat{\mc{V}})^{\dagger} + \hat{H} \hat{\mc{V}} $. In fact, the latter variation is algebraically inconsistent as the RHS contains terms proportional to the operator $\nabla^2$ (for example) whilst the LHS does not. To rectify this, we extend the operator $\hat{H}$ to the primary operator $\hat{G}$, which includes terms of this type
\begin{align}
\hat{G}=1+\sum_{s=0}^{\infty}\hat{G}^{(s)}~,\qquad \hat{G}^{(s)}=\frac{1}{2}\sum_{l=0}^{\lfloor s/2 \rfloor}\Big(\hat{H}^{(s-2l)}_{[s,l]}\Xi^l + \Xi^l\hat{H}^{(s-2l)}_{[s,l]}\Big)~. \label{ChiralGop}
\end{align}
The compensating operator $\Xi$ is self-adjoint, $\Xi^{\dagger}=\Xi$,\footnote{Operator $\Xi$ is self-adjoint with respect to the definition $\llangle \Xi\O,\Theta \rrangle =\llangle \O,\Xi^{\dagger}\Theta \rrangle $ for $\O,\Theta \in \mf{S}_{(1,-\frac{2}{3})}$  and hence also with respect to each definition in \eqref{chiralAdjoint}. See appendix \ref{AppendixAdjoint} for an explanation of the notation.} and hence by construction each $\hat{G}^{(s)}$ is also self-adjoint, $\big(\hat{G}^{(s)}\big)^{\dagger}=\hat{G}^{(s)}$. In \eqref{ChiralGop} we have introduced the operator
\begin{align}
\hat{H}^{(s-2l)}_{[s,l]}&=\sum_{k=0}^{s-2l}d_{(s-2l,k)}\Big\{\Big(\nabla_{\a\ad}-\frac{\ri}{2}\frac{s-2l-k}{s-2l+1}\nabla_{\a}\bar{\nabla}_{\ad}\Big)\nabla_{\a\ad}^{s-2l-k-1}H_{[s,l]}^{\a(s-2l)\ad(s2-l)}\nabla_{\a\ad}^k\notag\\
&\phantom{=\sum_{k=0}^{s}d_{(s,k)}\Big\{\Big(\nabla_{\a\ad}}+\frac{\ri}{2}\frac{k}{s-2l+1}\bar{\nabla}_{\ad}\nabla_{\a\ad}^{s-2l-k}H_{[s,l]}^{\a(s-2l)\ad(s-2l)}\nabla_{\a\ad}^{k-1}\nabla_{\a}\Big\}~.
\end{align}
which is essentially \eqref{ChiralHop} with $s\mapsto s-2l$ and satisfies properties \eqref{ChiralHprop}.
The gauge fields which parametrise this operator are primary and have the following weights and charges
\begin{align}
\mathbb{D} H_{[s,l]}^{\a(s-2l)\ad(s-2l)}=-\big(s-2l\big)H_{[s,l]}^{\a(s-2l)\ad(s-2l)}~,\qquad \mathbb{Y}H_{[s,l]}^{\a(s-2l)\ad(s-2l)}=0~.
\end{align}

From this construction it follows that the manifestly real and superconformal action
\begin{align}
\mathcal{S}[\F,H]=\int \text{d}^{4|4} z \, E \, \bar{\F}\phantom{.}\hat{G}\phantom{.}\F~,\label{SCgenact}
\end{align}
is invariant under the gauge transformations
\begin{align}
\delta \F = -\hat{\mc{V}}\F~, \qquad \delta \hat{G} =  \hat{G} \hat{\mc{V}}+\big(\hat{G} \hat{\mc{V}}\big)^{\dagger} ~, \label{ChiralNLgt}
\end{align}
and that this forms an algebraically consistent system. Here $\hat{\mc{V}}$ is defined as in \eqref{genvarChiral}, and the adjoint operator $\big(\hat{G} \hat{\mc{V}}\big)^{\dagger}$ should be computed in accordance with definition \eqref{chiralAdjointA}.\footnote{We note that, due to the subtleties described in appendix \ref{AppendixAdjoint}, the operator  $\big(\hat{G} \hat{\mc{V}}\big)^{\dagger}$ computed with respect  to \eqref{chiralAdjointA} will not generally coincide with the operator $ \hat{\mc{V}}^{\dagger}\hat{G}$ with $ \hat{\mc{V}}^{\dagger}$ computed with respect to \eqref{chiralAdjointC}.} 

The gauge variation of the original SCHS fields $H^{\a(s)\ad(s)}\equiv H^{\a(s)\ad(s)}_{[s,0]}$ may be extracted from the formal transformations \eqref{ChiralNLgt}. In particular, in appendix \ref{Appendix B} we show that the physical sub-sector (the $l=0$ part) of \eqref{ChiralNLgt} is equivalent to
\begin{subequations}\label{SCHSdeformedGT}
\begin{align}
\delta H_{\a(s)\ad(s)}&= \bar{\nabla}_{\ad}\L_{\a(s)\ad(s-1)}-\nabla_{\a}\bar{\L}_{\a(s-1)\ad(s)}+\mc{O}(H)~,\label{SCHSdeformedGTa}\\
\L_{\a(s)\ad(s-1)}&\equiv c_s \z^{[s,0]}_{\a(s)\ad(s-1)}~,\label{SCHSdeformedGTb}
\end{align}
\end{subequations} 
where $c_s$ are defined in \eqref{CubicGI}. Hence the free gauge transformations \eqref{SCHSgt} are recovered, plus non-linear and non-abelian corrections (whose explicit form we do not determine here). As in the non-supersymmetric case, it should be possible to show that the $H_{[s,l]}^{\a(s-2l)\ad(s-2l)}$ with $l>0$ are pure gauge.

To conclude we point out that, after integrating by parts, action \eqref{SCgenact} takes the form
\begin{subequations}
	\begin{align}
		\mathcal{S}[\Phi,H]&=\int \text{d}^{4|4} z \, E \, \Big\{ \bar{\Phi} \Phi + \sum_{s=0}^{\infty}\sum_{l=0}^{\lfloor s/2 \rfloor} H_{[s,l]}^{\a(s-2l)\ad(s-2l)}J_{\a(s-2l)\ad(s-2l)}^{[s,l]} \Big\}~,\\
		J^{[s,l]}_{\a(s-2l)\ad(s-2l)}&=\frac{(2\ri)^s}{2}\sum_{k=0}^{s-2l} (-1)^k \binom{s-2l}{k}^2 \Big \{ \nabla_{\a\ad}^k\Phi\nabla_{\a\ad}^{s-k-2l} \Xi^l \bar{\Phi} + \nabla_{\a\ad}^k \Xi^l \Phi\nabla_{\a\ad}^{s-k-2l}\bar{\Phi} \non \\  -\frac{\text{i}}{2}&\frac{s-2l-k}{k+1} \Big (\nabla_{\a\ad}^{k}\nabla_{\a}\Phi\nabla_{\a\ad}^{s-2l-k-1}\bar{\nabla}_{\ad} \Xi^l \bar{\Phi} + 
		\nabla_{\a\ad}^{k}\nabla_{\a} \Xi^l \Phi\nabla_{\a\ad}^{s-2l-k-1}\bar{\nabla}_{\ad}\bar{\Phi} \Big ) \Big\}~.
	\end{align}
\end{subequations}
For $l=0$, the currents $J_{\a(s-2l)\ad(s-2l)}^{[s,l]}$ coincide with \eqref{supercur}. Thus we see that, through this procedure, we have introduced a coupling between the auxiliary gauge fields and trivial supercurrents (trivial in the sense that they vanish on the free equations of motion, and are hence trivially conserved on-shell).


\subsection{Rigid symmetries of the free matter action}

It remains to provide some comments on the rigid symmetries of the free action. 
From \eqref{CubicGI} one may deduce that
$\delta_{\L}\mc{S}_{0}=-\delta_{\L}\mc{S}_{1}\big|_{H=0}$,
from which it follows that 
\begin{align}
\delta_{\L}H_{\a(s)\ad(s)}=0 \qquad \implies \qquad \delta_{\L}\mc{S}_{0}=0~. \label{RigidSym}
\end{align}
The subclass of gauge parameters $\L_{\a(s)\ad(s-1)}$ satisfying
\begin{align}
\chi_{\a(s)\ad(s)}=\bar{\chi}_{\a(s)\ad(s)}~,\qquad \chi_{\a(s)\ad(s)}:=\ri \bar{\nabla}_{\ad}\L_{\a(s)\ad(s-1)}~, \label{SCKTreality}
\end{align}
lead to \eqref{RigidSym}. The physical significance of such configurations become apparent upon observing that $\chi_{\a(s)\ad(s)}$ satisfies the constraints 
\begin{align}
\bar{\nabla}_{\ad}\chi_{\a(s)\ad(s)}=0\qquad \Longleftrightarrow \qquad \nabla_{\a}\chi_{\a(s)\ad(s)}=0~,
\end{align}
and is hence a real conformal Killing tensor superfield of the background superspace
\cite{HL1,KR19}.
 We observe that when $\L$ satisfies \eqref{SCKTreality}, the gauge transformations $\delta\Phi=-\hat{\mc{U}}^{(s)}\Phi$, with $\hat{\mc{U}}^{(s)}$ given by \eqref{ChiralPrimalOp} (with $\L$ and $\z$ identified), are completely parametrised by $\chi_{\a(s)\ad(s)}$
\begin{align}
\delta\Phi&=-\frac{\ri}{2}\sum_{k=1}^{s}a_{(s,k)}\Big\{2\nabla_{\a\ad}^{s-k}\chi^{\a(s)\ad(s)}\nabla_{\a\ad}^k+\frac{\ri s}{s+1}\nabla_{\a\ad}^{s-k}\bar{\nabla}_{\ad}\chi^{\a(s)\ad(s)}\nabla_{\a\ad}^{k-1}\nabla_{\a}\notag\\
&\phantom{=}+\frac{2k}{s+k+1}\Big(\nabla_{\a\ad}-\frac{\ri}{2}\frac{s}{s+1}\nabla_{\a}\bar{\nabla}_{\ad}\Big)\nabla_{\a\ad}^{s-k}\chi^{\a(s)\ad(s)}\nabla_{\a\ad}^{k-1}\Big\}\Phi~.
\end{align}
 These transformations constitute rigid symmetries of the free action, and consequently of its equation of motion.

As was done in section \ref{secCSrigid},  it is possible to deduce that these are the most general (non-trivial) rigid symmetries of the free action by direct computation. Let us suppose that the matter superfield transforms according to the rule $\delta \F = -\hat{\mc{U}}^{(s)}\F$, with
\begin{align}
\hat{\mc{U}}^{(s)} &= \sum_{k=1}^{s}a_{(s,k)}\Big\{ \nabla^{s-k}_{\b\bd}\bar{\nabla}^{\ad}\z^{\a(k)\b(s-k)\bd(s-k)\ad(k-1)}\nabla^k_{\a\ad}\notag \\
&\phantom{=}+\frac{\ri}{4}\bar{\nabla}^2\nabla^{s-k}_{\b\bd}\z^{\a(k)\b(s-k)\bd(s-k)\ad(k-1)}\nabla^{k-1}_{\a\ad}\nabla_{\a}\notag\\
&\phantom{=}-\frac{\ri}{4}\frac{k}{s+k+1}\bar{\nabla}^2\nabla_{\b}\nabla^{s-k}_{\b\bd}\z^{\a(k-1)\b(s-k+1)\bd(s-k)\ad(k-1)}\nabla^{k-1}_{\a\ad}\Big\}~, \qquad s\geq 1 \label{ChiralPrimalOpRigid}
\end{align}
and $a_{(s,k)}$ as defined in \eqref{chiralConstas}.
Then the condition that the corresponding variation of the free matter action \eqref{WZact} vanishes is
\begin{align}
0= \int \text{d}^{4|4}z \, E \, \bar{\F}\Big\{ \hat{\mc{U}}^{(s)}+ \big(\hat{\mc{U}}^{(s)}\big)^{\dagger}\Big\}\F~. \label{ChiralrigidVar}
\end{align}
The adjoint of $\hat{\mc{U}}^{(s)}$ should be computed in accordance with definition \eqref{chiralAdjointB}.  As mentioned in appendix \ref{AppendixAdjoint}, this does not determine $\big(\hat{\mc{U}}^{(s)}\big)^{\dagger}$ uniquely, and one finds the two expressions\footnote{The two expressions are equivalent up to integration by parts with respect to definition \eqref{chiralAdjointB}, and hence one could also take suitable linear combinations $\big(\hat{\mc{U}}^{(s)}\big)^{\dagger}=a\big(\hat{\mc{U}}^{(s)}_1\big)^{\dagger}+b\big(\hat{\mc{U}}^{(s)}_2\big)^{\dagger}$ with $a+b=1$.} 
\begin{subequations}
\begin{align}
\big(\hat{\mc{U}}^{(s)}_1\big)^{\dagger} = -\sum_{k=1}^{s}a_{(s,k)}\Big\{&\frac{k+1}{s+1}\nabla_{\b\bd}^{s-k}\nabla^{\a}\bar{\z}^{\a(k-1)\b(s-k)\ad(k)\bd(s-k)}\nabla_{\a\ad}^k\non\\
-&\frac{\ri}{4}\frac{k}{s+k+1}\nabla^2\bar{\nabla}_{\bd}\nabla_{\b\bd}^{s-k}\bar{\z}^{\a(k-1)\b(s-k)\ad(k-1)\bd(s-k+1)}\nabla_{\a\ad}^{k-1}\non\\
+&\frac{k}{s+1}\nabla^{\gamma}\nabla_{\g\bd}\nabla_{\b\bd}^{s-k}\bar{\z}^{\a(k-1)\b(s-k)\ad(k-1)\bd(s-k+1)}\nabla_{\a\ad}^{k-1}\non\\
-&\frac{\ri}{2}\frac{s}{s+1}\nabla^{\a}\bar{\nabla}_{\bd}\nabla_{\b\bd}^{s-k}\bar{\z}^{\a(k-1)\b(s-k)\ad(k-1)\bd(s-k+1)}\nabla_{\a\ad}^{k-1}\nabla_{\a}\non\\
+&\frac{\ri}{2}\frac{s-k}{s+1}\nabla_{\b}\bar{\nabla}_{\bd}\nabla^{\a}{}_{\bd}\nabla_{\b\bd}^{s-k-1}\bar{\z}^{\a(k-1)\b(s-k)\ad(k-1)\bd(s-k+1)}\nabla_{\a\ad}^{k-1}\nabla_{\a}\non\\
+&\frac{k}{s+1}\nabla^{\a}{}_{\bd}\nabla_{\b\bd}^{s-k}\bar{\z}^{\a(k-1)\b(s-k)\ad(k-1)\bd(s-k+1)}\nabla_{\a\ad}^{k-1}\nabla_{\a}\Big\}~,\label{adjoint1}\\
\big(\hat{\mc{U}}^{(s)}_2\big)^{\dagger} = -\sum_{k=1}^{s}a_{(s,k)}\Big\{&\frac{s+1}{k+1}\nabla_{\b\bd}^{s-k}\bar{\z}^{\a(k-1)\b(s-k)\ad(k)\bd(s-k)}\nabla_{\a\ad}^{k-1}\nabla_{\g\ad}\nabla^{\g}\non\\
-&\frac{\ri}{2}\frac{(s+1)(s-k)}{(s+2)(k+1)}\nabla_{\b}\bar{\nabla}_{\bd}\nabla_{\b\bd}^{s-k-1}\bar{\z}^{\a(k-1)\b(s-k)\ad(k)\bd(s-k)}\nabla_{\a\ad}^{k-1}\nabla_{\g\ad}\nabla^{\g}\non\\
+&\frac{\ri}{4}\frac{s+1}{s+2}\bar{\nabla}_{\bd}\nabla_{\b\bd}^{s-k}\bar{\z}^{\a(k-1)\b(s-k)\ad(k-1)\bd(s-k+1)}\nabla_{\a\ad}^{k-1}\nabla^2 \Big\}
\end{align}
\end{subequations}
for $s\geq 1$. It is important to point out that these expressions for the adjoint will not agree with those in \eqref{ChiralNLgt}, since the latter are computed in accordance with the definition \eqref{chiralAdjointA}. 

In order to find non-trivial solutions to \eqref{ChiralrigidVar} we use the expression \eqref{adjoint1} for the adjoint, since it has the same functional type as $\hat{\mc{U}}^{(s)}$, and denote $\big(\hat{\mc{U}}^{(s)}_1\big)^{\dagger} \equiv \big(\hat{\mc{U}}^{(s)}\big)^{\dagger}$. Then, after some algebra one finds that condition \eqref{ChiralPrimalOpRigid} is equivalent to
\begin{align}\label{BigEquation}
0= &\sum_{k=1}^sa_{(s,k)} \Big\{\Big[\nabla^{s-k}_{\b\bd}\bar{\nabla}^{\ad}\z^{\a(k)\b(s-k)\bd(s-k)\ad(k-1)}-\frac{k+1}{s+1}\nabla_{\b\bd}^{s-k}\nabla^{\a}\bar{\z}^{\a(k-1)\b(s-k)\ad(k)\bd(s-k)}\non\\
&-\frac{\ri}{4}\frac{s-k}{k}\bar{\nabla}^2\nabla_{\b}\nabla^{s-k-1}_{\b\bd}\z^{\a(k)\b(s-k)\bd(s-k-1)\ad(k)}+\frac{\ri}{4}\frac{s-k}{k}\nabla^2\bar{\nabla}_{\bd}\nabla_{\b\bd}^{s-k-1}\bar{\z}^{\a(k)\b(s-k-1)\ad(k)\bd(s-k)}\non\\
&-\frac{(s-k)(s+k+2)}{k(s+1)}\nabla^{\gamma}\nabla_{\g\bd}\nabla_{\b\bd}^{s-k-1}\bar{\z}^{\a(k)\b(s-k-1)\ad(k)\bd(s-k)}\Big]\nabla_{\a\ad}^k\non\\
&+\Big[~\frac{\ri}{4}\bar{\nabla}^2\nabla^{s-k}_{\b\bd}\z^{\a(k)\b(s-k)\bd(s-k)\ad(k-1)}+\frac{\ri}{2}\frac{s}{s+1}\nabla^{\a}\bar{\nabla}_{\bd}\nabla_{\b\bd}^{s-k}\bar{\z}^{\a(k-1)\b(s-k)\ad(k-1)\bd(s-k+1)}\non\\
&~~~-\frac{\ri}{2}\frac{s-k}{s+1}\nabla_{\b}\bar{\nabla}_{\bd}\nabla^{\a}{}_{\bd}\nabla_{\b\bd}^{s-k-1}\bar{\z}^{\a(k-1)\b(s-k)\ad(k-1)\bd(s-k+1)}\\
&~~~-\frac{k}{s+1}\nabla^{\a}{}_{\bd}\nabla_{\b\bd}^{s-k}\bar{\z}^{\a(k-1)\b(s-k)\ad(k-1)\bd(s-k+1)}\Big]\nabla_{\a\ad}^{k-1}\nabla_{\a}\Big\}\non\\
 & -\frac{\ri}{4}\frac{1}{s+2}a_{(s,1)}\Big(\bar{\nabla}^2\nabla_{\b}\nabla_{\b\bd}^{s-1}\z^{\b(s)\bd(s-1)}-\nabla^2\bar{\nabla}_{\bd}\nabla_{\b\bd}^{s-1}\bar{\z}^{\b(s-1)\bd(s)}-4\ri\frac{s+2}{s+1}\nabla^{\g}\nabla_{\g\bd}\nabla_{\b\bd}^{s-1}\bar{\z}^{\b(s-1)\bd(s)}\Big)~.\non
\end{align}
Each term in the square brackets above must vanish for each value of $k$ independently. At $k=s$, the first $[\cdots]$ term tells us that (cf. \eqref{SCKTreality})
\begin{align}
\chi^{\a(s)\ad(s)}=\bar{\chi}^{\a(s)\ad(s)}~,\qquad \chi^{\a(s)\ad(s)}:=\ri\bar{\nabla}^{\ad}\z^{\a(s)\ad(s-1)}~,\label{Reality2}
\end{align} 
and hence $\chi^{\a(s)\ad(s)}$ is a conformal Killing tensor superfield. Equation \eqref{Reality2} implies that 
\begin{align}
\bar{\nabla}^2\z^{\a(s)\ad(s-1)}=-\frac{2s}{s+1}\ri\bar{\nabla}_{\bd}\bar{\chi}^{\a(s)\ad(s-1)\bd}\quad \implies \quad \nabla^2\bar{\z}^{\a(s-1)\ad(s)}=\frac{2s}{s+1}\ri\nabla_{\b}\chi^{\b\a(s-1)\ad(s)}
\end{align}
which in turn implies 
\begin{align}
\nabla_{\b}\bar{\nabla}^2\z^{\b\a(s-1)\ad(s-1)}-\bar{\nabla}_{\bd}\nabla^2\bar{\z}^{\a(s-1)\ad(s-1)\bd}=-4\frac{s}{s+1}\nabla_{\b\bd}\chi^{\a(s-1)\b\ad(s-1)\bd}~.
\end{align}
The above identities may be used to show that all other terms in \eqref{BigEquation} vanish identically without having to impose any conditions on $\z^{\a(s)\ad(s-1)}$ other than \eqref{Reality2}. 
Thus we are lead to the same conclusion that all (non-trivial) rigid symmetries of $\mc{S}_0[\F,\bar{\F}]$ are parametrised by real conformal Killing tensor superfields $\chi^{\a(s)\ad(s)}$. If we denote the operator  \eqref{ChiralPrimalOpRigid} with $\z$ satisfying \eqref{Reality2} by $\hat{\mc{U}}_{\text{sckt}}^{(s)}$, then we have shown that for $s\geq 1$ 
\begin{align}
0= \hat{\mc{U}}_{\text{sckt}}^{(s)}+ \big(\hat{\mc{U}}_{\text{sckt}}^{(s)}\big)^{\dagger}~. \label{ChiralrigidOpEq}
\end{align}
 
For $s=0$ one finds that $-\delta \F = \mc{U}^{(0)}=\bar{\nabla}^2\z$ is a symmetry of  $\mc{S}_0[\F,\bar{\F}]$ only if $\bar{\nabla}^2\z = -\nabla^2\bar{\z}$, and hence $\bar{\nabla}^2 \z $ is an imaginary constant.
 
Finally, we note that transformations of the type $\delta \F = \ri\, \Xi^l \F$, for integer $l>0$, are trivial symmetries of $\mc{S}_0[\F,\bar{\F}]$. This follows from \eqref{chiralEoMcomp} and the fact that $\Xi$ is self-adjoint. Consequently, the order-$s$ transformations of the type 
\begin{align}
\delta \F = \big[~\hat{\mc{U}}_{\text{sckt}}^{(s-2l)}, ~\ri\, \Xi^l ~ \big] \F~,\qquad 1\leq l \leq \lfloor s/2 \rfloor~, \label{ChiralTrivSym}
\end{align}
are also trivial symmetries of $\mc{S}_0[\F,\bar{\F}]$. This may be checked directly as follows. First, one makes use of \eqref{chiralEoMcomp} and the results of \cite{KR19} (i.e. that $\hat{\mc{U}}_{\text{sckt}}^{(s)}$ maps solutions of \eqref{ChiralEoM} to solutions) to show that \eqref{ChiralTrivSym} satisfies $\delta \F \approx 0$. Second, one should show that \eqref{ChiralTrivSym} is a symmetry of \eqref{WZact}, $\delta \mc{S}_0=0$, by employing \eqref{ChiralrigidOpEq}.

We now briefly comment on the algebraic structure of the rigid symmetries described above. As discussed in appendix \ref{AppRigid}, they form a Lie algebra with respect to the commutator. Specifically, given two rigid symmetry operators $\hat{\mathcal{U}}^{(m)}_{\text{sckt}}$ and $\hat{\mathcal{U}}^{(n)}_{\text{sckt}}$ parametrised by the SCKTs $\chi^{\a(m)\ad(m)}$ and $\mu^{\a(n)\ad(n)}$, their commutator $\big[\hat{\mathcal{U}}^{(m)}_{\text{sckt}},\,\hat{\mathcal{U}}^{(n)}_{\text{sckt}}\big]$ also defines a rigid symmetry of the matter action. It is expected that this new symmetry will be parametrised by several SCKTs, with the highest rank parameter defined via the supersymmetric even Schouten-Nijenhuis bracket \cite{HL1,KR19}
\begin{align}
	&[ \chi , \mu ]_{\text{SSN}}^{\a(m+n-1) \ad(m+n-1)} :=  m \chi^{\a(m-1)\b \ad(m-1)\bd} \nabla_{\b \bd } \mu^{\a(n) \ad(n)}
	- n \mu^{\a(n-1)\b\ad(n-1)\bd} \nabla_{\b \bd } \chi^{\a(m) \ad(m)} \non \\
	+ &\frac{\ri m n}{2(m+1)(n+1)} \left( \bar{\nabla}_{\bd} \chi^{\a(m)\bd\ad(m-1)} \nabla_{\b} \mu^{\a(n-1)\b\ad(n)} - \bar{\nabla}_{\bd} \mu^{\a(n)\bd\ad(n-1)} \nabla_{\b} \chi^{\a(m-1)\b\ad(m)} \right) ~.
	\label{SN2}
\end{align}
The algebra of (non-trivial) rigid symmetries of $\mc{S}_0$, identified with the quotient \eqref{Quotient},  should correspond to the $\mc{N}=1$ superconformal higher-spin algebra $\mathfrak{shsc}^\infty (4|1) $ constructed in \cite{FL-4D, FL-algebras}.

To conclude, we describe the symmetries of the superconformal Wess-Zumino 
equation \eqref{ChiralEoM}, see \cite{HL2,KR19} for earlier discussions. 
Consider an operator $\hat \cV$ acting on the space of primary weight-1 chiral scalar superfields. It is a symmetry of the superconformal Wess-Zumino operator if 
$\nabla^2 \hat{\cV} \F_0 =0$ for every solution $\F_0$  
of the equation \eqref{ChiralEoM}. It is said to be a trivial symmetry of the superconformal Wess-Zumino operator if $\hat{\cV} = \hat{\cF} \nabla^2$ for some operator $\hat{\cF}$.
Let us consider a general symmetry operator $\hat{\mc{V}}^{(s)} $ 
with order $s$ given by eq. \eqref{genvarChiral}. The only non-trivial contribution  in \eqref{genvarChiral.b} corresponds to $l=0$. The remaining operator 
$\hat{\mc{U}}^{(s)}_{[s,0]} \equiv \hat{\mc{U}}^{(s)}$ is a symmetry of the superconformal Wess-Zumino operator if its top component $\z^{\a(s)\ad(s-1)}$ satisfies the condition
\eqref{SCKTreality}, which defines a conformal Killing tensor superfield, i.e. $\hat{\mc{U}}^{(s)}\equiv \hat{\mc{U}}^{(s)}_{\text{sckt}}$ .


\section{Matter coupled to integer superspin SCHS multiplets} \label{section4}

This section is devoted to analysing the superconformal cubic interactions of certain $\mc{N}=2$ matter multiplets with $\mc{N}=1$ SCHS gauge prepotentials $\{\Psi_{\a(s)\ad(s-1)},\bar{\Psi}_{\a(s-1)\ad(s)}\}$ carrying integer superspin-$s$. Here we only complete the Noether procedure to first order in the gauge fields. In principle, one may adapt the techniques employed in the previous sections to promote the prepotentials $\Psi_{\a(s)\ad(s-1)}$ to background superfields, but we leave this for future work. 

\subsection{Coupling to the superconformal gravitino multiplet}
In this subsection we restrict our attention to the $\mathcal{N}=1$ superconformal gravitino multiplet. It is described by a spinor superfield $\Psi_{\a}$, defined modulo the gauge transformations\footnote{In Minkowski superspace, this gauge transformation was introduced in Ref. \cite{GS}, which
proposed the off-shell formulation for the massless gravitino multiplet 
in terms of the gauge 
spinor prepotential $\J_\a$ in conjunction with two compensators,  
an unconstrained real scalar and a chiral scalar.
The model for the superconformal gravitino multiplet in Bach flat backgrounds was studied 
in \cite{KMT,Kuzenko:2020jie}.
} 
\begin{align}
	\label{GMGT}
	\d_{\O, \l} \Psi_\a  = \nabla_\a \O + \l_\a ~, \qquad \bar{\nabla}_\ad \l_\a = 0~,
\end{align}
and possesses the following superconformal properties
\begin{align}
	K_B \Psi_{\a}=0~, \qquad
	\mb{D}\Psi_{\a}= - \frac{1}{2} \Psi_{\a}~,\qquad \mb{Y}\Psi_{\a}=\frac{1}{3}\Psi_{\a}~.
\end{align}
Our goal is to describe cubic interactions between $\Psi_\a$ and several matter superfields, which we will collectively denote $\S^I$. The corresponding actions take the general form
\begin{align}
	\label{GMaction}
	\mc{S}_{\text{cubic}}[\S,\Psi]=\mc{S}_{0}[\S]+ c_1 \int \text{d}^{4|4}z \, E \, \Big ( \Psi^{\a}J_{\a}+  \bar{\Psi}_{\ad}\bar{J}^{\ad} \Big ) 
\end{align}
for some undetermined constant $c_1$, and where $J^\a$ is a primary descendant of $\S^I$ carrying dimension $5/2$ and $\sU(1)_R$ charge $-1/3$
\begin{align}
	K_B J^\a = 0 ~, \qquad \mathbb{D} J^\a = \frac 5 2 J^\a ~, \qquad \mathbb{Y} J^\a = - \frac 1 3 J^\a ~,
\end{align}
that is conserved on-shell:\footnote{In some degenerate cases these conservation equations will hold off-shell.}
\begin{align}
	\nabla^\b J_\b \approx 0 ~, \qquad \bar{\nabla}^2 J_\a \approx 0~.
\end{align}

Owing to the latter property, action \eqref{GMaction} is invariant under the gauge transformations \eqref{GMGT} on-shell. Our goal in this section is to extend this symmetry off the mass shell for certain matter multiplets.
Once the transformation laws of the matter superfields have been computed, it is possible to derive some rigid symmetries of the latter. Specifically, by imposing $\d_{\O,\l} \Psi_\a = 0$ the corresponding transformations of the matter multiplets will be parametrised by a scalar parameter $\ve$, defined by
\begin{align}
	\label{5.6}
	\nabla_\a \ve \equiv \nabla_\a \O = - \l_\a \quad \Longrightarrow \quad  \bar{\nabla}_\ad \nabla_\a \ve = 0~.
\end{align}
These conditions characterise $\nabla_\a \ve$ as a conformal Killing spinor superfield, which parametrises second conformal supersymmetries \cite{KR19}.

\subsubsection{The hypermultiplet (I)}
\label{GMHM1}

In this section we consider an $\cN=2$ massless hypermultiplet, which in $\mc{N}=1$ superspace may be described by two chiral scalar superfields $\Phi_+$ and $\Phi_-$. On an arbitrary background superspace, the free action is simply two copies of the Wess-Zumino action \eqref{WZact}
\begin{align}
	\mc{S}_{0}[\Phi_{\pm}]= \int \text{d}^{4|4}z \, E \, \Big\{ \bar{\Phi}_+ \Phi_+ + \bar{\Phi}_{-}\Phi_{-}\Big\}~,\qquad \bar{\nabla}_{\ad} \Phi_{\pm}=0~, \label{hypAct}
\end{align} 
with $\nabla^2\Phi_{\pm}\approx 0$ the corresponding equations of motion. This action is superconformal if each chiral superfield carries the properties
\begin{align}
	K_A\Phi_{\pm} = 0~, \qquad \mb{D}\Phi_{\pm}=\Phi, \qquad \mb{Y}\Phi_{\pm}=-\frac{2}{3}\Phi_{\pm}~. \label{hypPhiprop}
\end{align}

To couple this multiplet to $\Psi_\a$, the supercurrent of interest is
\begin{align}
	\label{HM1SC}
	J_\a =  - \F_{+} \nabla_\a \F_{-} + \nabla_\a \F_{+} \F_{-} ~,
\end{align}
which is transverse anti-linear on-shell and complex linear off-shell
\begin{align}
	\nabla^{\b} J_\b \approx 0 ~, \quad \bar{\nabla}^2 J_\a = 0~.  \label{DegenCons}
\end{align}
Owing to the latter condition, the cubic action \eqref{GMaction} is $\l$ gauge-invariant off-shell
\begin{align}
	\delta_{\l}\mc{S}[\Phi_{\pm},\Psi] = 0~.
\end{align}

Next, we seek to elevate the $\O$ gauge symmetry off the mass shell. To this end, we endow the matter multiplet with its own transformation rule. It may be shown that the most general transformation parametrised by $\O$ preserving all off-shell properties of $\F_{\pm}$ is
\begin{subequations}
	\label{5.10}
	\begin{align}
		\delta_{\O}\Phi_{+} &=\hat{\mc{U}}^{(1)}\bar{\Phi}_{-} ~,\qquad \delta_{\O}\Phi_{-}=-\hat{\mc{U}}^{(1)}\bar{\Phi}_{+}, \\
		\hat{\mathcal{U}}^{(1)} &= \bar{\nabla}^2 \bar{\O} + 2 \bar{\nabla}_\ad \bar{\O} \bar{\nabla}^\ad + \bar{\O} \bar{\nabla}^2 ~.
	\end{align}
\end{subequations}
Employing this result, it may be readily shown that the cubic action \eqref{GMaction} is invariant under $\O$ gauge transformations provided that $c_1 = 1$
\begin{align}
	c_1=1\qquad \implies \qquad \delta_{\O}\mc{S}[\Phi_{\pm},\Psi] = \mc{O}\big(\Psi\big) ~.
\end{align}

Now, we consider the special class of transformations \eqref{5.10} characterised by the condition $\d_{\O,\l} \Psi_\a = 0$, or equivalently \eqref{5.6}. These take the form
\begin{align}
	\delta_{\ve}\Phi_{+} &= \bar{\nabla}^2(\bar{\ve} \bar{\Phi}_{-}) ~, \qquad \delta_{\ve}\Phi_{-} = - \bar{\nabla}^2(\bar{\ve} \bar{\Phi}_{+}) ~,
\end{align}
which are exactly the conformal supersymmetries described in \cite{KR19}. Furthermore, due to the (in some sense degenerate) off-shell $\l_{\a}$ gauge symmetry, the reducibility parameters defined by $\delta_{\O}\Psi_{\a}=0~~\implies~~\nabla_{\a}\O=0$ also lead to rigid symmetries. However, the latter are easily shown to be trivial symmetries of $\mc{S}_0$, as $\delta_{\O}\F_{\pm}\approx 0$.

\subsubsection{The hypermultiplet (II)}
\label{GMHM2}

As is well known, the $\mathcal{N}=2$ hypermultiplet may also be realised in terms of a chiral and complex linear superfield, denoted $\F$ and $\G$, respectively. On generic superspace backgrounds, its free action is given by
\begin{align}
	\mc{S}_{0}[\Phi,\G]= \int \text{d}^{4|4}z \, E \, \Big\{ \bar{\Phi} \Phi - \bar{\G} \G \Big\}~, \qquad \bar{\nabla}_\ad \F = 0~, \quad \bar{\nabla}^2 \G = 0~. \label{hypAct2}
\end{align}
The corresponding equations of motion take the form
\begin{align}
	\nabla^2 \F \approx 0 ~, \qquad \nabla_\a \G \approx 0~.
\end{align}
We note that \eqref{hypAct2} is superconformal provided $\F$ obeys \eqref{Phiprop} and $\G$ has the properties
\begin{align}
	K_A \G = 0~, \qquad \mb{D}\G = \G, \qquad \mb{Y}\G=\frac{2}{3}\G~. \label{CLprop}
\end{align}

To couple this multiplet to $\Psi_\a$, it is necessary to make use of the supercurrent
\begin{align}
	J_\a =  \ri \big ( \nabla_\a \F \bar{\G} - \F \nabla_\a \bar{\G} \big )~,
\end{align}
which is conserved only on-shell (in contrast to \eqref{DegenCons})
\begin{align}
	\nabla^{\b} J_\b \approx 0 ~, \quad \bar{\nabla}^2 J_\a \approx 0~. 
\end{align}
As a consequence, action \eqref{GMaction} enjoys gauge invariance under \eqref{GMGT} on-shell. Thus, in keeping with our goal to extend gauge invariance off-shell, we require the matter multiplets to also transform. It may be shown that the most general variations parameterised by $\O$ and $\l_{\a}$ are
\begin{subequations}
	\label{5.14}
	\begin{align}
		\d_\O \Phi &= \hat{\mathcal{U}}^{(1)} \G ~, \qquad \hat{\mathcal{U}}^{(1)} = \ri \Big ( \bar{\nabla}^2 \bar{\O} + 2 \bar{\nabla}_\ad \bar{\O} \bar{\nabla}^\ad \Big )~, \\
		\d_\l \G &= \hat{\mathcal{V}}^{(1)} \Phi ~, \qquad
		\hat{\mathcal{V}}^{(1)} = \ri \Big ( \nabla^\a \lambda_\a + 2 \lambda^\a \nabla_\a \Big ) ~.
	\end{align}
\end{subequations}
One may then readily show that the cubic action \eqref{GMaction} is gauge-invariant provided that $c_1 = 1$
\begin{align}
	c_1=1\qquad \implies \qquad \delta_{\O,\l}\mc{S}[\Phi,\G,\Psi] = \mc{O}\big(\Psi\big) ~.
\end{align}

Next, we consider the matter transformations \eqref{5.14} constrained by the rigidity condition \eqref{5.6}. They take the form
\begin{align}
	\delta_{\ve}\Phi &= - \ri \bar{\nabla}^2(\bar{\ve} \G) ~, \qquad \delta_{\ve}\G = \ri (\nabla^2 \ve + 2 \nabla^\a \ve \nabla_\a) \Phi ~,
\end{align}
which coincide with the conformal supersymmetries derived in \cite{KR19}.

\subsubsection{The $\mathcal{N}=2$ vector multiplet} \label{section4.1.3}

To conclude our analysis for $\Psi_\a$, we analyse its cubic couplings to a massless $\mc{N}=2$ vector multiplet. As is well known, the latter is described in $\mathcal{N}=1$ superspace via a chiral scalar $\Phi$ and spinor $W_\a$. The superfield $W_\a$ satisfies the off-shell constraints
\begin{align}
	\label{VMconstraints}
	\bar{\nabla}_\ad W_\a = 0 ~, \qquad \nabla^\a W_\a = \bar{\nabla}_\ad \bar{W}^\ad ~,
\end{align}
which means that it describes the $\mathcal{N}=1$ vector multiplet. These constraints may be solved, allowing $W_\a$ to be expressed in terms of a real prepotential\footnote{It should be noted that $W_\a$ is invariant under gauge transformations \eqref{3.19b}.}
\begin{align}
	W_\a = - \frac 1 4 \bar{\nabla}^2 \nabla_\a H ~, \qquad H = \bar{H}~.
\end{align} 
Further, consistency of \eqref{VMconstraints} with the superconformal algebra implies the properties
\begin{align}
	K_B W_\a = 0 ~, \qquad \mathbb{D} W_\a = \frac{3}{2} W_\a ~, \qquad \mathbb{Y} W_\a = - W_\a ~. \label{VMprop}
\end{align}

Returning to the $\mathcal{N}=2$ vector multiplet, its free action takes the form
\begin{align}	
	\mc{S}_{0}[\F,W_\a]= \int \text{d}^{4|4}z \, E \, \bar{\Phi} \Phi ~+~ \bigg \{ \frac{1}{4} \int\text{d}^{4}x \, \text{d}^{2}\q \, \mathcal{E} \,  	W^\a W_\a + \text{c.c.} \bigg \}~, \label{VMN=2act}
\end{align}
where we recall that chiral actions are related to those over the full superspace via the rule
\begin{align}	
	\int \text{d}^{4|4}z \, E \, \mathcal{L} = - \frac 1 4 \int\text{d}^{4}x \, \text{d}^{2}\q \, \mathcal{E} \,  \bar{\nabla}^2 \mathcal{L}~,
\end{align}
and that $\cE$ denotes the chiral measure. Varying \eqref{VMN=2act}, we obtain the dynamical equations
\begin{align}
	\nabla^2 \Phi \approx 0~, \qquad \nabla^\a W_\a \approx 0 ~. \label{VMN=2EoM}
\end{align}

The spinor supercurrent needed to couple this multiplet to $\Psi_\a$ is
\begin{align}
	J_\a =  \ri W_\a \bar{\F}~,
\end{align}
which is conserved on-shell
\begin{align}
	\nabla^\b J_\b \approx 0 ~, \qquad \bar{\nabla}^2 J_\a \approx 0~.
\end{align}
As a consequence, the corresponding cubic action is only gauge invariant when the equations of motion are imposed.
To extend gauge invariance off-shell, we require the matter superfields to have the non-trivial linear variations. The most general ones preserving the properties \eqref{Phiprop}, \eqref{VMconstraints} and \eqref{VMprop}, and which are parametrised by $\O$ and $\l_{\a}$, are 
\begin{subequations}
	\label{5.18}
	\begin{align}
		\d_\O W_\a &= \frac \ri 4 \bar{\nabla}^2 \nabla_\a \Big ( \O \bar{\F} - \bar{\O} \F \Big )~, \\
		\d_\l \Phi &= - \ri \lambda^\a W_\a ~.
	\end{align}
\end{subequations}
These variations allow one to shown that the cubic action \eqref{GMaction} is gauge-invariant if $c_1 = 1$
\begin{align}
	c_1=1\qquad \implies \qquad \delta_{\O,\l}\mc{S}[\Phi,W,\Psi] = \mc{O}\big(\Psi\big)~.
\end{align}

To conclude, we derive the second (conformal) supersymmetries of this model by imposing the constraint \eqref{5.6} on the matter transformations \eqref{5.18}. We find that these take the form
\begin{align}
	\delta_{\ve}\Phi = \ri \nabla^\a \ve W_\a ~, \qquad \delta_{\ve}W_\a = \frac \ri 4 \bar{\nabla}^2 \nabla_\a \Big ( \ve \bar{\F} - \bar{\ve} \F \Big ) ~,
\end{align}
and should be compared with the second supersymmetries derived via an $\mc{N}=2 \rightarrow \mc{N}=1$ reduction of the vector multiplet.

\subsection{Hypermultiplet coupled to integer superspin SCHS multiplets}\label{Section2.2}

This subsection is aimed at extending the analysis of cubic interactions of a massless hypermultiplet (realised as two chiral superfields $\Phi_{\pm}$) with $\Psi_\a$ conducted in section \ref{GMHM1}. Specifically, we examine its cubic coupling to an infinite tower of integer superspin SCHS prepotentials. 

From $\Phi_{+}$ and $\Phi_{-}$ one may construct the following fermionic composite tensors:
\begin{align}
	J_{\a(s)\ad(s-1)}= - \sum_{k=0}^{s-1}(-1)^k\binom{s-1}{k}\binom{s}{k}&\Big\{\nabla_{\a\ad}^k\Phi_{+}\nabla_{\a\ad}^{s-k-1}\nabla_{\a}\Phi_{-} \notag\\
	&-\frac{s-k}{k+1}\nabla_{\a\ad}^k\nabla_{\a}\Phi_{+}\nabla_{\a\ad}^{s-k-1}\Phi_{-}\Big\} ~,\label{hypSupercur}
\end{align}
for any integer $s\geq 1$. We note that for $s=1$ it coincides with \eqref{HM1SC}.
It may be shown that $J_{\a(s)\ad(s-1)}$ possesses the following crucial features:
\begin{enumerate}[label=(\roman*)]
	\item (Anti-)symmetry under the exchange of $\Phi_{+}$ and $\Phi_{-}$
	\begin{subequations}
		\label{4.35}
		\begin{align}
			\Phi_{+} \leftrightarrow \Phi_{-} \qquad \implies \qquad J_{\a(s)\ad(s-1)}\mapsto (-1)^sJ_{\a(s)\ad(s-1)}~; \label{hypSym}
		\end{align}
		\item Transverse anti-linear on-shell and transverse linear off-shell for $s>1$,
		\begin{align}
			\nabla^{\b}J_{\b\a(s-1)\ad(s-1)}&\approx 0~, \\
			\bar{\nabla}^{\bd}J_{\a(s)\ad(s-2)\bd}&=0~, \label{hypTOS}
		\end{align}
		whilst for $s=1$ it is transverse anti-linear on-shell and linear off-shell,
		\begin{align}
			\nabla^{\b}J_{\b}&\approx 0~, \\
			\bar{\nabla}^2J_{\a}&=0~; 
		\end{align}
		\item Superconformal covariance on an arbitrary background superspace
		\begin{align}
			&K_B J_{\a(s)\ad(s-1)} = 0~, \notag\\
			\mb{D}J_{\a(s)\ad(s-1)}=\Big(s+\frac{3}{2}\phantom{.}\Big)&J_{\a(s)\ad(s-1)}, \qquad \mb{Y}J_{\a(s)\ad(s-1)}=-\frac{1}{3}J_{\a(s)\ad(s-1)}~. \label{hypcurrentprop}
		\end{align}
	\end{subequations}
\end{enumerate}
In Minkowski superspace, the supercurrents \eqref{hypSupercur} were introduced in \cite{KMT}. They were extended to AdS superspace in \cite{BHK18}.

The supercurrents $J_{\a(s)\ad(s-1)}$ naturally couple to the integer superspin-$s$ conformal superfields $\Psi_{\a(s)\ad(s-1)}$ via a Noether coupling
\begin{align}	
	\mc{S}_{1}[\Phi_{\pm},\Psi] =\sum_{s=1}^{\infty}  \mc{S}^{(s)}_{1}[\Phi_{\pm},\Psi]~,\qquad \mc{S}^{(s)}_{1}=c_s\int \text{d}^{4|4}z \, E \, \Psi^{\a(s)\ad(s-1)}J_{\a(s)\ad(s-1)}+\text{c.c.} ~,\label{hypNoether}
\end{align}
for some undetermined real coefficients $c_s$. Under the superconformal gauge group, the superfield $\Psi_{\a(s)\ad(s-1)}$ transforms according to
\begin{align}
	&K_B \Psi_{\a(s)\ad(s-1)}=0~,\notag\\
	\mb{D}\Psi_{\a(s)\ad(s-1)}=-\Big(s-\frac{1}{2}\Big)&\Psi_{\a(s)\ad(s-1)}~,\qquad \mb{Y}\Psi_{\a(s)\ad(s-1)}=\frac{1}{3}\Psi_{\a(s)\ad(s-1)}~.\label{hypSHCSprop}
\end{align}
By virtue of the above properties, the cubic action
\begin{align}
	\mc{S}_{\text{cubic}}[\Phi_{\pm},\Psi]=\mc{S}_{0}[\Phi_{\pm}]+\mc{S}_{1}[\Phi_{\pm},\Psi] \label{hypCubicact}
\end{align}
is invariant under the superconformal gauge group.
Furthermore, on account of \eqref{hypTOS}, it is also invariant under the gauge transformations
\begin{subequations} \label{hypSCHSgtAll}
	\begin{align}
		\d_{\O, \l} \Psi_\a  &= \nabla_\a \O + \l_\a ~, \qquad \bar{\nabla}_\ad \l_\a = 0~, \\
		\delta_{\Lambda,\O} \Psi_{\a(s)\ad(s-1)}  &=\nabla_{\a}\O_{\a(s-1)\ad(s-1)}+\bar{\nabla}_{\ad}\L_{\a(s)\ad(s-2)}~, \quad s > 1 \label{hypSCHSgt}
	\end{align}
\end{subequations}
where invariance with respect to $\O_{\a(s-1)\ad(s-1)}$ with $s\geq 1$ holds only if $\Phi_{\pm}$ are on-shell, $\delta_{\O}\mc{S}\approx 0$, whilst $\L_{\a(s)\ad(s-2)}$ and $\l_{\a}$ invariance holds identically, $\delta_{\L,\lambda}\mc{S}= 0$. The gauge parameters $\Lambda_{\a(s)\ad(s-2)}$ and $\O_{\a(s-1)\ad(s-1)}$ are complex unconstrained primary superfields.

The gauge transformation law \eqref{hypSCHSgt}
was introduced in  \cite{KS,HK2,BHK18}
 in the framework of the longitudinal  formulations for the massless integer superspin-$s$ multiplet.
The massless actions of   \cite{KS,HK2,BHK18} involve not only 
the gauge prepotential $\J_{\a(s) \ad(s-1)}$ and its conjugate
but also certain compensators.

In order to lift the $\O$-gauge symmetry off the mass shell, we require the matter superfields to transform according to general rule
\begin{align}
	\delta_{\O}\Phi_{+}= - \sum_{s=1}^{\infty}\hat{\mc{U}}^{(s)}\bar{\Phi}_{-}~,\qquad \delta_{\O}\Phi_{-}= -\sum_{s=1}^{\infty}(-1)^s\hat{\mc{U}}^{(s)}\bar{\Phi}_{+}~.\label{hypTrans}
\end{align}
This off-diagonal transformation rule is required in order for the variations of the kinetic \eqref{hypAct} and cubic \eqref{hypNoether} sectors to communicate.
Furthermore, the transformation rule of $\Phi_-$ is immediately fixed upon specifying that of $\Phi_+$ (or vice versa). This is because the variation of the kinetic sector must scale by a factor of $(-1)^s$ under the exchange $\Phi_+ \leftrightarrow \Phi_-$, which 
is required since the variation of the cubic sector inherits this symmetry property from \eqref{hypSym}.

Once again, the $\hat{\mc{U}}^{(s)}$ in \eqref{hypTrans} represents some differential operator of maximal order $s$. 
To preserve the chirality of $\Phi_{\pm}$, we require that each $\hat{\mc{U}}^{(s)}$ maps an anti-chiral superfield to a chiral one, $0=\bar{\nabla}_{\ad}\,\hat{\mc{U}}^{(s)}\bar{\F}_{\pm}$. Ensuring this constraint, and expressing the result in terms of unconstrained prepotentials, one is eventually led to the operator\footnote{Other transformation rules may be obtained from \eqref{hypchiraltrans} via field redefinitions of the form $\Theta_{\a(k-1)\ad(k-1)}\rightarrow \Theta_{\a(k-1)\ad(k-1)}+\bar{\nabla}_{\ad}\Gamma_{\a(k-1)\ad(k-2)}+\bar{\nabla}^2\chi_{\a(k-1)\ad(k-1)}$ etc. However, such terms are irrelevant for this model as $\Gamma$ and $\chi$ do not have the correct weights and charges to make them identifiable with (descendants of) $\O$ in \eqref{hypSCHSgt}.}
\begin{align}
	\hat{\mc{U}}^{(s)}=&\sum_{k=1}^{s}\Big\{ \Theta^{\a(k-1)\ad(k-1)}\nabla_{\a\ad}^{k-1}\bar{\nabla}^2+\frac{\ri}{2}\frac{k-1}{k}\bar{\nabla}_{\bd}\Theta^{\a(k-1)\ad(k-2)\bd}\nabla_{\a\ad}^{k-2}\nabla_{\a}\bar{\nabla}^2~\notag\\
	&\phantom{\sum_{k=1}^{s}}-2\bar{\nabla}^{\ad}\Theta^{\a(k-1)\ad(k-1)}\nabla_{\a\ad}^{k-1}\bar{\nabla}_{\ad}+\bar{\nabla}^2\Theta^{\a(k-1)\ad(k-1)}\nabla_{\a\ad}^{k-1}\notag\\
	&\phantom{\sum_{k=1}^{s}}+\bar{\nabla}^2\ell^{\a(k-1)\ad(k)}\nabla_{\a\ad}^{k-1}\bar{\nabla}_{\ad}+\bar{\nabla}_{\bd}\ell^{\a(k-1)\ad(k-1)\bd}\nabla_{\a\ad}^{k-1}\bar{\nabla}^2 \Big\} ~. \label{hypchiraltrans}
\end{align}
All terms involving the parameter $\ell$ may be eliminated via the field redefinition $\Theta^{\a(k-1)\ad(k-1)}\rightarrow \Theta^{\a(k-1)\ad(k-1)}-\bar{\nabla}_{\bd}\ell^{\a(k-1)\ad(k-1)\bd}$, though we prefer to keep these terms explicit for now. 

As $\hat{\mc{U}}^{(s)}$ must preserve the superconformal properties of $\Phi_{\pm}$, we see that $\Theta$ and $\ell$ satisfy
\begin{subequations}\label{hypprepotprop}
	\begin{align}
		\mb{D}\Theta_{\a(k-1)\ad(k-1)}&=-k\Theta_{\a(k-1)\ad(k-1)}~,\qquad ~~\mb{Y}\Theta_{\a(k-1)\ad(k-1)}=\frac{2}{3}\Theta_{\a(k-1)\ad(k-1)}~,\\
		\mb{D}\ell_{\a(k-1)\ad(k)}&=-\Big(k+\hf\Big)\ell_{\a(k-1)\ad(k)}~,\qquad \mb{Y}\ell_{\a(k-1)\ad(k)}=\frac{5}{3}\ell_{\a(k-1)\ad(k)}~.
	\end{align}
\end{subequations}
Upon imposing the primality constraint, 
\begin{align}
	K_{A}\phantom{.}\hat{\mc{U}}^{(s)}\bar{\Phi}_{\pm}=0~, \label{hypprimop}
\end{align}
we find that we must have $K_{B}\Theta_{\a(s-1)\ad(s-1)}=0$, allowing us to make the identification $\Theta_{\a(s-1)\ad(s-1)}\equiv \bar{\O}_{\a(s-1)\ad(s-1)}$, where $\bar{\O}_{\a(s-1)\ad(s-1)}$ is the conjugate of the gauge parameter appearing in \eqref{hypSCHSgt}.  We also find that all other coefficients must be descendants of $\bar{\O}_{\a(s-1)\ad(s-1)}$, and the only ans\"atze consistent with the properties \eqref{hypprepotprop} are\footnote{One could include terms of the form $c_k\bar{\nabla}_{\bd}\nabla_{\b}\nabla_{\b\bd}^{s-k-1}\bar{\O}^{\a(k-1)\b(s-k)\ad(k-1)\bd(s-k)}$ in \eqref{hypansatzsa}, but this may be shown to be equivalent to redefining the coefficients $b_k$ in \eqref{hypansatzsb}.  }
\begin{subequations} \label{hypansatzs}
	\begin{align}
		\Theta^{\a(k-1)\ad(k-1)}&=a_k\nabla_{\b\bd}^{s-k}\bar{\O}^{\a(k-1)\b(s-k)\ad(k-1)\bd(s-k)}~\qquad \qquad \qquad 1 \leq k \leq s-1~,\label{hypansatzsa}\\
		\ell^{\a(k-1)\ad(k)}&=b_{k} \nabla_{\b}\nabla_{\b\bd}^{s-k-1}\bar{\O}^{\a(k-1)\b(s-k)\ad(k)\bd(s-k-1)}~~~~~~~\qquad 1\leq k \leq s-1~, \label{hypansatzsb}
	\end{align}
\end{subequations}
for some undetermined coefficients $a_k$ and $b_k$, and where we have set $\ell_{\a(s-1)\ad(s)}=0$. Upon inserting \eqref{hypansatzs} into \eqref{hypchiraltrans} and enforcing \eqref{hypprimop}, the coefficients $a_k$ and $b_k$ may be determined uniquely (up to an overall normalisation), and the final expression for $\hat{\mc{U}}^{(s)}$ is 
\begin{align}
	\hat{\mc{U}}^{(s)}&= - \sum_{k=1}^{s}\binom{s-1}{s-k}\binom{s+k}{s+1}\Big\{\notag\\
	&\phantom{= +}\Big(\nabla_{\b\bd}+\frac{\ri}{2}\frac{s-k}{k}\bar{\nabla}_{\bd}\nabla_{\b}\Big)\nabla_{\b\bd}^{s-k-1}\bar{\O}^{\a(k-1)\b(s-k)\ad(k-1)\bd(s-k)}\nabla_{\a\ad}^{k-1}\bar{\nabla}^2\notag\\
	&\phantom{=}+\frac{\ri}{2}\frac{k-1}{k}\bar{\nabla}_{\bd}\nabla_{\b\bd}^{s-k}\bar{\O}^{\a(k-1)\b(s-k)\ad(k-2)\bd(s-k+1)}\nabla_{\a\ad}^{k-2}\nabla_{\a}\bar{\nabla}^2\notag\\
	&\phantom{=}-2\bar{\nabla}^{\ad}\nabla_{\b\bd}^{s-k}\bar{\O}^{\a(k-1)\b(s-k)\ad(k-1)\bd(s-k)}\nabla_{\a\ad}^{k-1}\bar{\nabla}_{\ad}\notag\\
	&\phantom{=}+\frac{\ri}{2}\frac{s-k}{k}\bar{\nabla}^2\nabla_{\b}\nabla_{\b\bd}^{s-k-1}\bar{\O}^{\a(k-1)\b(s-k)\ad(k)\bd(s-k-1)}\nabla_{\a\ad}^{k-1}\bar{\nabla}_{\ad}\notag\\
	&\phantom{=}+\bar{\nabla}^2\nabla_{\b\bd}^{s-k}\bar{\O}^{\a(k-1)\b(s-k)\ad(k-1)\bd(s-k)}\nabla_{\a\ad}^{k-1}\Big\}~.\label{hypFinalop}
\end{align}
We note that this is not the most general differential operator which preserves the superconformal properties of $\F_{\pm}$ and which  maps an anti-chiral field to chiral one. For this one should also include terms which vanish on the equations of motion in the ansatz \eqref{hypchiraltrans}, along the lines of what was done in sections \ref{CSsec2} and \ref{secChiralBack}. 

Computing the variation of \eqref{hypNoether} under $\O$ gauge transformations (i.e. \eqref{hypSCHSgtAll} and also  \eqref{hypTrans} with \eqref{hypFinalop}), it may be shown that invariance holds up to order $\Psi$ for the following choice of $c_s$
\begin{align}
	c_s = \frac{2s}{s+1} \qquad \implies \qquad \delta_{\O}\mc{S}_{\text{cubic}}[\Phi_{\pm},\Psi] = \mc{O}\big(\Psi\big)~, \label{hypCubicGI}
\end{align}
with $\Phi_{\pm}$ off-shell. Gauge invariance with respect to $\L$ and $\l$ holds identically, $\delta_{\L,\l}\mc{S}_{\text{cubic}}=0$. 
The above results are all in agreement with the ones obtained in section \ref{GMHM1} for the case $s=1$. 

In the flat superspace case, the cubic vertices \eqref{hypNoether} were introduced in \cite{KMT}. However, the gauge transformation of the chiral matter, eq. \eqref{hypTrans}, was not considered and, therefore, the coefficients \eqref{hypCubicGI} were not fixed.

As regards rigid symmetries of the free matter action, there are two separate types of reducibility parameters to consider, which are those satisfying: (i) $\delta_{\O,\L}\Psi_{\a(s)\ad(s-1)}=0$; or (ii) $\delta_{\O}\Psi_{\a(s)\ad(s-1)}=0$. For type (i) one arrives at the condition
\begin{align}
\nabla_\a \O_{\a(s-1) \ad(s-1)} = - \bar{\nabla}_\ad \L_{\a(s) \ad(s-2)}~, \label{hypRigid}
\end{align}
which implies an additional differential constraint on $\O$ (cf. \eqref{5.6})
\begin{align}
\bar{\nabla}_\ad \nabla_\a \O_{\a(s-1) \ad(s-1)} = 0~.
\end{align}
Consequently, it is clear that $\nabla_\a \O_{\a(s-1) \ad(s-1)}$ is a conformal Killing tensor \cite{KR19}.
In case (ii) it is clear the complex parameter $\O_{\a(s-1)\ad(s-1)}$ is longitudinal anti-linear, 
	\begin{align}
		\nabla_{\a}\O_{\a(s-1)\ad(s-1)}=0~.\label{LongLin}
	\end{align}
 Further, it may be shown that $\delta_{\O}\Phi_{\pm}$ with $\O$ satisfying \eqref{LongLin} corresponds to a trivial symmetry,
 since the variation \eqref{hypTrans} with \eqref{hypFinalop} vanishes on-shell, $\delta_{\O}\Phi_{\pm}\approx 0 $.

Finally, we note that in the limiting case $\Phi_-\mapsto \Phi_+ \equiv \Phi$, we obtain consistent superconformal cubic interactions between the chiral matter $\Phi$ and the integer superspin-$s$ conformal superfields $\Psi_{\a(s)\ad(s-1)}$. However, due to property \eqref{hypSym}, such couplings exist only for even $s$.


\section{Discussion} \label{secDiscuss}

Conformal higher-spin theory, sketched in \cite{Tseytlin} and fully developed in \cite{Segal}, is a rare example of a theory involving interacting higher-spin fields with a consistent Lagrangian formulation. 
When viewed as an induced action \cite{Tseytlin, Segal, BJM}, the crucial ingredient in defining CHS theory is the action $\mc{S}[\varphi,h]$ for a complex scalar field $\varphi$ interacting with  an infinite tower of background CHS fields $h$.
In this paper we have combined ideas advocated in \cite{Segal, BJM} with the framework of conformal geometry to construct $\mc{S}[\varphi,h]$ in a way which renders the background conformal symmetry manifest. 

Such a formulation was lacking in the literature. To be more specific, both Refs. \cite{Segal,BJM} work in terms of the spin-$s$ gauge field $h^{a(s)}$ having
\begin{align}
\delta^0_{\xi,\l}h^{a(s)}=\pa^{a}\xi^{a(s-1)}+\eta^{a(2)}\l^{a(s-2)} \label{tracefulCHSgt}
\end{align}
as its linearised gauge symmetry. Here, indices denoted by the same letter are to be symmetrised over, and hence all fields and parameters are totally symmetric and traceful. The fields $h^{a(s)}$ couple to on-shell traceless and conserved currents $\mc{J}_{a(s)}$ via the Noether interaction
\begin{align}
\mc{S}_{\text{NI}}[\varphi,h]=\sum_{s=0}^{\infty}\ri^s\int \text{d}^dx \, h^{a(s)}\mc{J}_{a(s)}~,\qquad \pa^b\mc{J}_{a(s-1)b}\approx 0~,\qquad \eta^{bc}\mc{J}_{a(s-2)bc}\approx 0~.\label{BekaertNI}
\end{align}
In four dimensions, the currents $\mc{J}_{a(s)}$ used in \cite{BJM} are 
\begin{subequations}
\begin{align}
\mc{J}_{a(s)}=\binom{2s}{s}^{-1}\sum_{n=0}^{\lfloor s/2 \rfloor} \sum_{k=0}^{n}\frac{(-1)^{k+n}}{2^{2n}}&\binom{n}{k}\binom{2s-2n}{s-n}\binom{s-n}{n}\big(\eta_{a(2)}\Box\big)^k\big(\partial_{a}\big)^{2n-2k}\tilde{\mc{J}}_{a(s-2n)}~,\\
\tilde{\mc{J}}_{a(s)}&=\sum_{l=0}^{s}\frac{(-1)^l}{2^s}\binom{s}{l}\big(\partial_{a}\big)^l\varphi\big(\partial_{a}\big)^{s-l}\bar{\varphi}~,
\end{align}
\end{subequations}
which are built from the currents $\tilde{\mc{J}}_{a(s)}$ constructed in \cite{BerendsCurrent}. 

After minimally lifting to conformal space (i.e. replacing $\pa_a\rightarrow \nabla_a$), the currents $\mc{J}_{a(s)}$ with $s=0$ and $s=1$  are simply $\mc{J}=\tilde{\mc{J}}=\vf\bar{\vf}$ and $\mc{J}_a=\tilde{\mc{J}}_a=\frac{1}{2}(\vf\nabla_a\bar{\vf}-\bar{\vf}\nabla_a\vf)$ respectively, which are obviously primary. 
The current with  $s=2$, 
\begin{align}
\mc{J}_{a(2)}=\tilde{\mc{J}}_{a(2)} -\frac{1}{12}\big(\nabla_a\nabla_a-\eta_{a(2)}\Box\big)\tilde{\mc{J}}~,
\end{align} 
also happens to be primary, which can be readily seen after relating it to the primary currents $j^{[2,0]}_{\a\b\ad\bd}$ and $j^{[2,1]}=\eta~ \mf{j}^{[2,1]}$ from \eqref{GenCur}\footnote{Recall that the compensator $\eta$ is merely a book-keeping device and hence the term $h_{[2,1]}j^{[2,1]}$ in \eqref{GenCura} can be rewritten as $\tilde{h}\mf{j}^{[2,1]}$ with $\tilde{h} = \eta h^{[2,1]}$ a primary weight 0 field. The field $\tilde{h}$ can be regarded as the trace of the conformal graviton: $h_{ab} = \bm{h}_{ab} +\eta_{ab}\tilde{h}$ with $\eta^{ab}\bm{h}_{ab}=0$.}
\begin{align}
(\s^a)_{\a\ad}(\s^b)_{\b\bd} \mc{J}_{ab}= \frac{1}{6}j_{\a\b\ad\bd}-\frac{1}{2}\ve_{\a\b}\ve_{\ad\bd}\mf{j}^{[2,1]}~,\qquad \mf{j}^{[2,1]}:=\frac{1}{2}\big(\bar{\vf}\Box\vf + \vf\Box\bar{\vf}\big)~.
\end{align}
However, in general, for $s\geq 3$ neither $\mc{J}_{a(s)}$ nor $\tilde{\mc{J}}_{a(s)}$ are primary descendants of $\vf$ off the mass shell. 
For example, in the case $s=3$ one has
\begin{align}
\mc{J}_{a(3)}&=\tilde{\mc{J}}_{a(3)}-\frac{3}{20}(\nabla_a\nabla_a-\eta_{a(2)}\Box)\tilde{\mc{J}}_a~,\qquad K_c \mc{J}_{a(3)}= \frac{3}{5} \eta_{ca}\eta_{a(2)}\nabla^{b}\tilde{\mc{J}}_b\approx 0~, \label{6.4}
\end{align}
and primality holds only on-shell. 
This means that for the coupling \eqref{BekaertNI} to be conformally invariant, the gauge fields $h^{a(s)}$ must transform into one another under conformal transformations (i.e. are also non-primary), and hence the background conformal symmetry is hidden.\footnote{Segal argued \cite{Segal} that in the linearised limit, one can use the $\l$ transformation in \eqref{tracefulCHSgt} to gauge away the trace of $h$, $h^{a(s)}\equiv \bm{h}^{a(s)}$ with $\eta_{bc}\bm{h}^{a(s-2)bc}=0$, and then special conformal transformations act diagonally on $\bm{h}^{a(s)}$ (in other words they are primary, $K_c\bm{h}^{a(s)}=0$). In our context one then has e.g. $K_c(\bm{h}^{a(3)}\mc{J}_{a(3)})=0$  and coupling \eqref{BekaertNI} is primary for traceless $\bm{h}^{a(s)}$. However, no comment was made regarding the non-linear case.  } For the sake of comparison, we note that $\mc{J}_{a(3)}$ is related to the currents of section \ref{CSsec1} via
\begin{subequations}
\begin{align}
	(\s^a)_{\a\ad}(\s^b)_{\b\bd}(\s^c)_{\g\gd}\mc{J}_{abc}&=- \frac{\ri}{20} j_{\a\b\g\ad\bd\gd}+ \ri  ( \ve_{\a \b} \ve_{\ad \bd} \mathfrak{j}_{\gamma \dot{\gamma}} + \ve_{\g \a} \ve_{\gd \ad} \mathfrak{j}_{\bb} + \ve_{\b \g} \ve_{\bd \gd} \mathfrak{j}_{\aa}) ~, \\
	\mathfrak{j}_{\aa}:&= \frac{\ri}{15} \varphi \nabla_\aa \Box  \bar{\varphi} - \frac{\ri}{10} \nabla_\aa \varphi \Box \bar{\varphi} + \frac{\ri}{24} \nabla^{\bb} \varphi \nabla_{\aa} \nabla_{\bb} \bar{\varphi} + \text{c.c.}
\end{align}
\end{subequations}
The current $j_{\a\b\g\ad\bd\gd}$ is primary whilst $\mathfrak{j}_{\aa}$ is not (and hence the latter is not proportional to the primary trivial current $j_{\a\ad}^{[3,1]}$ in \eqref{GenCur} after absorbing $\eta$ terms into $h^{\a\ad}_{[3,1]}$).

In addition to having manifest conformal symmetry, our formalism differs to that of previous works \cite{ Segal, BJM} in several other important aspects.
Firstly, our model is formulated on arbitrary conformally-flat backgrounds, whilst those in \cite{ Segal, BJM} live on a flat background.
Secondly, we deal exclusively with $\sSL(2,\mathbb{C})$ irreducible (i.e. traceless and totally symmetric) tensor fields which are primary. Consequently, in order to make the system consistent we are forced to introduce `auxiliary' fields into the model, which play the role of the traces that are present in \cite{ Segal, BJM}. However, as argued in section \ref{CSsec3}, it appears these auxiliary fields may be gauged away (at the expense of complicating the residual gauge symmetry), and hence represent pure gauge degrees of freedom.  
Lastly, we deal strictly with differential operators whilst Refs. \cite{ Segal, BJM} make use of Weyl quantization techniques and hence the Weyl symbols of differential operators.\footnote{In the hope of streamlining computational aspects, it would be interesting to adapt and apply Weyl quantisation techniques to our formalism, and in particular make it compatible with conformal (super)space. } 

The last two points were necessary in order to achieve the second main goal of this paper, which was the generalisation of the results of section \ref{secCS} to the supersymmetric case. In particular we have constructed, for the first time, the $\cN=1$  superconformal model $\mc{S}[\Phi,H]$ for a superconformal chiral multiplet  $\Phi$ consistently coupled to an infinite tower of background SCHS gauge multiplets $H$. This is done in such a way that the superconformal symmetry is manifest and the corresponding gauge transformations of the SCHS fields is a non-abelian extension of their free gauge transformations. Having derived $\mc{S}[\Phi,H]$, it may be used to introduce an effective action $\Gamma[H]$ defined by 
\begin{align}
\re^{\ri\Gamma[H]}=\int\mc{D}\Phi\mc{D}\bar{\F}\, \re^{\ri \mc{S}[\F,H]}~. \label{SCHSeff}
\end{align} 
Then, as sketched in \cite{KMT}, an interacting SCHS theory may be formally identified with the logarithmically divergent part of $\Gamma[H]$,
i.e. as an induced action. 
In general, the superconformal symmetry of $\Gamma[H]$ is anomalous at the quantum level. However, 
 the logarithmically divergent part of $\Gamma[H]$ is superconformal, local and gauge invariant. 
In a forthcoming paper \cite{KLFP}, we plan to explicitly show that the action so defined begins, at quadratic order, with the linearised SCHS action introduced in \cite{KMT}. 

It would be interesting to extend our analysis of section \ref{masslesschiral} to the $\cN=2$ superconformal case. 
The free actions for $\cN=2$ SCHS multiplets  $\U^{\a(s) \ad(s) } =\bar \U^{\a(s) \ad(s) } $ have recently been constructed  on arbitrary conformally flat backgrounds in \cite{Kuzenko:2021pqm}. 
The same work also sketched the program of coupling  an off-shell superconformal hypermultiplet $q$ to $\U^{\a(s) \ad(s) }$   using an action of the form
\bea
\mc{S}[q,\bar q;\U]= \mc{S}_{\rm hyper} [q , \bar q] +\sum_{s=0}^{\infty}
\int \rd^4x\rd^4\theta\rd^4\bar\theta\, E\, \U^{\a(s) \ad(s) } J_{\a(s) \ad(s)} +\cdots~.
\label{6.6}
\eea
Here  $\mc{S}_{\rm hyper} $ denotes an off-shell action for $q$, $J_{\a(s) \ad(s)} $ are conserved higher-spin supercurrents, and the ellipsis represents any terms which should be 
included to make the system consistent (in accordance with the results of this paper).
Integrating out $q$ and $\bar q$ in the effective action
\bea
\re^{ \ri \,\G[\U] } = \int \cD q \cD \bar q \,\re^{ \ri \,\mc{S}[q, \bar q; \U]}~,
\label{6.7}
\eea
and computing the logarithmically divergent part of the effective action, at lowest order one is expected to end up with the $\cN=2$ superconformal higher-spin models described in \cite{Kuzenko:2021pqm}.

The supercurrents $J_{\a(s) \ad(s)}$ given in \cite{Kuzenko:2021pqm} were constructed in terms of an on-shell hypermultiplet.
In order to compute the path integral \eqref{6.7} these supercurrents must be extended off the mass shell. 
There exist two powerful superspace approaches which offer off-shell formulations for the charged hypermultiplet: the harmonic superspace \cite{GIKOS,GIOS} and the projective superspace \cite{KLR,LR1,LR2}.\footnote{See \cite{K2010} for a brief review of these approaches and their relationship.} 
Both formulations can be used to compute the path integral \eqref{6.7}.\footnote{Recently, off-shell formulations for massless $\cN=2$ higher-spin multiplets were constructed in harmonic superspace in \cite{BIZ1,BIZ2}. These results may be useful to uncover how the $\cN=2$ SCHS prepotentials of \cite{Kuzenko:2021pqm} 
are embedded in harmonic superfields.}
 Extending our $\cN=1$ analysis to the $\cN=2$ case should require the use $\cN=2$ conformal superspace \cite{ButterN=2} which is compatible with the concept of covariant projective multiplets \cite{KLRT-M2,Butter-projective1, Butter-projective2}.

So far, conformal matter has not yet been successfully put in a background of half-integer spin-$(s+\frac12)$ conformal fields $\psi_{\a(s+1)\ad(s)}$. This is required in order to define an interacting theory of fermionic CHS fields as an induced action.  Such a model is contained, at the component level, within the action $\mc{S}[\Phi,H]$ constructed in section \ref{secChiralBack}.
A component analysis of the non-linear model $\mc{S}[\Phi,H]$ is a non-trivial task and would require an independent study, however we can make some preliminary observations. First, the constituent matter fields of the chiral multiplet $\F$ are a weight one complex scalar $\varphi$, a weight $3/2$ Majorana fermion $\phi_{\a}$, and a weight two auxiliary complex scalar $f$, which we collectively denote $\Sigma^I$ (including conjugates). 
Each of these matter fields will transform into one another under the corresponding linear variation, i.e. $\delta^{1} \Sigma^I = -\hat{\mc{U}}^I{}_J\Sigma^J$ for some non-diagonal $\hat{\mc{U}}$.
Second, there will be a single copy of each fermionic CHS field and a double copy of each bosonic CHS field present.
Each of these CHS fields will also transform into one another under the non-abelian gauge transformations. 

These observations appear to be a consequence of the fact that we have started from a model in superspace, which therefore has supersymmetry built into it. 
However, we believe that one would arrive at the same conclusions regardless of this starting point. 
Indeed, beginning with some arbitrary matter multiplet $\Sigma^I$ in spacetime, if one wants to couple it to a background tower of fermionic CHS fields then, for cubic vertices to exist, $\Sigma^I$ must consist of both bosonic and fermionic matter. The multiplet $\Sigma^I=\{\varphi,\phi_{\a},f\}$ is the simplest candidate, and the fields must transform into one another.
It is also clear that the tower cannot consist of purely fermionic CHS fields since for transformations $\delta^{1}_{\xi} \Sigma^I = -\hat{\mc{U}}_{\xi}^I{}_J\Sigma^J$ parameterised by fermionic fields $\xi$,  one needs bosonic parameters and hence bosonic CHS fields to close the corresponding gauge algebra. 
Furthermore, if one adopts the point of view that models such as $\mc{S}[\Sigma^I,h,\psi]$ etc. arise from gauging all rigid symmetries of the free matter action,\footnote{See appendix \ref{AppRigid} for a summary of this perspective.} then (conformal) supergravity is forced upon them. This is because the matter action $\mc{S}_0[\Sigma^I]$ has (conformal) supersymmetry as a rigid symmetry, and one therefore ends up with local (conformal) supersymmetry.

As the above discussion suggests, different types of (super)conformal matter $\Sigma^I$ will generally give rise to conserved  (super)currents with differing properties (e.g. conservation equation and conformal representation). As a result, the spectrum of (super)conformal gauge fields $\mf{h}^A$ which they couple to in $\mc{S}[\Sigma, \mf{h}]$, as well as the associated non-linear induced action $\mc{S}_{\text{CHS}}[\mf{h}]$, changes accordingly. For example, the primary weight $(2-k)$ scalar field $\vf$ with Lagrangian density $\bar{\vf}\Box^k\vf$ has been conjectured \cite{BG} to couple to CHS fields $h^{(t)}_{\a(s)\ad(s)}$ with (linearised) depth-$t$ gauge transformations $\delta h^{(t)}_{\a(s)\ad(s)}= \nabla_{\a\ad}^t\xi_{\a(s-t)\ad(s-t)}$ introduced in \cite{Vasiliev2}.  
Using the formalism developed in this work, it would be interesting to work out the details of this story, and also its supersymmetric analogue, for which the results of \cite{Kuzenko:2019eni} should be useful.

 

\noindent
{\bf Acknowledgements:}\\
The authors are grateful to X. Bekaert, I. L. Buchbinder, E. Joung, A. Yu. Segal and A. A. Tseytlin for useful comments. 
The work of SK and MP is supported in part by the Australian 
Research Council, project No. DP200101944.
The work of ER is supported by the Hackett Postgraduate Scholarship UWA,
under the Australian Government Research Training Program. Both MP and ER thank the MATRIX Institute in Creswick for hospitality and support during part of this project.

\appendix


\section{Geometry of conformal (super)gravity}\label{AppCSS}

This appendix is aimed at briefly reviewing the geometric approach to conformal gravity in $d > 3$ dimensions and its $\cN=1$ supersymmetric extension for $d=4$.

\subsection{Geometry of conformal gravity in $d > 3$ dimensions}
\label{ConfGeoAppendix}

Here we review the salient details of conformal geometry in $d>3$ dimensions pertinent to this work. Our construction follows Refs. \cite{BKNT-M1,BKNT}, which employ the gauging procedure of the conformal group advocated in the original works \cite{KTvanN77,KTvanN78}.

We consider a curved spacetime $\mathcal{M}^{d}$ parametrised by local coordinates $x^m$, $m = 0,~\dots~,d-1$. The structure group is chosen to be the conformal group, $\sSO(d,2)$, whose Lie algebra is spanned by the translation $P_a$, Lorentz $M_{ab}$, dilatation $\mathbb{D}$ and the special conformal $K_a$ generators. They obey the commutation relations:
\begin{subequations}
\label{confAlgebra}
\begin{align}
[M_{ab},M_{cd}] &= 2 \eta_{c[a} M_{b]d} - 2 \eta_{d[a} M_{b]c} ~, \\
[M_{ab}, P_{c}] &= 2 \eta_{c[a} P_{b]} ~, \qquad \qquad \qquad \qquad [\mathbb{D}, P_a] = P_a ~, \\
[M_{ab}, K_{c}] &= 2 \eta_{c[a} P_{b]} ~, \qquad \qquad \qquad \qquad [\mathbb{D}, K_a] = -K_a ~, \\
[K_{a}, P_b] &= 2 \eta_{ab} \mathbb{D} + 2 M_{ab} ~,
\end{align}
\end{subequations}
where all other commutators vanish.

In order to define the covariant derivatives $\nabla_a$, we associate with each non-translational generator $(M_{ab}, \mathbb{D}, K_a)$ a connection one form $(\o_{a}{}^{bc}, \mathfrak{b}_a, \mathfrak{f}_{a}{}^{b})$ and with $P_a$ the vielbein $e_{a}{}^{m}$. These play the role of gauge fields for their corresponding gauge symmetries. The covariant derivatives are then given by
\begin{align}
\label{A.1}
\nabla_{a}=e_{a}{}^{m}\partial_m-\frac{1}{2}\omega_{a}{}^{bc}M_{bc}-\mathfrak{b}_a\mathbb{D}-\mathfrak{f}_{a}{}^{b}K_b~.
\end{align}
We note that the translation generators do not show up in \eqref{A.1}. It is assumed that they are replaced by the covariant derivatives, which obey the relations
\bea
\label{A.3}
\big[M_{ab},\nabla_c\big]=2\eta_{c[a}\nabla_{b]}~,\qquad \big[\mathbb{D},\nabla_{a} \big]=\nabla_{a}~,\qquad \big[K_{a},\nabla_{b}\big] &=& 2 \eta_{ab} \mathbb{D} + 2 M_{ab} ~.
\eea

To describe conformal gravity, the torsion and curvature tensors appearing in $[\nabla_a , \nabla_b]$ must be subject to 
certain covariant constraints such that the latter is expressed solely in terms of the Weyl tensor $C_{abcd} = C_{[ab][cd]}$, which is a primary field of weight $2$, and its covariant derivatives. We remind the reader that a field $\psi$ (with suppressed indices) is said to be primary and of weight $\D$ if it obeys 
\be
K_a \psi = 0 ~, \qquad \mathbb{D} \psi = \D \psi ~.
\ee
The solutions to the aforementioned constraints are as follows
\bea
\label{A.2}
\big[\nabla_{a},\nabla_{b} \big]&=&-\hf C_{abcd} M^{cd} - \frac{1}{2(d-3)} \nabla^d C_{abcd} K^{c} ~.
\eea

One can always make use of the special conformal gauge freedom to choose a vanishing dilatation connection, $\mf{b}_a = 0$. The conformal covariant derivative then takes the form\footnote{Strictly speaking, the $\omega_{abc}$ appearing in \eqref{A.1} is different to the one appearing in \eqref{fextract}. However the two only differ by expressions involving $\mf{b}_a$, which has been set to zero here.  }
\be \label{fextract}
\nabla_a = \cD_a - \frak{f}_a{}^b K_b \ , \quad \cD_a := e_a - \hf \omega_a{}^{bc} M_{bc} \ .
\ee 
Operator $\cD_a $ coincides with the usual (torsion-free) Lorentz covariant derivative. In this gauge the Lorentz curvature
\be
\cR_{ab}{}^{cd} := 2 e_{[a}{}^m e_{b]}{}^n \partial_m \omega_{n}{}^{cd}- 2 \omega_{[a}{}^{cf} \omega_{b]f}{}^{d}
\ee
may be expressed as
\be  \cR_{ab}{}^{cd} = C_{ab}{}^{cd} - 8 \d_{[a}^{[c} \mathfrak{f}_{b]}{}^{d]}  \ .
\ee
One can then solve the special conformal connection in terms of the Lorentz curvature
\be \mathfrak{f}_{ab} = - \frac{1}{2 (d - 2)} \cR_{ab} + \frac{1}{4 (d - 1) (d - 2)} \eta_{ab} \cR \ , 
\ee
where we have defined
\be \cR_{ac} := \eta^{bd} \cR_{abcd} \ , \quad \cR := \eta^{ab} \cR_{ab} \ .
\ee
This procedure of setting $\mf{b}_a = 0$ and introducing the derivative $\cD_a$ is known as degauging.

\subsection{Geometry of $\cN=1$ conformal supergravity in $d=4$}
\label{CSSappendix}

This appendix briefly reviews relevant aspects of $\mathcal{N}=1$ conformal superspace in four dimensions  \cite{ButterN=1}, which is a natural supersymmetric extension of the purely bosonic geometry discussed above. Further details can be found in the original paper \cite{ButterN=1} and in the review located in appendix A of \cite{Kuzenko:2020jie}. Our spinor conventions coincide with those of \cite{BK}.

To begin, we consider a curved superspace $\mathcal{M}^{4|4}$ parametrised by local coordinates 
$z^{M} = 
(x^{m},\theta^{\m},\bar \theta_{\dot{\mu}})$, $m = 0,1,2,3$, $\mu = 1,2$, $\dot{\mu} = \dot{1},\dot{2}$. Its structure group is chosen to be $\sSU(2,2|1)$, the $\cN=1$ superconformal group, whose Lie superalgebra is
spanned by the supertranslation $P_A=(P_a, Q_\a ,\bar Q^\ad)$, Lorentz $M_{ab}$, R-symmetry $\mathbb{Y}$, dilatation $\mathbb{D}$ and the special superconformal $K^A=(K^a, S^\a ,\bar S_\ad)$ generators. They obey, in addition to \eqref{confAlgebra}, the following (anti-)commutation relations
\begin{subequations}
	\begin{align}
	[\mathbb{Y}, Q_\a] &= Q_\a ~, \quad  [\mathbb{Y}, \bar{Q}^\ad] = - \bar{Q}^\ad~, \quad
	[\mathbb{D}, Q_\a] = \hf Q_\a ~,
	\quad [\mathbb{D}, \bar{Q}^\ad ] = \hf \bar{Q}^\ad ~, \\
	[\mathbb{Y}, S^\a] &= - S^\a ~,
	\quad [\mathbb{Y}, \bar{S}_\ad] = \bar{S}_\ad~, \quad
	[\mathbb{D}, S^\a] = - \hf S^\a~,
	\quad [\mathbb{D}, \bar{S}_\ad ] = - \hf \bar{S}_\ad ~, \\
	\{ S_\a , Q_\b \} &= \ve_{\a \b} \big( 2 \mathbb{D} - 3 \mathbb{Y} \big) - 4 M_{\a \b} ~, \qquad
	\{ \bar{S}_\ad , \bar{Q}_\bd \} = - \ve_{\ad \bd} \big( 2 \mathbb{D} + 3 \mathbb{Y}) + 4 \bar{M}_{\ad \bd}  ~, \\
	[S_\a , P_\bb] &= 2 \ri \ve_{\a \b} \bar{Q}_{\bd} \ , \qquad \quad [\bar{S}_\ad , P_\bb] =
	- 2 \ri \ve_{\ad \bd} Q_{\b} ~, \qquad \quad 	\{ S_\a , \bar{S}_\ad \} = 2 \ri  K_{\aa} ~, \\
	[K_{\a \ad}, Q_\b] &= - 2 \ri \ve_{\a \b} \bar{S}_{\ad} \ , \qquad \quad [K_\aa, \bar{Q}_\bd] =
	2 \ri  \ve_{\ad \bd} S_{\a} ~,
	\end{align}
\end{subequations}
where all other graded commutators vanish.

In analogy with the bosonic construction of the previous subsection, we associate with each non-translational generator $(M_{ab},\mathbb{Y},\mathbb{D},K^A)$ a connection one-form $(\O_A{}^{bc}, \F_A, B_A, \mathfrak{F}_{AB})$ and with $P_A$ the supervielbein $E_{A}{}^M$. These then play the role of gauge fields for the corresponding gauge symmetries and appear in the superspace covariant derivatives
\begin{align}
\label{CSSD}
\nabla_A &= (\nabla_a, \nabla_\alpha, \bar\nabla^\ad)	=E_A{}^M \pa_M - \hf \Omega_A{}^{bc} M_{bc} - \ri \Phi_A \mathbb{Y}
- B_A \mathbb{D} - \mathfrak{F}_{AB}K^B ~.
\end{align}
We note that the translation generators do not appear in \eqref{CSSD}. It is assumed that they are replaced by $\nabla_A$, which, in addition to \eqref{A.3}, obey
\begin{subequations}
	\label{A.8}
	\begin{align}
	[\mathbb{Y}, \nabla_\a] &= \nabla_\a ~, \quad  [\mathbb{Y}, \bar{\nabla}^\ad] = - \bar{\nabla}^\ad~, \quad
	[\mathbb{D}, \nabla_\a] = \hf \nabla_\a ~, \quad [\mathbb{D}, \bar{\nabla}^\ad ] = \hf \bar{\nabla}^\ad ~, \\
	\{ S_\a , \nabla_\b \} &= \ve_{\a \b} \big( 2 \mathbb{D} - 3 \mathbb{Y} \big) - 4 M_{\a \b} ~, \qquad
	\{ \bar{S}_\ad , \bar{\nabla}_\bd \} = - \ve_{\ad \bd} \big( 2 \mathbb{D} + 3 \mathbb{Y}) + 4 \bar{M}_{\ad \bd}  ~, \\
	[S_\a , \nabla_\bb] &= 2 \ri \ve_{\a \b} \bar{\nabla}_{\bd} \ , \qquad \quad [\bar{S}_\ad , \nabla_\bb] =
	- 2 \ri \ve_{\ad \bd} \nabla_{\b} ~,\\
	[K_{\a \ad}, \nabla_\b] &= - 2 \ri \ve_{\a \b} \bar{S}_{\ad} \ , \qquad \quad [K_\aa, \bar{\nabla}_\bd] =
	2 \ri  \ve_{\ad \bd} S_{\a} ~.
	\end{align}
\end{subequations}

To describe conformal supergravity, the torsion and curvature tensors appearing in $[\nabla_A, \nabla_B\}$ must be subject to certain covariant constraints such that the latter is expressed solely in terms of the super-Weyl tensor $W_{\a \b \g} = W_{(\a \b \g)}$, its conjugate $\bar{W}_{\ad \bd \gd}$ and their descendants. The solution to these constraints is as follows
\begin{subequations}
	\label{A.9}
	\bea
	\{ \nabla_{\a} , \nabla_{\b} \} & = & 0 ~, \quad \{\nabla_{\a} , \bar{\nabla}_{\ad} \} = - 2 \ri \nabla_{\a \ad} ~, \\
	\big[ \bar{\nabla}_{\ad} , \nabla_{\b \bd} \big] & = & - \ri \ve_{\ad \bd} \Big( 2 W_{\b}{}^{\g \d} M_{\g \d} + \frac{1}{2} \nabla^{\a} W_{\a \b \g} S^{\g} + \frac{1}{2} \nabla^{\a \gd} W_{\a \b}{}^{\g} K_{\g \gd} \Big) ~.
	\eea
\end{subequations}
The structure of this algebra leads to highly non-trivial implications. In particular, we consider a primary superfield $\J$ (with suppressed indices), $K_B \J = 0$. Its dimension $\D$ and $\sU(1)_R$ charge $q$ are defined as $\mathbb{D} \J = \D \J$ and $\mathbb{Y} \J = q \J$. Requiring $\J$ to be covariantly chiral, we find that it must not carry any dotted indices, $\J = \J_{\a(m)}$ and that its $\sU(1)_R$ charge is determined in terms of its dimension
\bea
\label{chiralDimChargeN=1}
	K^B \J_{\a(m)} = 0 ~, \quad \bar{\nabla}_\ad \J_{\a(m)} = 0 ~, \quad \implies \quad q = - \frac{2}{3} \D~.
\eea
In particular, $W_{\a \b \g}$ is a primary, covariantly chiral superfield of dimension $3/2$
\be
K^D W_{\a \b \g} =0~, \quad \bar \nabla^\dd W_{\a\b\g}=0 ~, \quad 
{\mathbb D} W_{\a\b\g} = \frac 32 W_{\a\b\g} \quad \implies \quad \mathbb{Y}W_{\a\b\g}=-W_{\a\b\g}~.
\ee


\section{Variations on themes by Noether and Segal}\label{AppRigid}

In section \ref{NPapp} we described the basic ideas of the Noether procedure, for which there are two main approaches. The second approach, which we only briefly mentioned, revolved around gauging rigid symmetries of the free matter action.\footnote{We refer the reader to \cite{BekaertRigid} for a detailed and rigorous account of rigid symmetries, and also a discussion on their gauging, which is complementary to the discussion given in this appendix. }
To construct the non-linear actions $\mc{S}[\varphi,h]$ and $\mc{S}[\Phi,H]$ in the main body, we first employed the (first) Noether approach, and then abandoned this in favour of Segal's method \cite{Segal}. 
Having the benefit of hindsight, it is instructive to \textit{sketch} a common formulation for both Segal's and Noether's (second) approach to this problem, with the aim of emphasising their similarities. 
In what follows we borrow the notation introduced in section \ref{NPapp}.  

One can view the free action $\mc{S}_0[\Sigma^I]$ of some matter multiplet $\Sigma^I$ as the starting point of both Segal and Noether's method. Let us express the free matter action in the form
 \begin{align}
 \mc{S}_0[\Sigma^I]=\int \text{d} \mu \, \bar{\Sigma}^{\bar{I}} \hat{K}_{\bar{I}J}\Sigma^J\label{NvSmatter}
 \end{align}
where the kinetic operator $\hat{K}_{\bar{I}J}$ is self-adjoint
 and $\bar{\Sigma}^{\bar{I}}$ is the complex conjugate of $\Sigma^I$. 
Consider the linear transformation $\delta^1 \Sigma^I=-\hat{\mc{V}}^I{}_J\Sigma^J$ of the matter fields. In what follows we will suppress indices with impunity. Varying \eqref{NvSmatter} we see that this is a rigid symmetry (i.e. $\delta\mc{S}_0=0$) if  $\hat{\mc{V}}\equiv \hat{\mc{V}}_0$ satisfies the equation
\begin{align}
0=\hat{\mc{V}}_0^{\dagger}\hat{K}+\hat{K}\,\hat{\mc{V}}_0~.\label{NvSRigidOpEq}
\end{align}
Clearly, this equation is automatically satisfied by any symmetry  
 of the form 
\begin{align}
\hat{\mc{V}}_0=\ri \,\hat{\o} \hat{K}~,\qquad \hat{\o}^{\dagger}=\hat{\o}~, \label{NvStrivSym}
\end{align}
with $\hat{\o}$ an arbitrary Hermitian differential operator. The symmetry associated with such an operator is therefore said to be a trivial symmetry. Operators satisfying \eqref{NvSRigidOpEq} which can not be expressed in the form \eqref{NvStrivSym} will be referred to as non-trivial rigid symmetries. 

Let us denote the linear spaces of differential operators associated with all rigid, trivial and non-trivial symmetries by $V^{\text{Diff}}_{\text{Rigid}},~V^{\text{Diff}}_{\text{Trivial}}$ and $V^{\text{Diff}}_{\text{Ntrivial}}\cong V^{\text{Diff}}_{\text{Rigid}}/V^{\text{Diff}}_{\text{Trivial}} $ respectively. 
The linear space $V^{\text{Diff}}_{\text{Rigid}}$, endowed with the commutator $[\cdot,\cdot]$ as the Lie bracket,
 forms a Lie algebra which we denote $\ms{A}^{\text{Diff}}_{\text{Rigid}}$. 
That $\big[\hat{\mc{V}}_0,\,\hat{\mc{U}}_0\big] \in V^{\text{Diff}}_{\text{Rigid}}$ for any $\hat{\mc{V}}_0,\hat{\mc{U}}_0 \in V^{\text{Diff}}_{\text{Rigid}}$ may be easily verified using \eqref{NvSRigidOpEq}. 
The subalgebra of trivial symmetries forms a two-sided ideal within $\ms{A}^{\text{Diff}}_{\text{Rigid}}$ which we denote by $\ms{A}^{\text{Diff}}_{\text{Trivial}}$. Indeed, for $\hat{\mc{V}}_0=\ri \hat{\o}\hat{K}\in V^{\text{Diff}}_{\text{Trivial}}$ and $\hat{\mc{U}}_0 \in V^{\text{Diff}}_{\text{Rigid}}$ we have 
\begin{align}
\big[\hat{\mc{U}}_0,\, \hat{\mc{V}}_0\big]=\ri\big(\hat{\mc{U}}_0\hat{\o}+\hat{\o}\hat{\mc{U}}^{\dagger}_0\big)\hat{K} \in  V^{\text{Diff}}_{\text{Trivial}}~. \label{NvSCommTriv}
\end{align}
The Lie algebra of non-trivial rigid symmetries of $\mc{S}_0$ is identified with the quotient algebra 
\begin{align} 
\ms{A}^{\text{Diff}}_{\text{Rigid}}/\ms{A}^{\text{Diff}}_{\text{Trivial}} \equiv \ms{A}^{\text{Diff}}_{\text{Ntrivial}}~. \label{Quotient}
\end{align}
Symmetries of the type \eqref{NvSCommTriv} will be of particular interest below.

Let us package all of the non-trivial rigid symmetries into the operator $\hat{\mc{U}}_{\xi_0} \in V^{\text{Diff}}_{\text{Ntrivial}}$,\footnote{Strictly speaking, the elements of $V^{\text{Diff}}_{\text{Ntrivial}}$ are cosets. In the following we always work with a representative. }
 which is parameterised by the set of fields $\xi^A_0$ (e.g. $\sum_{s=0}^{\infty}\hat{\mc{U}}^{(s)}_{\z_{\text{ckt}}}$ in section \ref{secCSrigid}).
Parameters $\xi^A_0$ must satisfy special constraints in order for \eqref{NvSRigidOpEq} to hold, which we denote
\begin{align}
0=\mf{D}^A{}_B\,\xi_0^B \label{NvSconstrainedPar}
\end{align}
for some differential operator $\mf{D}^A{}_B$.
For example, \eqref{NvSconstrainedPar} could be the conformal Killing equation. 
The first step in gauging the non-trivial rigid symmetries $\hat{\mc{U}}_{\xi_0}$ is to relax the condition \eqref{NvSconstrainedPar}, and hence \eqref{NvSRigidOpEq}, and replace the parameters $\xi^A_0$ with unconstrained ones $\xi^A$. The variation of the matter action then  takes either of the two equivalent forms
\begin{align}
\delta_{\xi}\mc{S}_0[\Sigma^I] = -\int \text{d} \mu \, \mf{D}^A{}_B\,\xi^B\mf{J}_A = -\int \text{d} \mu \, \bar{\Sigma}^{\bar{I}}\big( \, \hat{\mc{U}}_{\xi}^{\dagger}\hat{K}+\hat{K}\,\hat{\mc{U}}_{\xi} \,\big)_{\bar{I}J}\Sigma^J ~. \label{NvSnzvar}
\end{align}
Here $\mf{J}_A$ is some tensor bilinear in $\Sigma^I$ which is conserved on-shell $\big(\mf{D}^A{}_B\big)^{\rm T}\mf{J}_A\approx 0$. 

To compensate for \eqref{NvSnzvar}, one introduces a gauge field $\mf{h}^A$ which transforms as
\begin{align}
\delta_{\xi}^0 \mf{h}^A= \mf{D}^A{}_B\xi^B~. \label{NvSzeroHgt1}
\end{align}
The latter couples to $\mf{J}_A$ via the Noether coupling and the matter action is altered accordingly
\begin{align}
\mc{S}_{\text{cubic}}[\Sigma^I,\mf{h}^A] = \mc{S}_0[\Sigma^I]+\mc{S}_1[\Sigma^I,\mf{h}^A]~,\qquad \mc{S}_1[\Sigma^I,\mf{h}^A]=\int \text{d}\mu \, \mf{h}^A \mf{J}_A~.\label{NvScubicAct1}
\end{align}
It is informative to integrate by parts, after which one arrives at the equivalent expression
\begin{align}
\mc{S}_{\text{cubic}}[\Sigma^I,\mf{h}^A]=\int \text{d} \mu \, \bar{\Sigma}^{\bar{I}} \big(\hat{K}_{\bar{I}J} + \hat{H}_{\bar{I}J} \big)\Sigma^J \label{NvScubicAct2}
\end{align}
where $\hat{H}$ is a self-adjoint operator parameterised by (and linear in) the gauge fields $\mf{h}^A$.  From the second equality in \eqref{NvSnzvar}, we see that the operator $\hat{H}$ must vary according to 
\begin{align}
\delta^0_{\xi} \hat{H}_{\bar{I}J}= \big(\hat{\mc{U}}_{\xi}^{\dagger}\hat{K}+\hat{K}\,\hat{\mc{U}}_{\xi}\big)_{\bar{I}J}~. \label{NvSzeroHgt2}
\end{align} 
By construction, this operator equation is equivalent to the tensor equation \eqref{NvSzeroHgt1}.

At this point, gauge invariance holds only to first order in the gauge fields, $\delta_{\xi}\mc{S}_{\text{cubic}}=\mc{O}(\mf{h}^A)$. 
More precisely, under $\delta^1 \Sigma^I=-\hat{\mc{U}}_{\xi}^I{}_J\Sigma^J$ and \eqref{NvSzeroHgt2}, the cubic action \eqref{NvScubicAct2} varies as
\begin{align}
\delta_{\xi}\mc{S}_{\text{cubic}}=-\int \text{d}\mu\, \bar{\Sigma}^{\bar{I}}\big(\hat{\mc{U}}^{\dagger}\hat{H}+\hat{H}\,\hat{\mc{U}}\big)_{\bar{I}J}\Sigma^J~.
\end{align}
This may be compensated for by deforming the gauge transformation of $\hat{H}_{\bar{I}J}$ according to 
\begin{align}
\delta_{\xi}\hat{H}_{\bar{I}J}=\delta_{\xi}^0\hat{H}_{\bar{I}J}+\delta_{\xi}^1\hat{H}_{\bar{I}J}~,\qquad \delta^1_{\xi} \hat{H}_{\bar{I}J}= \big(\hat{\mc{U}}_{\xi}^{\dagger}\hat{H}+\hat{H}\,\hat{\mc{U}}_{\xi}\big)_{\bar{I}J}~.
\end{align} 
This is of course equivalent to $\delta_{\xi}\mf{h}^A=\delta_{\xi}^0\mf{h}^A+\delta_{\xi}^1\mf{h}^A$ for some $\delta_{\xi}^1\mf{h}^A$.\footnote{For the examples considered in this paper, the variation $\delta_{\xi}\mf{h}^A$ can be recovered from $\delta_{\xi}\hat{H}_{\bar{I}J}$ via the (un)dressing maps provided in appendix \ref{Appendix B}.} 
At first sight, it appears as though we have formed a system consistent at all orders, since \eqref{NvScubicAct2} is invariant under $\delta^1 \Sigma^I=-\hat{\mc{U}}_{\xi}^I{}_J\Sigma^J$ and $\delta_{\xi}\hat{H}_{\bar{I}J}$. However, as discussed in the main body, the aforementioned gauge transformations do not generally form a closed algebra, as they ought to (see e.g. \cite{BBvD}). 

The problem is that so far we have only considered the non-trivial rigid symmetries (recall that we specified $\mc{\hat{U}}_{\xi_0}\in V^{\text{Diff}}_{\text{Ntrivial}}$).
 In general, once these are gauged (i.e. when $\xi_0^A$ is promoted to be unconstrained), they do not form a closed algebra.\footnote{We are unaware of a general result asserting whether the space $V^{\text{Diff}}_{\text{Ntrivial}}$ endowed with $[\cdot,\cdot]$ forms a subalgebra within $\ms{A}^{\text{Diff}}_{\text{Rigid}}$, or in other words, if for $\hat{\mc{V}}_{\xi_0},\hat{\mc{V}}_{\z_0}\in V^{\text{Diff}}_{\text{Ntrivial}}$ one has $[\hat{\mc{V}}_{\xi_0},\hat{\mc{V}}_{\z_0}]=\hat{\mc{V}}_{\l_0}$ for some $\hat{\mc{V}}_{\l_0} \in V^{\text{Diff}}_{\text{Ntrivial}}$. } To ensure the algebra closes, one should gauge the trivial symmetries of $\mc{S}_0$ too. 
 The trivial symmetries that can be gauged are the ones generated by taking the commutator of $\mc{\hat{U}}_{\xi_0}$ and $\hat{\mc{V}}_{\omega}=\ri \hat{\o}\hat{K}\in V^{\text{Diff}}_{\text{Trivial}}$ (cf. \eqref{CStrivckt}):
\begin{align}
\big[\hat{\mc{U}}_{\xi_0},\, \hat{\mc{V}}_{\o}\big]=\ri\big(\hat{\mc{U}}_{\xi_0}\hat{\o}\hat{K}-\hat{\o}\hat{K}\hat{\mc{U}}_{\xi_0}\big)\equiv \hat{\mc{V}}_{\xi_0}~. \label{blahblah}
\end{align}
 Note that we have not expressed this commutator as we did in \eqref{NvSCommTriv}, since we will now promote $\xi_0^A$ to be unconstrained, and hence relax condition \eqref{NvSRigidOpEq} for $\mc{\hat{U}}_{\xi_0}$. This means that the corresponding $\hat{\mc{V}}_{\xi}$ in \eqref{blahblah} is no longer a trivial symmetry of $\mc{S}_0$, and we must introduce another set of gauge fields $\mf{h}_{\text{aux}}^A$ to compensate for this. 
 
  Repeating the same procedure as above, one ends up with the action
 \begin{align}
 \mc{S}[\Sigma^I,\mf{h}^A,\mf{h}^A_{\text{aux}}]=\int \, \text{d}\mu \, \bar{\Sigma}^{\bar{I}}\hat{G}_{\bar{I}J}\Sigma^J ~,\qquad \hat{G}_{\bar{I}J}=\big(\hat{K} + \hat{H} +\hat{H}_{\text{aux}}\big)_{\bar{I}J}
 \end{align}
 where $\hat{H}_{\text{aux}}$ is a self-adjoint operator parameterised by (and linear in) the gauge fields $\mf{h}_{\text{aux}}^A$.
 The matter fields $\Sigma^I$ and the operator $\hat{G}_{\bar{I}J}$ transform according to
 \begin{align}
 \delta \Sigma^I=-\big(\hat{\mc{U}}_{\xi}+\hat{\mc{V}}_{\xi}\big)^I{}_{J}\Sigma^J~,\qquad \delta\hat{G}_{\bar{I}J} =\Big(\big(\hat{\mc{U}}_{\xi}+\hat{\mc{V}}_{\xi}\big)^{\dagger}\hat{G} +\hat{G}\big(\hat{\mc{U}}_{\xi}+\hat{\mc{V}}_{\xi}\big) \Big)_{\bar{I}J}~.
 \end{align}
This forms a consistent system to all orders in the gauge fields, and hence the rigid symmetries have been successfully gauged.
 The configuration $\hat{G}_0\equiv \hat{G}\big|_{\mf{h}^A=\mf{h}^A_{\text{aux}}=0}=\hat{K}$ may be referred to as the `vacuum' of the theory, and then the rigid symmetries have the usual interpretation as those transformations preserving the vacuum.


\section{Direct computation of general primary operator} \label{Appendix A}

Rather than using the compensator $\eta$ to build primary operators, one can instead proceed by direct computation and employ the same procedure that allowed us to deduce \eqref{CSOp}. More precisely, the most general differential operator of maximal order $s$ has the form
\begin{align}
\hat{\mathcal{W}}^{(s)}&=\sum_{l=0}^{\lfloor s/2 \rfloor}\sum_{k=0}^{s-2l}\xi^{\a(k)\ad(k)}_{(s,l)}\nabla_{\a\ad}^k\Box^l~,\label{CSgenOPbf}
\end{align}
for some coefficients $\xi^{\a(k)\ad(k)}_{(s,l)}$ with weight $-(k+2l)$. Upon imposing the constraint $K_{\b\bd}\hat{\mathcal{W}}^{(s)}\vf =0$ one finds that for each fixed $l$ the lowest weight (highest rank) parameter $\xi^{\a(s-2l)\ad(s-2l)}_{(s,l)}\equiv \xi^{\a(s-2l)\ad(s-2l)}_{[s,l]}$ is primary and hence has the properties
\begin{subequations}
\begin{align}
\mathbb{D}\xi_{[s,l]}^{\a(s-2l)\ad(s-2l)}=-s\xi_{[s,l]}^{\a(s-2l)\ad(s-2l)}~,\qquad  K_{\b\bd}\xi_{[s,l]}^{\a(s-2l)\ad(s-2l)}=0~,\qquad 0\leq l \leq \lfloor s/2 \rfloor ~. \label{ConfpropCSgpv2}
\end{align}
In addition, all other parameters are descendants of $\xi_{[s,l]}^{\a(s-2l)\ad(s-2l)}$  satisfying the equation
\begin{align}
0&=K_{\b\bd}\xi_{(s,l)}^{\a(k)\ad(k)}-4\frac{k+1}{k+2}(k+l+1)(k+l+2)\xi^{\a(k)}_{(s,l)}{}_{\b\bd}{}^{\ad(k)}+4l(l+1)\xi^{\a(k-1)\ad(k-1)}_{(s,l+1)}\delta_{\b}^{\a}\delta_{\bd}^{\ad}  \label{genopCon}
\end{align}
\end{subequations}
for $0\leq l \leq \lfloor s/2 \rfloor$ and $0\leq k \leq s-2l-1$. For $l=0$ the last term in \eqref{genopCon} is not present and we recover \eqref{CSSymm2}. Hence the $l=0$ sector forms a closed subsector and is solved by the expressions \eqref{l=0subsector} with $\xi_{(s,0)}\mapsto \z_{(s)}$.  For $1\leq l \leq \lfloor s/2 \rfloor$ all parameters in \eqref{genopCon} with distinct $l$ and $k$ are mixed, and a suitable ansatz for the descendants $\xi^{\a(k)\ad(k)}_{(s,l)}$ is 
\begin{align}
\xi^{\a(k)\ad(k)}_{(s,l)}=\sum_{p=0}^{k}\sum_{q=l+p}^{\lceil \frac{s-k+p}{2}\rceil} a_{(s,l,k,p,q)}\Box^{q-l-p}&\big(\nabla^{\a\ad}\big)^p\nabla_{\b\bd}^{s-2q+p-k} \non\\
&\times\xi_{[s,q]}^{\a(k-p)\b(s-2q+p-k)\ad(k-p)\bd(s-2q+p-k)}~, \label{Ansatzgenop}
\end{align}
for $1 \leq l \leq \lfloor s/2 \rfloor$ and $0 \leq k \leq s-2l -1 $. Upon inserting \eqref{Ansatzgenop} into \eqref{genopCon}, it should be possible to determine the numerical coefficients $a_{(s,l,k,p,q)}$ uniquely, up to an overall normalisation. However, in the general case, this proves to be a difficult task. 

For example, the expressions for the operator $\hat{\mathcal{W}}^{(s)}$ with $1 \leq s \leq 4$ are
\begin{subequations}
\begin{align}
\hat{\mathcal{W}}^{(1)}&= \hat{\mathcal{U}}^{(1)}_{[1,0]}~,\\
\hat{\mathcal{W}}^{(2)}&= \hat{\mathcal{U}}^{(2)}_{[2,0]} + \xi_{[2,1]}\Box~, \\
\hat{\mathcal{W}}^{(3)}&= \hat{\mathcal{U}}^{(3)}_{[3,0]} + \xi^{\a\ad}_{[3,1]}\nabla_{\a\ad}\Box
+\frac{1}{2} \nabla_{\b\bd}\xi^{\b\bd}_{[3,1]}\Box~,\\
\hat{\mathcal{W}}^{(4)}&= \hat{\mathcal{U}}^{(4)}_{[4,0]} + \xi^{\a(2)\ad(2)}_{[4,1]}\nabla_{\a\ad}^2\Box + \nabla_{\b\bd}\xi^{\a\b\ad\bd}_{[4,1]}\nabla_{\a\ad}\Box+\frac{3}{14}\nabla_{\b\bd}^2\xi^{\b(2)\bd(2)}_{[4,1]}\Box~\non\\
&\phantom{=}+\xi_{[4,2]}\Box^2+\frac{3}{10}\Box\xi_{[4,2]}\Box-\frac{1}{2}\nabla^{\a\ad}\xi_{[4,2]}\nabla_{\a\ad}\Box~, 
\end{align}
\end{subequations}
where $\hat{\mathcal{U}}^{(s)}_{[s,0]}$ is given by \eqref{newGPop}.\footnote{In both the compensator and the direct computation methods, the $l=0$ subsector is exactly the same and so we denote it by the same notation $\hat{\mathcal{U}}^{(s)}_{[s,0]}$.} 
To clarify the relation between the two approaches (compensator and direct), we recall the corresponding expressions for $\hat{\mathcal{V}}^{(s)}$,\footnote{One could instead place the powers of $\eta\Box$ to the left, for example $\hat{\mc{V}}^{(s)}:=\sum_{l=0}^{\lfloor s/2 \rfloor}\big(\eta\overrightarrow{\square}\big)^l\hat{\mc{U}}^{(s-2l)}_{[s,l]} $. However, these expressions may also be shown to be equivalent in a similar sense to \eqref{Equivalence} (except in this case the parameter in $\hat{\mathcal{U}}^{(s)}_{[s,0]}$ will also have to be redefined). }
\begin{subequations}
\begin{align}
\hat{\mathcal{V}}^{(1)}&=\hat{\mathcal{U}}^{(1)}_{[1,0]} ~,\\
\hat{\mathcal{V}}^{(2)}&=\hat{\mathcal{U}}^{(2)}_{[2,0]} + \z_{[2,1]}\eta\Box ~,\\
\hat{\mathcal{V}}^{(3)}&=\hat{\mathcal{U}}^{(3)}_{[3,0]} + \Big(\z^{\a\ad}_{[3,1]}\overrightarrow{\nabla}_{\a\ad}+\frac{1}{4}\nabla_{\b\bd}\z^{\b\bd}_{[3,1]}\Big)\eta\Box~,\\
\hat{\mathcal{V}}^{(4)}&=\hat{\mathcal{U}}^{(4)}_{[4,0]} + \Big(\z^{\a(2)\ad(2)}_{[4,1]}\overrightarrow{\nabla}^2_{\a\ad}+\frac{2}{3}\nabla_{\b\bd}\z^{\a\b\ad\bd}_{[4,1]}\overrightarrow{\nabla}_{\a\ad}+\frac{1}{15}\nabla_{\b\bd}^2\z^{\b(2)\bd(2)}_{[4,1]}\Big)\eta\Box\non\\
&\phantom{=}+\z_{[4,2]}(\eta\overrightarrow{\square})(\eta\overrightarrow{\square})~.
\end{align}
\end{subequations}
In the case $s=1$ it is clear that $\hat{\mathcal{V}}^{(1)}=\hat{\mathcal{W}}^{(1)}$, whilst for $s=2$ we see that $\hat{\mathcal{V}}^{(2)}=\hat{\mathcal{W}}^{(2)}$ upon making the identification $\xi_{[2,1]}=\eta\z_{[2,1]}$. For $s>2$, to compare the two methods it is necessary to compare the two truncated towers of operators $\hat{\mathcal{V}}_{\text{trunc}}^{(s)}:=\sum_{k=0}^{s}\hat{\mathcal{V}}^{(s)}$ and $\hat{\mathcal{W}}_{\text{trunc}}^{(s)}:=\sum_{k=0}^{s}\hat{\mathcal{W}}^{(s)}$ for fixed values of $s$. For $s=3$ one may show that $\hat{\mathcal{V}}_{\text{trunc}}^{(3)}=\hat{\mathcal{W}}_{\text{trunc}}^{(3)}$ upon making the identifications $\xi^{\a\ad}_{[3,1]}=\eta \z^{\a\ad}_{[3,1]}$ and $\xi_{[2,1]}= \eta \z_{[2,1]}+\l$ where $\l=\frac{1}{2}\nabla_{\a\ad}\eta\z^{\a\ad}_{[3,1]}-\frac{1}{4}\eta\nabla_{\b\bd}\z^{\b\bd}_{[3,1]}$ is a primary scalar field of weight $-2$. Similarly, for $s=4$ one may show that $\hat{\mathcal{V}}_{\text{trunc}}^{(4)}=\hat{\mathcal{W}}_{\text{trunc}}^{(4)}$ upon identifying 
\begin{subequations} \label{Equivalence}
\begin{align}
\xi^{\a(2)\ad(2)}_{[4,1]}&=\eta\z_{[4,1]}^{\a(2)\ad(2)}~,\\
\xi_{[4,2]}&=\eta^2\z_{[4,2]}~,\\
\xi^{\a\ad}_{[3,1]}&=\eta\z^{\a\ad}_{[3,1]}+\nabla_{\b\bd}\eta\z^{\a\b\ad\bd}_{[4,1]}+\frac{1}{3}\eta\nabla_{\b\bd}\z^{\a\b\ad\bd}_{[4,1]}-\frac{1}{2}\eta^2\nabla^{\a\ad}\z_{[4,2]}~,\\
\xi_{[2,1]}&=\eta\z_{[2,1]}+ \frac{2}{7}\nabla_{\a\ad}^2\eta\z^{\a(2)\ad(2)}_{[4,1]}-\frac{3}{7}\nabla_{\a\ad}\eta\nabla_{\a\ad}\z^{\a(2)\ad(2)}_{[4,1]}-\frac{11}{35}\eta\nabla_{\a\ad}^2\z^{\a(2)\ad(2)}_{[4,1]}\non\\
&\phantom{=}+\frac{2}{5}\eta\Box\eta\z_{[4,2]}+\frac{11}{10}\eta\nabla_{\a\ad}\eta\nabla^{\a\ad}\z_{[4,2]}-\frac{4}{5}\eta^2\Box\z_{[4,2]}+\frac{3}{10}\nabla^{\a\ad}\eta\nabla_{\a\ad}\eta\z_{[4,2]} \non\\
&\phantom{=} +\frac{1}{2} \nabla_{\a\ad}\eta\z^{\a\ad}_{[3,1]}-\frac{1}{4} \eta\nabla_{\a\ad}\z^{\a\ad}_{[3,1]}~.
\end{align}
\end{subequations}
These definitions are of course consistent with the properties \eqref{ConfpropCSgp} and \eqref{ConfpropCSgpv2}.
In the general case, the two towers are related via complicated field identifications which begin
\begin{align}
\xi_{[s,l]}^{\a(s-2l)\ad(s-2l)}=\eta^{l}\z_{[s,l]}^{\a(s-2l)\ad(s-2l)}+\cdots~.
\end{align}
Hence the two methods are equivalent when considering a tower  of transformations  like \eqref{genvarCS}.


\section{Hermitian adjoint of a primary operator} \label{AppendixAdjoint}

In order to promote the (S)CHS fields from infinitesimal sources to background fields, in sections \ref{secCS} and \ref{masslesschiral} we have employed a formalism which makes use of the Hermitian adjoint of various operators. The latter operation involves integrating by parts, which in conformal (super)space is a subtle and technical issue.\footnote{For a detailed discussion of this, we refer the reader to appendix B of \cite{Kuzenko:2020jie}. In summary, beginning with a primary Lagrangian, naive integration by parts (IBP) is valid only when the Lagrangian one ends with is also primary, which should be checked by hand.  Although this IBP rule holds in all cases known to the authors, it still remains a conjecture, and we assume it is valid throughout this work.  } Furthermore, in superspace there is a certain amount of ambiguity in integrating by parts when constrained (e.g. chiral) superfields  are involved. 
Due to these subtle points it is necessary to give a careful definition of what is meant by the Hermitian adjoint of an operator in this context, which is the purpose of this appendix. 

\subsection{Complex scalar field}

Let $\mf{S}_{\Delta}$ denote the space of complex primary scalar fields with conformal weight $\Delta$. In section \ref{secCS} we encountered differential operators of the type $\hat{H}: \mf{S}_1 \rightarrow \mf{S}_3 $ and $\hat{\mc{U}}:\mf{S}_1 \rightarrow \mf{S}_1$. To compute the Hermitian adjoint of these operators we introduce the two pairings
\begin{align}
\langle \cdot, \cdot \rrangle :\qquad \mf{S}_1 \times \mf{S}_3 \rightarrow \mathbb{C}~,\qquad \qquad \langle \phi, \theta \rrangle = \int \text{d}^4x \, e \, \bar{\phi} ~\theta~,\\
\llangle \cdot, \cdot \rangle :\qquad \mf{S}_3 \times \mf{S}_1 \rightarrow \mathbb{C}~,\qquad \qquad \llangle \theta ,  \psi\rangle = \int \text{d}^4x \, e \, \bar{\theta} ~\psi~.
\end{align}
Here and in the remainder of this appendix $\phi,\psi \in \mf{S}_1$ and $\theta \in \mf{S}_3$ .

We define the adjoint of $\hat{H}$ to be the operator $\hat{H}^{\dagger}:\mf{S}_1 \rightarrow \mf{S}_3$ satisfying
\begin{align}
\llangle \hat{H}\psi,\phi \rangle =\langle  \psi, \hat{H}^{\dagger}\phi \rrangle \label{CSadjointA}
\end{align}
In particular this definition means that $\hat{H}^{\dagger}$ satisfies $\mb{D} \hat{H}^{\dagger} \phi = 3\hat{H}^{\dagger} \phi$ and $ K_{\b\bd}\hat{H}^{\dagger}\phi =0$.
We define the adjoint of $\hat{\mc{U}}$ to be the operator $\hat{\mc{U}}^{\dagger}:\mf{S}_3 \rightarrow \mf{S}_3$ satisfying
\begin{align}
\langle \hat{\mc{U}}\phi,\theta \rrangle =\langle  \phi, \hat{\mc{U}}^{\dagger}\theta \rrangle~.\label{CSadjointB}
\end{align}
According to this definition the operator $\hat{\mc{U}}^{\dagger}$ must satisfy  $\mb{D} \hat{\mc{U}}^{\dagger} \theta = 3\hat{\mc{U}}^{\dagger} \theta$ and $ K_{\b\bd}\hat{\mc{U}}^{\dagger}\theta =0$.

Unlike the supersymmetric case, discussed below, there is no ambiguity in the above definitions of the Hermitian adjoint of an operator.

\subsection{Chiral scalar superfield}

Let $\mf{S}_{(\Delta,q)}$ denote the space of primary scalar superfields with conformal weight $\Delta$ and $\sU(1)_R$ charge $q$. The subspace of chiral superfields with the same properties will be denoted  $\mf{S}^{\text{chiral}}_{(\Delta,q)}$, and in this case $\Delta$ and $q$ are related via $q=-\frac{2}{3}\Delta$. 
In the section \ref{masslesschiral} we encountered operators of the type $\hat{H}:\mf{S}^{\text{chiral}}_{(1,-\frac{2}{3})} \rightarrow \mf{S}_{(1,-\frac{2}{3})}$ and $\hat{\mc{U}}: \mf{S}^{\text{chiral}}_{(1,-\frac{2}{3})} \rightarrow \mf{S}^{\text{chiral}}_{(1,-\frac{2}{3})}$.
To compute the corresponding Hermitian adjoint we introduce the following three pairings
\begin{subequations}\label{Pairing}
\begin{align}
\langle \cdot,\cdot \rangle:&\qquad  \mf{S}^{\text{chiral}}_{(1,-\frac{2}{3})}\times \mf{S}^{\text{chiral}}_{(1,-\frac{2}{3})}\rightarrow \mathbb{C}~,\qquad\qquad  \langle \Psi,\F \rangle := \int \text{d}^{4|4}z\, E\, \bar{\Psi}\F~,\label{PairingA}\\
\langle \cdot,\cdot \rrangle:&\qquad  \mf{S}^{\text{chiral}}_{(1,-\frac{2}{3})}\times \mf{S}_{(1,-\frac{2}{3})}\rightarrow \mathbb{C}~,\qquad \qquad \langle \Psi, \Theta \rrangle := \int \text{d}^{4|4}z\, E\, \bar{\Psi}\Theta~,\label{PairingB}\\
\llangle \cdot,\cdot \rangle:&\qquad  \mf{S}_{(1,-\frac{2}{3})}\times \mf{S}^{\text{chiral}}_{(1,-\frac{2}{3})}\rightarrow \mathbb{C}~,\qquad\qquad  \llangle \Theta,\F \rangle := \int \text{d}^{4|4}z\, E\, \bar{\Theta}\F~.\label{PairingC}
\end{align}
\end{subequations}
Here and in the remainder of this appendix $\Psi,\Phi \in \mf{S}^{\text{chiral}}_{(1,-\frac{2}{3})}$ and $\Theta \in  \mf{S}_{(1,-\frac{2}{3})}$. 

For $\hat{H}$ we will encounter only one definition for its adjoint; as the operator $\hat{H}^{\dagger}$ satisfying 
\begin{subequations}\label{chiralAdjoint}
\begin{align}
\llangle \hat{H}\Psi,\F \rangle =\langle  \Psi, \hat{H}^{\dagger}\F \rrangle ~.\label{chiralAdjointA}
\end{align}
In particular, this means that $\hat{H}^{\dagger}: \mf{S}^{\text{chiral}}_{(1,-\frac{2}{3})} \rightarrow \mf{S}_{(1,-\frac{2}{3})}$. 
On the other-hand, we will encounter two sensible definitions for the adjoint of $\hat{\mc{U}}$; as the operator $\hat{\mc{U}}^{\dagger}: \mf{S}^{\text{chiral}}_{(1,-\frac{2}{3})} \rightarrow \mf{S}_{(1,-\frac{2}{3})} $ satisfying
\begin{align}
\langle  \hat{\mc{U}}\Psi,\F \rangle &=\langle  \Psi, \hat{\mc{U}}^{\dagger}\F \rrangle ~,\label{chiralAdjointB}
\end{align}
or as the operator $\hat{\mc{U}}^{\dagger}: \mf{S}_{(1,-\frac{2}{3})} \rightarrow \mf{S}_{(1,-\frac{2}{3})} $ satisfying
\begin{align}
\langle  \hat{\mc{U}}\Psi, \Theta \rrangle &=\langle  \Psi, \hat{\mc{U}}^{\dagger}\Theta \rrangle ~. \label{chiralAdjointC}
\end{align}
\end{subequations}
Definition \eqref{chiralAdjointA} proves to determine $\hat{H}^{\dagger}$ uniquely. However, as we will see below, in general neither of the definitions \eqref{chiralAdjointB} nor \eqref{chiralAdjointC} determine $\hat{\mc{U}}^{\dagger}$ uniquely (though they seem to be unique up to integration by parts, see below).

The adjoint must be defined carefully because there is an ambiguity associated with integration by parts when constrained superfields are involved, which we now illustrate via some simple examples. 
To begin, consider the first-order operator $\hat{H}:~\mf{S}^{\text{chiral}}_{(1,-\frac{2}{3})} \rightarrow \mf{S}_{(1,-\frac{2}{3})}$ where\footnote{This operator is proportional to $\hat{H}^{(s)}$ with $s=1$, see \eqref{ChiralHop}.} 
\begin{align}
\hat{H}=\ri H^{\a\ad}\nabla_{\a\ad}-\frac{1}{4}\bar{\nabla}_{\ad}H^{\a\ad}\nabla_{\a}+\frac{1}{12}\nabla_{\a}\bar{\nabla}_{\ad}H^{\a\ad}+\frac{\ri}{3}\nabla_{\a\ad}H^{\a\ad}~,
\end{align} 
and $H^{\a\ad}$ is an unconstrained real primary superfield with weight $-1$ and charge $0$. 
When computing its adjoint according to \eqref{chiralAdjointA}, upon integrating by parts naively one might find
\begin{align}
\hat{H}^{\dagger}=\ri H^{\a\ad}\nabla_{\a\ad}+a\bar{\nabla}_{\ad}H^{\a\ad}\nabla_{\a}+ b\nabla_{\a}\bar{\nabla}_{\ad}H^{\a\ad}+\frac{\ri}{3}\nabla_{\a\ad}H^{\a\ad}
\end{align}
 with $a=0$ and $b=-1/6$. However, these values of $a$ and $b$ do not yield a primary operator; $K_A\hat{H}^{\dagger}\Phi \neq 0$. Due to the freedom of integrating the $\nabla_{\a}$ by parts (it passes through $\bar{\Psi}$ since $\bar{\Psi}$ is anti-chiral), it appears that $a$ and $b$ can take any value satisfying $a+b=-1/6$. 
However, definition \eqref{chiralAdjointA} resolves this ambiguity, since it requires $\hat{H}^{\dagger}$ to map into $\mf{S}_{(1,-\frac{2}{3})}$. The condition $K_A\hat{H}^{\dagger}\Phi =0$ uniquely determines
 $a=-1/4$ and $b=1/12$, which in this case implies $\hat{H}^{\dagger}=\hat{H}$.
 
 Next, let us consider the first order operator $\hat{\mc{U}}:\mf{S}^{\text{chiral}}_{(1,-\frac{2}{3})} \rightarrow \mf{S}^{\text{chiral}}_{(1,-\frac{2}{3})} $ where
 \begin{align}
 \hat{\mc{U}}= 2\ri\bar{\nabla}^{\ad}\z^{\a}\nabla_{\a\ad}-\frac{1}{2}\bar{\nabla}^2\z^{\a}\nabla_{\a}+\frac{1}{6}\bar{\nabla}^2\nabla_{\b}\z^{\b}~, \label{AdjointExample00}
 \end{align}
 and $\z^{\a}$ is an unconstrained primary superfield with weight $-3/2$ and charge $1$. Computing the adjoint according to definition \eqref{chiralAdjointB} gives either of the following operators
 \begin{subequations}\label{AdjointExample2}
 \begin{align}
 \hat{\mc{U}}_1^{\dagger}&= -2\ri\nabla^{\a}\bar{\z}^{\ad}\nabla_{\a\ad}-\frac{1}{2}\nabla^{\a}\bar{\nabla}_{\bd}\bar{\z}^{\bd}\nabla_{\a}-\ri\nabla^{\a}{}_{\bd}\bar{\z}^{\bd}\nabla_{\a} -\frac{1}{6}\nabla^2\bar{\nabla}_{\bd}\bar{\z}^{\bd}-\ri\nabla^{\g}\nabla_{\g\bd}\bar{\z}^{\bd}~,\\
 \hat{\mc{U}}_2^{\dagger}&=-2\ri\bar{\z}^{\ad}\nabla_{\g\ad}\nabla^{\g}+\frac{1}{3}\bar{\nabla}_{\bd}\bar{\z}^{\bd}\nabla^2~,
 \end{align}
 \end{subequations}
 which are equivalent up to integration by parts (recall that both $\Psi$ and $\F$ in \eqref{chiralAdjointB} are chiral).\footnote{Consequently, for the adjoint one could have any linear combination $\hat{\mc{U}}^{\dagger}=a\hat{\mc{U}}_1^{\dagger}+b\hat{\mc{U}}_2^{\dagger}$ such that $a+b=1$.}
 Computing the adjoint with respect to \eqref{chiralAdjointC} gives
\begin{subequations}\label{AdjointExample3}
 \begin{align}
 \hat{\mc{U}}_1^{\dagger}&= -2\ri\nabla^{\a}\bar{\z}^{\ad}\nabla_{\a\ad}-\frac{1}{2}\nabla^{\a}\bar{\nabla}_{\bd}\bar{\z}^{\bd}\nabla_{\a}-\ri\nabla^{\a}{}_{\bd}\bar{\z}^{\bd}\nabla_{\a} -\frac{1}{6}\nabla^2\bar{\nabla}_{\bd}\bar{\z}^{\bd}-\ri\nabla^{\g}\nabla_{\g\bd}\bar{\z}^{\bd} \\
 &\phantom{....}+\frac{\ri}{4}\nabla^2\bar{\z}^{\ad}\bar{\nabla}_{\ad}~,\notag\\
 \hat{\mc{U}}_2^{\dagger}&=-2\ri\bar{\z}^{\ad}\nabla_{\g\ad}\nabla^{\g}+\frac{1}{3}\bar{\nabla}_{\bd}\bar{\z}^{\bd}\nabla^2-\frac{1}{2}\bar{\z}^{\ad}\nabla^2\bar{\nabla}_{\ad}~,
 \end{align}
 \end{subequations}
 which are also equivalent up to integration by parts. The only difference between  expressions \eqref{AdjointExample2} and \eqref{AdjointExample3} are terms which vanish on chiral superfields, which holds in the general case. 
 
 There seem to be several, albeit superficial, ways to address the non-uniqueness of the adjoint $\hat{\mc{U}}^{\dagger}$ associated with both definitions. First, one could further specify that $\hat{\mc{U}}^{\dagger}$ should be of the same functional type as $\hat{\mc{U}}$.
 Second, one could require that $\hat{\mc{U}}^{\dagger}$ should not contain any terms which, when acting on the matter field, vanish on the free equations of motion. Finally, for definition \eqref{chiralAdjointB}, one could require that upon imposing condition \eqref{Reality2} on $\z$, the resulting $\hat{\mc{U}}^{\dagger}$ should preserve chirality (which should be the case due to \eqref{ChiralrigidVar}). 
Each of these select $\hat{\mc{U}}_1^{\dagger}$ as the correct expression for the adjoint of \eqref{AdjointExample00} which, for our purposes, is the preferred one.


\section{Extraction of free (S)CHS gauge transformations} \label{Appendix B}
The purpose of this appendix is to analyse the free (i.e. lowest-order) sector of the formal gauge transformations
\begin{align}
\delta \hat{G} = \hat{G}\hat{\mc{V}}+\big(\hat{G}\hat{\mc{V}}\big)^{\dagger}~.\label{B1}
\end{align}
In particular, by honing in on the `physical' sub-sector 
of \eqref{B1}, we show that the `physical' (S)CHS fields, i.e. those with $l=0$ in \eqref{Hboxop} and \eqref{ChiralGop}, have the correct free gauge transformations plus non-linear and non-abelian corrections.

\subsection{Complex scalar field}

To isolate the lowest-order part of the physical sub-sector, in the LHS of of \eqref{B1} we insert 
\begin{align}
\hat{G} = \Box+\hat{H}~,\qquad \hat{H}=\sum_{s=0}^{\infty}\sum_{k=0}^{s}b_{(s,k)}\nabla^{s-k}_{\b\bd} h^{\a(k)\b(s-k)\ad(k)\bd(s-k)}\nabla_{\a\ad}^k
\end{align}
with $b_{(s,k)}=(-\ri)^s\binom{s}{k}\binom{s+k}{k}$, cf. \eqref{CScubicOp}. In the RHS of \eqref{B1} we insert $\hat{G}=\Box$ and 
\begin{align}
\hat{\mc{V}}=\sum_{s=0}^{\infty}\hat{\mc{U}}^{(s)}~,\qquad \hat{\mc{U}}^{(s)} = \sum_{k=0}^{s}a_{(s,k)}\nabla_{\bb}^{s-k} \z^{\a(k) \b(s-k) \ad(k) \bd(s-k)} \nabla_\aa^k~,
\end{align}
with $a_{(s,k)}=\ri^s \binom{s}{k}^2\binom{2s+2}{s-k}^{-1} $, cf. \eqref{CSOp}. Therefore, we arrive at the equation
\begin{align}
\delta_0 \hat{H} = \sum_{s=0}^{\infty}\Big(\big(\hat{\mc{U}}^{(s)}\big)^{\dagger}\Box+\Box\hat{\mc{U}}^{(s)}\Big)~,\label{B.5}
\end{align}
where we have denoted the variation corresponding to this truncation scheme by $\delta_0$.
If we split $\z^{\a(s)\ad(s)}$ into two real fields,  $\z^{\a(s)\ad(s)}= R^{\a(s)\ad(s)}+\ri I^{\a(s)\ad(s)}$, then from the first term in \eqref{RigidVarCS} we see that $R^{\a(s)\ad(s)}$ contributes to the variation of the auxiliary gauge fields. Therefore, to isolate the physical sub-sector, we restrict $\z^{\a(s)\ad(s)}$ to be pure imaginary,  $\z^{\a(s)\ad(s)}= \ri I^{\a(s)\ad(s)}$.

 In order to extract the variation of the `undressed' field $ h^{\a(s)\ad(s)}$ from $\delta_0 \hat{H}$, it proves useful to introduce the (non-primary) `dressed' field $g^{\a(s)\ad(s)}$ where\footnote{The relations \eqref{DressingMap} and \eqref{UndressingMap} are similar to the dressing and undressing (or reconstruction) maps of \cite{Segal}.}
 \begin{subequations}
\begin{align}
g^{\a(s)\ad(s)}=\sum_{k=0}^{\infty}b_{(s+k,s)}\nabla^{k}_{\b\bd} h^{\a(s)\b(k)\ad(s)\bd(k)}~, \label{DressingMap}
\end{align}and hence $\hat{H}=\sum_{s=0}^{\infty}g^{\a(s)\ad(s)}\nabla_{\a\ad}^s$. The field $g^{\a(s)\ad(s)}$ may be undressed according to 
\begin{align}
h^{\a(s)\ad(s)}= \sum_{k=0}^{\infty}d_{(s,k)}\nabla^{k}_{\b\bd} g^{\a(s)\b(k)\ad(s)\bd(k)}~,\qquad d_{(s,k)} = \frac{(-1)^k\binom{s+k}{k}^2\ri^s}{\binom{2s+k+1}{k}\binom{2s}{s}}~. \label{UndressingMap}
\end{align}
\end{subequations}
The constants $d_{(s,k)}$ are defined as the solutions to the system of equations
\begin{align}
\delta_{0,l} = \sum_{k=0}^{l}d_{(s,k)}b_{(s+l,s+k)}~,\qquad ~\text{ for all }~ l \in \mathbb{Z}_{\geq 0}~. \label{Orthogonality}
\end{align}

From \eqref{RigidVarCS} and \eqref{B.5} one may deduce that the variation of $g^{\a(s)\ad(s)}$ is
\begin{align}
\delta_0 g^{\a(s)\ad(s)} = \sum_{k=0}^{\infty}&a_{(s+k,s)}\nabla_{\b\bd}^k\Big[\ri \Box I^{\a(s)\b(k)\ad(s)\bd(k)} \non\\
&-\frac{s^2(2s+2k+2)(2s+2k+1)}{(s+k)^2(2s+k+1)(2s+k+2)}\nabla^{\a\ad}I^{\a(s-1)\b(k)\ad(s-1)\bd(k)}\Big]~. \label{B.8}
\end{align}
Miraculously, when translating this into a variation of the undressed fields via \eqref{UndressingMap},\footnote{Unfortunately, one cannot simply read off $\delta h $ by comparing  \eqref{B.8} with the variation of \eqref{DressingMap}. The proper way to proceed is via the undressing map \eqref{UndressingMap}. } all terms of the type $ \Box I$ in  \eqref{B.8} prove to cancel exactly. Then, after some algebra, one obtains the expression
\begin{align}
\delta_0 h^{\a(s)\ad(s)} =\sum_{l=0}^{\infty} (-1)^{s+l+1}\frac{l!s^2}{(s+l)^2}\binom{2s+2l}{s+l}^{-1}&\Big[\sum_{k=0}^{l}d_{(s,k)}b_{(s+l,s+k)}\Big] \non\\
&\times\nabla_{\b\bd}^l\nabla^{\a\ad}I^{\a(s-1)\b(l)\ad(s-1)\bd(l)}~.
\end{align}
Finally, by employing relation \eqref{Orthogonality}, we arrive at the intended result
\begin{align}
\delta h_{\a(s)\ad(s)}&= \nabla_{\a\ad}\ell_{\a(s-1)\ad(s-1)}+\mc{O}(h)~
\end{align}
with $\ell_{\a(s-1)\ad(s-1)}= (-1)^{s+1}\binom{2s}{s}^{-1}I_{\a(s-1)\ad(s-1)}$, which is equivalent to \eqref{CHSdeformedGTb}.
The notation $\mc{O}(h)$ indicates first-order terms involving $h_{\a(s)\ad(s)}$, other CHS fields $h_{\a(s')\ad(s')}$ with $s'\neq s$, and perhaps also the auxiliary ones $h^{[s,l]}_{\a(s-2l)\ad(s-2l)}$ with $0<l\leq \lfloor s/2 \rfloor$ and arbitrary $s$. In general the presence of such terms is guaranteed outside of the truncation scheme described above.

\subsection{Chiral scalar superfield}

To isolate the lowest order variation of the physical sub-sector for $s\geq 0
$, in the LHS of \eqref{B1} we insert $\hat{G} =1+\hat{H}$ with
\begin{align}
\hat{H}&=\sum_{s=0}^{\infty}\sum_{k=0}^{s}d_{(s,k)}\Big\{\Big(\nabla_{\b\bd}-\frac{\ri}{2}\frac{s-k}{s+1}\nabla_{\b}\bar{\nabla}_{\bd}\Big)\nabla_{\b\bd}^{s-k-1}H^{\a(k)\b(s-k)\ad(k)\bd(s-k)}\nabla_{\a\ad}^k\notag\\
&\phantom{=\sum_{k=0}^{s}d_{(s,k)}\Big\{\Big(\nabla_{\a\ad}}+\frac{\ri}{2}\frac{k}{s+1}\bar{\nabla}_{\bd}\nabla_{\b\bd}^{s-k}H^{\a(k)\b(s-k)\ad(k-1)\bd(s-k+1)}\nabla_{\a\ad}^{k-1}\nabla_{\a}\Big\}~, \label{E53}
\end{align}
 and $d_{(s,k)}=(-2\ri)^s\binom{s}{k}\binom{s+k+1}{k}$, cf. \eqref{ChiralHop}. In the RHS of \eqref{B1} we insert $\hat{G}=1$ and $\hat{\mc{V}}=\sum_{s=0}^{\infty}\hat{\mc{U}}^{(s)}$ with
\begin{subequations}
\begin{align}
\hat{\mc{U}}^{(s)} &= \sum_{k=1}^{s}a_{(s,k)}\Big\{ \nabla^{s-k}_{\b\bd}\bar{\nabla}^{\ad}\z^{\a(k)\b(s-k)\bd(s-k)\ad(k-1)}\nabla^k_{\a\ad}+\frac{\ri}{4}\bar{\nabla}^2\nabla^{s-k}_{\b\bd}\z^{\a(k)\b(s-k)\bd(s-k)\ad(k-1)}\nabla^{k-1}_{\a\ad}\nabla_{\a}\notag \\
&\phantom{=}-\frac{\ri}{4}\frac{k}{s+k+1}\bar{\nabla}^2\nabla_{\b}\nabla^{s-k}_{\b\bd}\z^{\a(k-1)\b(s-k+1)\bd(s-k)\ad(k-1)}\nabla^{k-1}_{\a\ad}\Big\}~,\qquad s\geq 1\\
\hat{\mc{U}}^{(0)} &= \bar{\nabla}^2\z
\end{align}
\end{subequations}
 and $a_{(s,k)}=(2\ri)^s\binom{s}{k}\binom{s-1}{k-1}\binom{2s+1}{s-k}^{-1}$, cf. \eqref{ChiralPrimalOp}.  Then, we arrive at the variational formula
\begin{align}
\delta_0 \hat{H} = \sum_{s=0}^{\infty}\Big(\hat{\mc{U}}^{(s)}+\big(\hat{\mc{U}}^{(s)}\big)^{\dagger}\Big)~,\label{B.35}
\end{align}
where $\big(\hat{\mc{U}}^{(s)}\big)^{\dagger}$ is given by \eqref{adjoint1},\footnote{In \eqref{B1} the definition \eqref{chiralAdjointA} is used to compute $(\hat{G}\hat{\mc{V}})^{\dagger}$, whilst for \eqref{adjoint1} definition \eqref{chiralAdjointB} is used. This is consistent since in this section we have set $\hat{G}=1$ in the RHS of \eqref{B1}. } and we have denoted the variation corresponding to this truncation scheme by $\delta_0$.

One can express the operator \eqref{E53} as  $\hat{H}=\sum_{s=0}^{\infty}G^{\a(s)\ad(s)}\nabla_{\a\ad}^s+\sum_{s=1}^{\infty}G^{\a(s)\ad(s-1)}\nabla_{\a\ad}^{s-1}\nabla_{\a}$ where the (non-primary) dressed superfields are defined according to
\begin{subequations}
\begin{align}
G^{\a(s)\ad(s)}&=\sum_{k=0}^{\infty}d_{(s+k,s)}\Big(\nabla_{\b\bd}-\frac{\ri}{2}\frac{k}{s+k+1}\nabla_{\b}\bar{\nabla}_{\bd}\Big)\nabla_{\b\bd}^{k-1}H^{\a(s)\b(k)\ad(s)\bd(k)}~,\\
G^{\a(s)\ad(s-1)}&= \frac{\ri}{2}\sum_{k=0}^{\infty}\frac{s}{s+k+1}d_{(s+k,s)}\bar{\nabla}_{\bd}\nabla_{\b\bd}^kH^{\a(s)\b(k)\ad(s-1)\bd(k+1)}~.
\end{align}
We note that $G^{\a(s)\ad(s-1)}$ is completely determined by $G^{\a(s)\ad(s)}$ through the relation\footnote{This is a consequence of $\hat{H}=\sum_{s=0}^{\infty}G^{\a(s)\ad(s)}\nabla_{\a\ad}^s+\sum_{s=1}^{\infty}G^{\a(s)\ad(s-1)}\nabla_{\a\ad}^{s-1}\nabla_{\a}$ having the properties \eqref{ChiralHprop} in addition to being self-adjoint.}
\begin{align}
G^{\a(s)\ad(s-1)} = \frac{\ri}{2}\frac{s}{s+1}\bar{\nabla}_{\bd}G^{\a(s)\ad(s-1)\bd}~.
\end{align}
\end{subequations}
The naked superfields $H^{\a(s)\ad(s)}$ may be recovered via the undressing map
\begin{subequations}\label{undressingSusy}
\begin{align}
H^{\a(s)\ad(s)}&=\sum_{k=0}^{\infty}f_{(s+k,s)}\Big(\nabla_{\b\bd}-\frac{\ri}{2}\frac{k}{s+k+1}\nabla_{\b}\bar{\nabla}_{\bd}\Big)\nabla_{\b\bd}^{k-1}G^{\a(s)\b(k)\ad(s)\bd(k)}~,\\
f_{(s,k)}&= (-1)^k\left(\frac{\ri}{2}\right)^{s}\frac{\binom{s+k}{k}^2}{\binom{2s+k+2}{k}\binom{2s+1}{s+1}}~.
\end{align}
\end{subequations}
 The constants $f_{(s,k)}$ are defined as the solutions to the system of equations
\begin{align}
\delta_{0,l} = \sum_{k=0}^{l}f_{(s,k)}d_{(s+l,s+k)}~,\qquad ~\text{ for all }~ l \in \mathbb{Z}_{\geq 0}~. \label{Orthogonality2}
\end{align}
These (un)dressing maps are valid for all $s\geq 0$.

Using \eqref{B.35}, for $s\geq 1$  one finds that 
\begin{align}
\delta_0G^{\a(s)\ad(s)}=\sum_{k=0}^{\infty}&f_{(s+k,k)}\Big(\nabla^{k}_{\b\bd}\bar{\nabla}^{\ad}\z^{\a(s)\b(k)\ad(s-1)\bd(k)}-\frac{k+1}{s+k+1}\nabla_{\b\bd}^{k}\nabla^{\a}\bar{\z}^{\a(s-1)\b(k)\ad(s)\bd(k)}\non\\
&-\frac{\ri}{4}\frac{k}{s}\bar{\nabla}^2\nabla_{\b}\nabla^{k-1}_{\b\bd}\z^{\a(s)\b(k)\ad(s)\bd(k-1)}+\frac{\ri}{4}\frac{k}{s}\nabla^2\bar{\nabla}_{\bd}\nabla_{\b\bd}^{k-1}\bar{\z}^{\a(s)\b(k-1)\ad(s)\bd(k)}\non\\
&-\frac{k(2s+k+2)}{s(s+k+1)}\nabla^{\gamma}\nabla_{\g\bd}\nabla_{\b\bd}^{k-1}\bar{\z}^{\a(s)\b(k-1)\ad(s)\bd(k)}\Big)~.
\end{align}
After converting this into a variation of the bare SCHS field via \eqref{undressingSusy} we find
\begin{align}
\delta_0 H^{\a(s)\ad(s)}&=\sum_{l=0}^{\infty}(-1)^{s+l}\binom{2s+l+1}{s+l+1}^{-1}\frac{s}{s+l}\Big[\sum_{k=0}^{l}f_{(s,k)}d_{(s+l,s+k)}\Big]\Big\{\nabla^{l}_{\b\bd}\bar{\nabla}^{\ad}\z^{\a(s)\b(l)\ad(s-1)\bd(l)}\non\\
&\phantom{=}-\frac{s+1}{s+l+1}\nabla_{\b\bd}^{l}\nabla^{\a}\bar{\z}^{\a(s-1)\b(l)\ad(s)\bd(l)}-\frac{l(s+1)}{s(s+l+1)}\nabla^{\gamma}\nabla_{\g\bd}\nabla_{\b\bd}^{l-1}\bar{\z}^{\a(s)\b(l-1)\ad(s)\bd(l)}\non\\
&\phantom{=}-\frac{\ri}{4}\frac{l}{s}\nabla_{\b}\bar{\nabla}^2\nabla^{l-1}_{\b\bd}\z^{\a(s)\b(l)\ad(s)\bd(l-1)}+\frac{\ri}{4}\frac{l}{s}\bar{\nabla}_{\bd}\nabla^2\nabla_{\b\bd}^{l-1}\bar{\z}^{\a(s)\b(l-1)\ad(s)\bd(l)}\Big\}~.
\end{align} 
Upon employing relation \eqref{Orthogonality2} we arrive at the intended result
\begin{align}
\delta H_{\a(s)\ad(s)}=\bar{\nabla}_{\ad}\L_{\a(s)\ad(s-1)}-\nabla_{\a}\bar{\L}_{\a(s-1)\ad(s)} + \mc{O}\big(H\big)
\end{align}
with $\L_{\a(s)\ad(s-1)}= (-1)^s\binom{2s+1}{s+1}^{-1}\z_{\a(s)\ad(s-1)}$, which is equivalent to \eqref{SCHSdeformedGT}.

Similarly, for $s=0$ one may show that
\begin{align}
\delta_0 G &= \bar{\nabla^2}\z + \nabla^2 \bar{\z} -\frac{\ri}{4}\sum_{s=1}^{\infty}\frac{1}{s+2}a_{(s,1)}\Big(\bar{\nabla}^2\nabla_{\b}\nabla_{\b\bd}^{s-1}\z^{\b(s)\bd(s-1)}-\nabla^2\bar{\nabla}_{\bd}\nabla_{\b\bd}^{s-1}\bar{\z}^{\b(s-1)\bd(s)}\non\\
&\phantom{=}-4\ri\frac{s+2}{s+1}\nabla^{\g}\nabla_{\g\bd}\nabla_{\b\bd}^{s-1}\bar{\z}^{\b(s-1)\bd(s)}\Big)~,\label{scalarvar}
\end{align}
In conjuction with \eqref{undressingSusy} this implies that 
\begin{align}
\delta H = \bar{\nabla}^2\z + \nabla^2\bar{\z} +\mc{O}\big(H\big)~,
\end{align}
where one should make use of the identity $\sum_{k=1}^{l}f_{(0,k)}d_{(l,k)}=-(-2\ri)^l$.


\section{Additional conformal supercurrents} \label{AppE}

In sections \ref{masslesschiral} and \ref{section4}, we studied the interactions of several matter multiplets with the SCHS gauge superfields $H_{\a(s)\ad(s)}$ and $\Psi_{\a(s)\ad(s-1)}$. At the cubic level, they enter the action by coupling to their dual Noether supercurrents. This appendix serves as a compilation of the latter for several important matter multiplets.

\subsection{The vector multiplet}
We recall from section \ref{section4.1.3} that the on-shell $\mathcal{N}=2$ vector multiplet is described in $\mathcal{N}=1$ superspace via two chiral superfields; the scalar $\F$ and spinor $W_\a$. Their equations of motion are given in \eqref{VMN=2EoM} and superconformal properties in \eqref{Phiprop} and \eqref{VMprop}. Our goal is then to construct all possible conformal supercurrents as higher derivative descendants of these matter superfields. We note however that the currents constructed solely in terms of $\F$ were introduced in sections \ref{masslesschiral} and \ref{section4}, thus we will exclude them here. 

First, from $W_\a$, and its conjugate $\bar{W}_\ad$, we may construct the following real descendants
\begin{align}
J_{\a(s) \ad(s)}(W,\bar{W}) &= \ri^{s+1} \sum_{k=0}^{s} (-1)^k \binom{s}{k} \Big \{ \binom{s}{k+1} \nabla_\aa^k W_\a \nabla_\aa^{s-k-1} \bar{W}_{\ad} \non \\
&\qquad \qquad \qquad \qquad + \frac{\ri}{2} \binom{s}{k+2} \nabla_{\aa}^k \nabla_\a W_\a \nabla_\aa^{s-k-2} \bar{\nabla}_\ad \bar{W}_\ad \Big \} ~, \quad s \geq 1~.
\end{align}
They may be shown to be primary off-shell, 
\begin{subequations}
\begin{align}
K_B J_{\a(s)\ad(s)}(W,\bar{W}) = 0~,
\end{align}
and conserved on-shell
\begin{align}
\nabla^{\b} J_{\b\a(s-1)\ad(s)}(W,\bar{W}) \approx 0 \quad \Longleftrightarrow \quad \bar{\nabla}^{\bd} J_{\a(s) \bd \ad(s-1)}(W,\bar{W}) \approx 0~,
\end{align}
\end{subequations}
thus they constitute conformal supercurrents. 

Next, from $W_\a$ and $\bar{\Phi}$, one may derive the fermionic descendants
\begin{align}
J_{\a(s) \ad(s-1)}(W,\bar{\F}) &= \ri^{s} \sum_{k=0}^{s-1} (-1)^k \binom{s-1}{k} \Big \{ \binom{s}{k+1} \nabla_\aa^k W_\a \nabla_\aa^{s-k-1} \bar{\F} \non \\
&\qquad \qquad \qquad \qquad + \frac{\ri}{2} \binom{s}{k+2} \nabla_{\aa}^k \nabla_\a W_\a \nabla_\aa^{s-k-2} \bar{\nabla}_\ad \bar{\F} \Big \} ~, \quad s \geq 1~.
\end{align}
They are primary off-shell,
\begin{subequations}
\begin{align}
K_B J_{\a(s)\ad(s-1)}(W,\bar{\F}) = 0~,
\end{align}
and conserved on-shell
\begin{align}
s > 1:& \qquad \qquad \nabla^{\b} J_{\b\a(s-1)\ad(s-1)}(W,\bar{\F}) \approx 0~,  \quad \quad \bar{\nabla}^{\bd} J_{\a(s) \bd \ad(s-1)}(W,\bar{\F}) \approx 0~,  \\
s= 1:& \qquad \qquad \qquad \qquad \;\; \nabla^{\b} J_{\b}(W,\bar{\F}) \approx 0~,  \quad \quad \bar{\nabla}^{2} J_{\a}(W,\bar{\F}) \approx 0 ~, 
\end{align}
\end{subequations}
Hence, they are also conformal supercurrents.

\subsection{Massless hypermultiplet} 

In the main body it was mentioned that the hypermultiplet may be realised in $\mathcal{N}=1$ superspace either in terms of two chiral superfields, see section \ref{GMHM1}, or as a chiral and complex linear superfield, see section \ref{GMHM2}. In principle one may also consider its realisation in terms of two complex linear superfields, but we exclude this case from consideration.

First, we will work with two chiral superfields $\Phi_{\pm}$. From them we may construct two real higher-derivative descendants
\begin{subequations}
\begin{align}
J^{(S)}_{\a(s) \ad(s)}(\F_{\pm},\bar{\F}_{\pm}) &= \ri^s \sum_{k=0}^{s} (-1)^k \binom{s}{k}^2 \Big \{ \nabla_\aa^k \F_{+} \nabla_\aa^{s-k} \bar{\F}_{-} - \frac{\ri}{2} \frac{s-k}{k+1} \nabla_\aa^k \nabla_\a \F_{+} \nabla_\aa^{s-k-1} \bar{\nabla}_\ad \bar{\F}_{-} \non \\
&\qquad + \nabla_\aa^k \F_{-} \nabla_\aa^{s-k} \bar{\F}_{+} - \frac{\ri}{2} \frac{s-k}{k+1} \nabla_\aa^k \nabla_\a \F_{-} \nabla_\aa^{s-k-1} \bar{\nabla}_\ad \bar{\F}_{+} \Big \} ~, \quad s \geq 0~, \\
J^{(A)}_{\a(s) \ad(s)}(\F_{\pm},\bar{\F}_{\pm}) &= \ri^{s-1} \sum_{k=0}^{s} (-1)^k \binom{s}{k}^2 \Big \{ \nabla_\aa^k \F_{+} \nabla_\aa^{s-k} \bar{\F}_{-} - \frac{\ri}{2} \frac{s-k}{k+1} \nabla_\aa^k \nabla_\a \F_{+} \nabla_\aa^{s-k-1} \bar{\nabla}_\ad \bar{\F}_{-} \non \\
&\qquad - \nabla_\aa^k \F_{-} \nabla_\aa^{s-k} \bar{\F}_{+} + \frac{\ri}{2} \frac{s-k}{k+1} \nabla_\aa^k \nabla_\a \F_{-} \nabla_\aa^{s-k-1} \bar{\nabla}_\ad \bar{\F}_{+} \Big \} ~, \quad s \geq 0~.
\end{align}
\end{subequations}
It is clear that the (latter) former are (anti-)symmetric under the interchange $\Phi_{+} \leftrightarrow \Phi_{-}$. They were derived for the first time in \cite{BHK18}. Further, they are primary off-shell 
\begin{subequations}
\label{E.6}
\begin{align}
K_B J^{(S)}_{\a(s)\ad(s)}(\F_\pm,\bar{\F}_\pm) = 0~,
\end{align}
and conserved on-shell
\begin{align}
s \geq 1:& \qquad \qquad \nabla^{\b} J^{(S)}_{\b\a(s-1)\ad(s)}(\Phi_\pm,\bar{\Phi}_\pm) \approx 0 \quad \Longleftrightarrow \quad \bar{\nabla}^{\bd} J^{(S)}_{\a(s) \bd \ad(s-1)}(\Phi_\pm,\bar{\Phi}_\pm) \approx 0~,  \\
s = 0:& \qquad \qquad \qquad \quad \nabla^{2} J^{(S)}(\Phi_\pm,\bar{\Phi}_\pm) \approx 0 \quad \Longleftrightarrow \quad \bar{\nabla}^{2} J^{(S)}(\Phi_\pm,\bar{\Phi}_\pm) \approx 0 ~.
\end{align}
\end{subequations}
It should be noted that the \eqref{E.6} also holds when $J^{(S)}$ is replaced by $J^{(A)}$.

Next, from the on-shell chiral superfield $\F$ and complex linear superfield $\G$, it is possible to construct two families of descendants. The first are bosonic and complex,
\begin{align}
J_{\a(s) \ad(s)}(\Phi,\G) &=  \sum_{k=0}^{s} (-1)^k \binom{s}{k}^2 \Big \{ \nabla_\aa^k \F \nabla_\aa^{s-k} \G - \frac{\ri}{2} \frac{s-k}{k+1} \nabla_\aa^k \nabla_\a \F \nabla_\aa^{s-k-1} \bar{\nabla}_\ad \G \non \\
&\qquad \qquad \qquad \qquad \qquad - \frac{\ri}{2} \frac{s-k}{s-k+1} \nabla_{\aa}^k \Phi \nabla_\aa^{s-k-1} \bar{\nabla}_\ad \nabla_\a \G \Big \} ~, \quad s \geq 0~,
\end{align}
while the second are fermionic,
\begin{align}
J_{\a(s) \ad(s-1)}(\Phi,\bar{\G}) &= \sum_{k=0}^{s-1} (-1)^k \binom{s}{k} \binom{s}{k+1} \Big \{ \nabla_\aa^k \nabla_\a \F \nabla_\aa^{s-k-1} \bar{\G} - (k+1) \nabla_\aa^k \F \nabla_\aa^{s-k-1} \nabla_\a \bar{\G} \non \\
& \qquad \qquad \qquad \qquad + \frac{\ri}{2} (s-k-1) \nabla_{\aa}^k \nabla_\a \Phi \nabla_\aa^{s-k-2} \bar{\nabla}_\ad \nabla_\a \bar{\G} \Big \} ~, \quad s \geq 1~,
\end{align}
which are primary off-shell
\begin{subequations}
\begin{align}
K_B J_{\a(s)\ad(s)}(\F, \G) = 0 ~, \qquad K_B J_{\a(s)\ad(s-1)}(\F, \bar{\G}) = 0~.
\end{align}
The bosonic currents satisfy the conservation equations
\begin{align}
s \geq 1:& \qquad \qquad \nabla^{\b} J_{\b\a(s-1)\ad(s)}(\F,\G) \approx 0~,  \quad \quad \bar{\nabla}^{\bd} J_{\a(s) \bd \ad(s-1)}(\F,\G) = 0 ~,  \\
s = 0:& \qquad \qquad \qquad \qquad \, \nabla^{2} J(\Phi,\G) \approx 0~,  \quad \quad \bar{\nabla}^{2} J(\Phi,\G) = 0 ~.
\end{align}
Similarly, the fermionic currents are also conserved
\begin{align}
s > 1:& \qquad \qquad \nabla^{\b} J_{\b\a(s-1)\ad(s-1)}(\F,\bar{\G}) \approx 0~,  \quad \quad \bar{\nabla}^{\bd} J_{\a(s) \bd \ad(s-2)}(\F,\bar{\G}) \approx 0 ~,  \\
s = 1:& \qquad \qquad \qquad \qquad \;\; \nabla^{\a} J_\a(\Phi,\bar{\G}) \approx 0~,  \quad \quad \bar{\nabla}^{2} J_\a(\Phi,\bar{\G}) \approx 0 ~.
\end{align}
\end{subequations}
Hence, they constitute conserved supercurrents.

Additionally, we construct the following real higher-derivative bilinears of $\G$ and $\bar{\G}$
\begin{align}
\label{GGSC}
J_{\a(s) \ad(s)}(\G,\bar{\G}) &= \ri^{s} \sum_{k=0}^{s} (-1)^k \binom{s}{k} \binom{s+1}{k+1} \Big \{\nabla_\aa^k \bar{\G} \nabla_\aa^{s-k} \G - \frac{\ri}{2} (s-k) \Big[ \nabla_\aa^k \nabla_\a \bar{\G} \nabla_\aa^{s-k-1} \bar{\nabla}_\ad \G \non \\
& \qquad \qquad - \nabla_\aa^k \bar{\nabla}_\ad \bar{\G} \nabla_\aa^{s-k-1} \nabla_\a \G  + \frac{k+1}{s-k+1} \nabla_\aa^k \bar{\G} \nabla_\aa^{s-k-1} \bar{\nabla}_\ad \nabla_\a \G \non \\
& \qquad \qquad + \frac{s-k}{k+2} \nabla_\aa^k \bar{\nabla}_\ad \nabla_\a \bar{\G} \nabla_\aa^{s-k-1} \G \Big] \Big \} ~, \qquad \qquad \qquad \qquad s \geq 0~,
\end{align}
which are also conformal supercurrents as they are primary off-shell
\begin{subequations}
\begin{align}
K_B J_{\a(s)\ad(s)}(\G,\bar{\G}) = 0 ~,
\end{align}
and conserved on-shell
\begin{align}
s \geq 1:& \qquad \qquad \nabla^{\b} J_{\b\a(s-1)\ad(s)}(\G,\bar{\G}) \approx 0 \quad \Longleftrightarrow \quad \bar{\nabla}^{\bd} J_{\a(s) \bd \ad(s-1)}(\G,\bar{\G}) \approx 0 ~,  \\
s = 0:& \qquad \qquad \qquad \qquad \nabla^{2} J(\G,\bar{\G}) \approx 0 \quad \Longleftrightarrow \quad \bar{\nabla}^{2} J(\G,\bar{\G}) \approx 0 ~.
\end{align}
\end{subequations}
It should be noted that in Minkowski superspace the supercurrent multiplets \eqref{GGSC} appeared for the first time in \cite{Koutrolikos2}.


\begin{footnotesize}

\end{footnotesize}

\end{document}